\documentclass{aa}
\pdfoutput=1 
\usepackage[T1]{fontenc} 
\usepackage{color}

\newcommand{\vx}{{\bf x}}

\newcommand{\vr}{{\bf r}}
\newcommand{\vn}{{\bf n}}

\newcommand{\vk}{{\bf k}}
\newcommand{\ii}{{\rm i}}
\newcommand{\dd}{{\rm d}}

\newcommand{\xib}{\overline{\xi}}
\newcommand{\xis}{\overline{\xi}_{s}}

\newcommand{\hrho}{\hat{\rho}}
\newcommand{\rhob}{\overline{\rho}}

\newcommand{\hlambda}{\hat{\lambda}}

\newcommand{\trees}{{\rm trees}}

\newcommand{\lines}{{\rm lines}}
\newcommand{\vertices}{{\rm vertices}}

\newcommand{\RL}{{\rm RL}}
\newcommand{\mF}{{\cal F}}
\newcommand{\mM}{{\cal M}}
\newcommand{\mN}{{\cal N}}
\newcommand{\mP}{{\cal P}}
\newcommand{\mA}{{\cal A}}
\newcommand{\mB}{{\cal B}}

\newcommand{\Nb}{\overline{N}}

\newcommand{\Cov}{{\rm Cov}}
\newcommand{\Poiss}{{\rm Poisson}}
\newcommand{\halo}{{\rm halo}}
\newcommand{\model}{{\rm model}}
\newcommand{\sign}{{\rm sign}}
\newcommand{\sym}{{\rm sym.}}

\newcommand{\shortd}{{\rm short\ dist.}}
\newcommand{\PBC}{{\rm PBC}}
\newcommand{\zRL}{{\zeta_{\rm RL}}}
\newcommand{\rhom}{{\rho_{\rm m}}}
\newcommand{\drho}{{\delta_{\!\rho}}}
\newcommand{\Dirac}{{\delta_{\rm Dirac}}}
\newcommand{\ex}{{\rm ex}}

\definecolor{Red}{rgb}{0.65,0.08,0.05}

\begin{document}

\title{Covariances of density probability distribution functions. Lessons from hierarchical models}

\author{Francis Bernardeau$^{1,2}$}
\institute{
$^1$Universit\'e Paris-Saclay, CNRS, CEA, Institut de physique th\'eorique, 91191, Gif-sur-Yvette, France\\
$^{2}$Institut d'Astrophysique de Paris, CNRS and Sorbonne Universit\'e, UMR 7095, 98 bis bd Arago, 75014 Paris, France}


 \abstract{Statistical properties of the cosmic density fields are to a large extent encoded in the shape of the one-point density probability distribution functions (PDF) as measured in surveys. In order to successfully exploit such observables, a detailed functional form of the covariance matrix of the one-point PDF is needed.}{The objectives are to model the properties of this covariance for general stochastic density fields and for stochastic fields that reproduce the properties expected in cosmology. The accuracy of the proposed forms is evaluated in specific cases.}{The study was conducted in a cosmological context and determined whether the density is defined absolutely or relatively to the sample mean density. Leading and subleading contributions were identified within a large class of models, the so-called hierarchical models. They come from either large or short separation contributions. The validity of the proposed forms for the covariance matrix was assessed with the help of a toy model, the minimum tree model, for which a corpus of exact results could be obtained (forms of the one- and two-point PDF, large-scale density-bias functions, and full covariance matrix of the one-point PDF).}{It is first shown that the covariance matrix elements are directly related to the spatial average of the two-point density PDF within the sample. The dominant contribution to this average is explicitly given for hierarchical models (coming from large scale contribution), which leads to the construction of specific density-bias functions. However, this contribution alone cannot be used to construct an operational likelihood function. Subdominant large-scale effects are found to provide corrective terms, but also a priori lead to limited information on the covariance matrix. Short distance effects are found to be more important but more difficult to derive as they depend more on the details of the model. However, a simple and generic form of these contributions is proposed. Detailed comparisons in the context of the {Rayleigh-Levy flight model} show that  the large-scale effects capture the bulk of the supersample effects and that, by adding the short-distance contributions, a qualitatively correct model of the likelihood function can be obtained.}{}

\date{\today}
\maketitle

\section{Introduction}

In the context of cosmological studies, the concept of counts-in-cells statistics has been put forward for a long time as a unique way to quantify the statistical properties of the cosmological fields \citep{1979MNRAS.186..145W,1995ApJS...96..401C,1989A&A...220....1B,1999A&A...349..697B}. It was then shown in particular that counts-in-cells statistics, which represents a discrete representation of the local density probability distribution function (PDF), could be directly related to the correlation hierarchy of the density field.

Interest in these types of observables was recently renewed for several reasons. The size of the surveys makes accurately measuring these quantities more realistic. This is already the case for surveys such as the Dark Energy Survey \citep[DES collaboration;][]{2018PhRvD..98d3526A}, the Kilo-Degree Survey \citep[KIDS;][]{2021A&A...646A.140H}, and the Hyper Suprime Cam \citep[HSC;][]{2019PASJ...71...43H}. The future promises even larger and more powerful surveys such as Euclid \citep{2011arXiv1110.3193L,2018LRR....21....2A} and the Rubin Observatory \citep{2019ApJ...873..111I}. Moreover, the theoretical foundations for these constructions (at least in the cosmological context) has been considerably strengthen with the realization that the large-deviation theory \citep[LDT; for a general review, see][]{2011arXiv1106.4146T} could successfully be invoked, as shown in \cite{2016PhRvD..94f3520B}. It clarifies the applicability of the theory to the cosmological density field and places previous works on a much more solid foundation \citep{2002A&A...382..412V,2014PhRvD..90j3519B}. The ability of density PDF to constrain cosmology was emphasized in \cite{2016MNRAS.460.1549C} and completed in \cite{2020MNRAS.498..464F} and in \cite{2020MNRAS.495.4006U}, who showed that these observable could efficiently constrain the neutrino mass or primordial non-Gaussianities.  Finally, although the matter PDF is not a direct observable, as is matter density, it can be closely approached with the help of luminous tracer statistics (\cite{2020MNRAS.498L.125R}), more convincingly in weak-lensing fields, as advocated in numerous recent papers \citep{2021MNRAS.503.5204B,2000A&A...364....1B}, or with combined approaches such as density-split statistics \citep{2018PhRvD..98b3507G,2018PhRvD..98b3508F,2018MNRAS.481.5189B}, which proved to be particularly promising.

The construction of a full theory of these observable requires a detailed analysis of its global error budget, however, due to finite-size surveys, imperfect tracers, and so on. Some of these aspects have been explored in early studies such as
\citet{1996ApJ...470..131S} and \citet{1999MNRAS.310..428S}, but a full theory is still lacking. The developments presented in this paper are made in this context. More precisely, the purpose of this study is to explore what determines the expression of the covariance of data vectors whose elements are local quantities, such as the density contrast and density profiles, in cosmological contexts, that is, in classical random fields with long-range correlations. Derivations were made furthermore assuming statistical homogeneity and isotropy. The domain of application encompasses both counts-in-cells statistics, basically 2D or 3D counts of density tracers, or proxies to projected densities such as mass maps for weak-lensing tomographic observations. 

In order to gain insights into the different contributions and the effects that might contribute to the covariance, we rely on the
use of the hierarchical models to derive results we think rather general. The immense advantage of using such models is that they naturally incorporate many of the features expected in density cosmological fields (e.g., the magnitude of the high-order correlation functions), and there are also models for which many exact results can be obtained in particular for counts-in-cells statistics. 
The goal of these constructions is to eventually infer precisely what the performance of PDF measurements would be on the determination of cosmological parameters, taking advantage of results such as those found in \cite{2021MNRAS.505.2886B}, which give the response function of these observable to various cosmological parameters

Section 2 is devoted to the presentation of the  general framework. The subsequent section explores different contributions, from large-scale effects with the derivation of several bias functions to short-distance contributions. Results are derived in a framework as general as possible, including discrete noise associated with the use of a finite number of tracers. Section 4 presents the general hierarchical models, and more specifically, the Rayleigh-Levy flight model that we use as a toy model to evaluate the performances of approximate schemes. In Section 5, simplified models for the covariance matrix are presented and evaluated with the help of a set of numerical experiments. Section 6 summarizes the results that have be found and specifies their expected range of application. 

The text is complemented by appendices that contain a large amount of material. They present the hierarchical models, their mathematical description, and the mean-field approximation that is used throughout for explicit computations. Appendix C is more specifically devoted to the minimal tree model and the construction of the exact mean-field covariance matrix.

\section{General framework. Construction of covariance matrices}

The purpose of this section is to show how the elements of the covariance matrix are related to the joint density PDFs within a given survey. We first formalize this relation in a general framework before we explore its consequence in the context we are interested in. 
We assume we are interested in the 
PDF of some local quantity, $\mu$, that can be evaluated within a survey, thus defining a field $\mu(\vx)$ throughout the survey. The a priori typical example of this quantity is the density (see below for a more precise illustration of what this quantity could be).
The value of $\mu$ is assumed to lie in some ensemble $\mM$ (that can be simply the real numbers), and the data vector we are interested in
consists of the probabilities $p_{i}$ that $\mu$ lie within the subsets $\mM_{i}$ (which are a priori nonzero within $\mM$). The one-point PDF of $\mu$ is then given by
\begin{equation}
p_{i}(\vx)=\int_{\mM_{i}} \dd \mu\ \mP(\mu,\vx),
\end{equation}
if $\mP(\mu,\vx)\dd\mu$ is the PDF of $\mu$ at location $\vx$. $p_{i}(\vx)$ is then assumed
to be independent  of $\vx$ in the context we are interested in, for which statistical homogeneity is assumed.
More formally, we can define the characteristic function $\chi_{i}(\vx),$ which takes the value $1$ where $\mu(\vx)\in\mM_{i}$ and $0$ otherwise. 

An estimation of $p_{i}$ would then be given by the volume fraction of the survey where $\mu(\vx)\in\mM_{i}$. We note this estimate as $P_{i}$
\footnote{This is an ideal estimate in the sense that $\mu$ is evaluated in an infinite number of locations. We therefore neglect here the impact of measuring $\mu$ on a finite number of locations when evaluating $P_{i}$. Regarding this aspect, a specific derivation that takes a finite number of measurements into account can be found in \citet{2016MNRAS.460.1598C}
}, 
\begin{equation}
P_{i}  =\frac{1}{V} \int\dd \vx\ \chi_{i}(\vx),
\end{equation}
which is then itself a random variable, the properties of which we are interested in. More precisely, we would like to derive an operational form for the likelihood function of a set of $P_{i}$ variable. We limit our investigation here to the construction of the likelihood
from the covariance matrix, assuming that the likelihood of $P_{i}$ is close enough to a Gaussian distribution\footnote{Whether this is a correct assumption is difficult to assess in general. It probably depends on the detailed properties of the setting. The Conclusion section contains further comments on this aspect.}.

The ensemble average of $P_{i}$ is 
\begin{equation}
\langle P_{i} \rangle =\frac{1}{V} \int\dd\vx\ \langle \chi_{i}(\vx)\rangle = \frac{1}{V} \int\dd\vx\ p_{i}(\vx)=p_{i.}
\end{equation}
We can further define a joint PDF of the same field, $\mP(\mu,\vx;\mu',\vx'),$ which is the joint PDF of $\mu$ and $\mu'$ 
in locations $\vx$ and $\vx'$. Defining
$p_{ij}(\vx,\vx')$ as the joint ensemble average of  $\mP(\mu,\vx;\mu',\vx'),$ we have
\begin{equation}
p_{ij}(\vx,\vx')=\int_{\mM_{i}} \dd \mu \int_{\mM_{j}} \dd \mu'\ \mP(\mu,\vx;\mu',\vx').
\end{equation}
The elements of the covariance matrix of $P_{i}$ are then formally 
\begin{eqnarray}
\langle P_{i} P_{j} \rangle &=& \frac{1}{V^{2}} \int\dd\vx \int\dd\vx'\ \langle \chi_{i}(\vx)\chi_{j}(\vx')\rangle \nonumber \\
&=& \frac{1}{V^{2}} \int\dd\vx \int\dd\vx'\ p_{ij}(\vx,\vx')\equiv \overline{p}_{ij}\label{basicrelation}.
\end{eqnarray}
This gives the relation between the covariances and joint PDF. If $p_{ij}(\vx,\vx')$ depends only on the relative distance
between $\vx$ and $\vx'$, this expression can be recast in terms of the distribution of such distances, $P_{s}(r_d)$, in the form
\begin{eqnarray}
\langle P_{i} P_{j} \rangle &=&\int\dd r_d\,P_{s}(r_d) p_{ij}(r_d).
\label{keyrelationCov}
\end{eqnarray}
The precise form of $P_{s}(r_d)$ depends on the detail of the survey. Explicit forms can be given in case of simple regular surveys such as square surveys\footnote{For a square survey of unit size (with nonperiodic boundary conditions), the distance 
distribution function $P_{s}(r_{d})$ is given by
\begin{equation}
P_{s}(r_{d})=
\begin{cases}
 2 r_{d} ((r_{d}-4) r_{d}+\pi ) & 0<r_{d}<1 \\
 -2 r_{d} \Big(2_{d}+r_{d}^2-4 \sqrt{r_{d}^2-1}  & \nonumber\\
 \hspace{.1cm}-2 \sec ^{-1}\left(\frac{r_{d}}{\sqrt{r_{d}^2-1}}\right)+2 \sec ^{-1}(r_{d})\Big) & 1<r_{d}<\sqrt{2}
\end{cases},
\label{PdExpression}
\end{equation}
as can be obtained after integrating over three of four of the position coordinates.
}.
In the context of statistically homogeneous and isotropic random fields, this latter expression is used. In particular, we wish to determine the configurations that contribute most to $\overline{p}_{ij}$. They obviously depend on both the random processes we consider and on the definition of $\mM_{i}$ and $\mM_{j}$. In order to be more specific, we assume in the following that $\mu$ is a local density assigned to be in bins $(i)$ centered on $\rho_{i}$ and with width $\dd\rho_{i}$ (assumed a priori to be arbitrarily small),  so that 
\begin{equation}
P_{i}=P(\rho_{i})\dd\rho_{i}.
\end{equation}
If necessary, local densities could be obtained after the field $\mu(\vx)$ has been convolved with a window function $W_{R}(\vx)$, associated with a scale $R$ that is
\begin{equation}
\rho(\vx)=\int\dd\vx'\,\mu(\vx-\vx')W_{R}(\vx').
\end{equation}
It is then assumed $R$ is small compared to the sample size in order to identify what the leading contributions to 
the joint PDFs might be. In practice, $W_{R}$ might also be a simple top-hat window function, but this is not necessarily the case. It could be more elaborated filters, such as compensated filters (of zero average), such as those introduced for cosmic shear analysis \citep{1996MNRAS.283..837S,1998ApJ...498...26K,2000A&A...364....1B}.

We furthermore allow the estimated densities $\rho_{i}$ to be defined with respect the overall density of the sample $\rho_{s}$,
\begin{equation}
\rho_{s}=\frac{1}{V}\int\dd\vx\ \mu(\vx).
\end{equation}
For instance, we could be interested in $\hrho_{i}\equiv\rho_{i}/\rho_{s}$ or $\rhob_{i}\equiv\rho_{i}-(\rho_s-1),$ which are frequently encountered situations in praxis. Then $\rho_{s}$ is itself a random variable whose correlation with $\rho(\vx)$ ought to be taken into account. We then need to explore the properties of either $P(\rho_{i},\rho_{j};\vx,\vx')$ or $P(\rho_{s},\rho_{i},\rho_{j};\vx,\vx')$ from which functions of interest can be built, that is,
\begin{eqnarray}
P(\hrho,\hrho')&=&\int\dd\rho_{s}\,\rho_{s}^{2}\,P(\rho_{s},\hrho\rho_{s},\hrho'\rho_{s};\vx,\vx')\\
P(\rhob,\rhob')&=&\int\dd\rho_{s}\,P(\rho_{s},\rhob_{i}-1+\rho',\rhob'-1+\rho_{s};\vx,\vx'),
\end{eqnarray}
from which the covariance elements such as 
\begin{eqnarray}
\Cov(\rho_{i},\rho_{j})\dd\rho_{i}\dd\rho_{j}&=&
\int\dd r_{d}\,P_{s}(r_d) P(\rho_{i},\rho_{j};r_d)\dd\rho_{i}\dd\rho_{j}\nonumber\\
&&-P(\rho_{i})P(\rho_{j})\dd\rho_{i}\dd\rho_{j}
\end{eqnarray}
can be derived and whose properties we wish to explore. We wish in particular to build a model of the likelihood function from such a 
covariance, requiring full knowledge of its eigenvalues and eigendirections. In this respect, it is implicit that the number 
of bins $(i)$ to be
used is finite. We nonetheless present at least in this first section the results in the continuous limit for $\rho_{i}$.
It is finally to be noted that as stated before, we restrict our analysis to covariance matrices, but higher-order correllators might also be considered by generalizing the relation (\ref{basicrelation}) to a higher number of variables.  

\section{PDF covariances in the context of  cosmological models}
\label{sec:PDFcovariances}

\subsection{Modeling the joint PDF}

To make progress, we need to make further assumptions about the mathematical structure of the joint PDF. In the following, we assume in particular that joint PDFs can be obtained from their cumulant generating functions (CGF)\footnote{This is not necessarily so, as exemplified in
\citet{2011ApJ...738...86C,2012ApJ...750...28C}.}, $\varphi(\lambda_{i},\lambda_{j};r_{d})$. The latter is defined as
\begin{eqnarray}
\exp\left(\varphi(\lambda_{i},\lambda_{j};r_{d})\right)&=&\langle \exp\left({\lambda_{i}\rho_{i}+\lambda_{j}\rho_{j}}\right)\rangle\nonumber\\
&&\hspace{-2.5cm}
=\int \dd \rho_{i}\dd\rho_{j}\,P(\rho_{i},\rho_{j};r_d)\,\exp\left({\lambda_{i}\rho_{i}+\lambda_{j}\rho_{j}}\right),
\end{eqnarray}
and it is assumed that this relation can be inverted to give the joint PDF from Laplace inverse transformations,
\begin{equation}
P(\rho_{i},\rho_{j};r_{d})=\int\frac{\dd\lambda_{i}}{2\pi\ii}\frac{\dd\lambda_{j}}{2\pi\ii}e^{-\lambda_{i}\rho_{i}-\lambda_{j}\rho_{j}+\varphi(\lambda_{i},\lambda_{j};r_{d})},
\end{equation}
where the integrations are made a priori along the imaginary axis.
The CGFs themselves are closely related to the averaged correlation functions of the underlying field. In the following, we develop models for which these correlation functions can be computed precisely.

\subsection{Large-distance contributions to the joint density PDF}

We start by assuming that covariances are dominated by long-range correlation and not by proximity effects (e.g., densities taken in nearby regions). Whether this assumption is correct obviously depends on the particular model and setting we consider, as we detail below.
There are large sets of models for which general expressions can be given in this regime. They are the so-called hierarchical models, originally introduced in \citet{1980lssu.book.....P}, discussed in more detail in
\citet{1984ApJ...279..499F,1984ApJ...277L...5F,1989A&A...220....1B,1992A&A...255....1B}, and further formalized in \citet{1999A&A...349..697B} as described below; it is also true in the quasilinear regime as originally pointed out in \citet{1996A&A...312...11B} and derived in more detail in \cite{2016MNRAS.460.1598C}. In these regimes, we obtain the following
functional form (see previous references and the detailed derivation in Appendix B):
\begin{eqnarray}
\varphi(\lambda_{s},\lambda_{i},\lambda_{j})&=&
\lambda_{s}+\varphi_{0}(\lambda_{i})+\varphi_{0}(\lambda_{j})
\nonumber\\
&&
\hspace{-2.5cm}+\frac{\lambda_{s}^{2}}{2}\int\dd\vx_{s}\dd\vx'_{s}\,\xi(\vx_{s},\vx'_{s})
+\lambda_{s}\int\dd\vx_{s}\,\xi(\vx_{s},\vx_{1})\,\varphi_{1}(\lambda_{i})
\nonumber\\
&&\hspace{-2.5cm}+\lambda_{s}\int\dd\vx_{s}\,\xi(\vx_{s},\vx_{2})\,\varphi_{1}(\lambda_{j})
+\varphi_{1}(\lambda_{i})\,\xi(\vx_{1},\vx_{2})\,\varphi_{1}(\lambda_{j})\label{gen3ptbias}
,\end{eqnarray}
where 
$\xi(\vx,\vx')$ is the two-point correlation function of the density field at positions $\vx$ and $\vx'$ , and $\varphi_{0}(\lambda)$
and $\varphi_{1}(\lambda)$ are specific functions of $\lambda$ that depend on the details of the model.

Then, setting $\lambda_{s}$ to zero, we can easily obtain the expression of the joint PDF at leading order in $\xi(r_{d})$,
\begin{equation}
P(\rho_{i},\rho_{j};r_{d})=P(\rho_{i})P(\rho_{j})\left(1+b(\rho_{i})\xi(r_d)\,b(\rho_{j})\right).\label{jointpdflinear}
\end{equation}
Here $P(\rho_{i})$ is the one-point density PDF (i.e., implicitly at scale $R$), and $b(\rho_{i})$ is the density-bias function (to be distinguished from the standard halo-bias function). It also depends on $\rho_{i}$ (and on the scale $R$) so that in the previous expression, the dependence on $\rho_{i}$, $\rho_{j}$ , and $r_{d}$ can be factorized out.

To be more precise, $P(\rho_{i})$ is given by the inverse Laplace transform of the CGF (see, e.g., \cite{1989A&A...220....1B} and \cite{2013arXiv1311.2724B} for a detailed derivation of this inversion),
\begin{equation}
P(\rho_{i})=\int\frac{\dd \lambda}{2\pi \ii}\exp\left(-\ii\lambda\rho_{i}+\varphi_{0}(\lambda)\right),\label{Prho}
\end{equation}
where $\varphi_{0}(\lambda)$ is the CGF of the density taken at scale $R$ (i.e., for the filter $W_{R}$). The function $b(\rho_{i})$ is defined through a similar relation,
\begin{equation}
b(\rho_{i})P(\rho_{i})=\int\frac{\dd \lambda}{2\pi \ii}\,\varphi_{1}(\lambda)\,\exp\left(-\ii\lambda\rho_{i}+\varphi_{0}(\lambda)\right).\label{biasdefinition}
\end{equation}

The function $\varphi_{1}(\lambda)$ can be explicitly computed in the context of perturbation theory calculations~(\cite{2016MNRAS.460.1598C}). This is the case also for models in the so-called hierarchical models (see appendices). In the latter case, these calculations can be extended to higher order, as we describe below,  providing ways to better assess the domain of validity of this expansion.

According to the previous relation, this implies that this form translates into the expression of the covariance coefficients for the density PDF. More precisely, we expect 
\begin{equation}
\Cov(\rho_{i},\rho_{j})=b(\rho_{i})P(\rho_{i})\,\xib_{s}\,b(\rho_{j})P(\rho_{j})\label{NaiveCov}
,\end{equation}
where $\xis$ is the average value of the two-point correlation function $\xi(r_{d})$ within the sample.

It is to be noted, however, that this is true if
\begin{itemize}
\item the term in $\xib_{s}$ is indeed the leading contribution of the expansion (\ref{gen3ptbias}). This is obviously not the case for samples with periodic boundary conditions, for which $\xib_{s}$ vanishes by construction;
\item the density is defined regardless of the density of the sample. Its expectation value therefore does not coincide with $\rho_{s}$ for a given sample.
\end{itemize}

It can also be noted that in the Gaussian limit, we have $b(\rho_{i})=\delta_{i}/\xi$. Applying the relation (\ref{jointpdflinear}) to the density within one cell and to the sample density $\rho_{s}=1+\delta_{s}$ leads then to the following expression for the conditional expression of density PDF,
\begin{equation}
P(\rho_{i}\vert \rho_{s})=P(\rho_{i})\left(1+\delta_{s}\,b(\rho_{i})\right).
\end{equation}
This leads to the interpretation of the function $b(\rho_{i})$ as the response function of the density PDF to the
sample density. This means that although the density-bias function cannot be derived from the density PDF alone, we should be able to derive it if we are in possession of an operational method to compute the density PDF for arbitrary cosmological parameters (in a way similar to the derivation of halo-bias function as pioneered in \citet{1996MNRAS.282..347M}). Undoubtedly, this result is closely related to the so-called supersample effects \citep[as described for the power spectra covariance in][]{2013PhRvD..87l3504T}, that is, the impact of modes of scale comparable to or larger than the sample size.  This is not necessarily their only contribution
(because subdominant large-scale contributions can also contribute to the covariance), however, but likely to be the most important contribution, as described below. 

The density-bias function obeys the following consistency relations: 
\begin{eqnarray}
\int\dd \rho\, b(\rho)P(\rho)&=&0,\label{bconsistency1}\\
\int\dd \rho\,  \rho\ b(\rho)P(\rho)&=&1,\label{bconsistency2}
\end{eqnarray}
as initially pointed out in \cite{1992A&A...255....1B}.

\subsection{Case of relative densities}

The previous formula applies to the local densities, evaluated regardless of the sample density.
It does not apply in particular to standard settings (e.g., densities measured out of galaxy counts) where the density is defined with respect to the mean density of the sample. To address this case in particular, we should consider 
\begin{equation}
\hrho_{i}=\frac{\rho_{i}}{\rho_{\rm s}}
\end{equation}
as the observable for which the covariance is to be computed. In this case, the formal derivation of the PDFs is presented in the appendix, and it leads to
 \begin{eqnarray}
P(\hrho_{i})&=&\int 
\frac{\dd\lambda_{i}}{2\pi\ii}
\left[
\frac{\partial\varphi}{\partial\lambda_{s}}
\right]_{\big\vert_{\lambda_{s}=-\lambda_{i}\hrho_{i}}}
\exp\left[\varphi(-\lambda_{i}\hrho_{i},\lambda_{i})\right]\\
P(\hrho_{i},\hrho_{j})&=&\int 
\frac{\dd\lambda_{i}}{2\pi\ii}\, \frac{\dd\lambda_{j}}{2\pi\ii}
\left[
\left(
\frac{\partial\varphi}{\partial\lambda_{s}}
\right)^{2}+\frac{\partial^{2}\varphi}{\partial\lambda_{s}^{2}}
\right]_{\big\vert_{\lambda_{s}=-\lambda_{i}\hrho_{i}-\lambda_{j}\hrho_{j}}}
\nonumber\\
&&\times\exp\left[\varphi(-\lambda_{i}\hrho_{i}-\lambda_{j}\hrho_{j},\lambda_{i},\lambda_{j})\right].
\label{joinhrhoPDF}
\end{eqnarray}
We then use relation (\ref{gen3ptbias}) to compute the form of this function. At this stage, it is to be noted that the 
expressions
$\int\dd\vx_{0}\dd\vx'_{0}\,\xi(\vx_{0},\vx'_{0})$, $\int\dd\vx_{0}\,\xi(\vx_{0},\vx_{1})$ and $\xi_{12}$ all take the same averaged value when integrated over the sample.  We note here this common value as $\xis$. 
Inserting the resulting expressions of the CGF in both the expressions of $P(\hrho_{i})$ and $P(\hrho_{i}, \hrho_{j})$ and expanding all terms at linear oder in $\xis$ , we obtain\begin{eqnarray}
P(\hrho_{i},\hrho_{j})-P(\hrho_{i})\,P(\hrho_{j})&=&
\xis\,\int
\frac{\dd\lambda_{i}}{2\pi\ii}\, \frac{\dd\lambda_{j}}{2\pi\ii}
\nonumber
\\
&& \hspace{-3.5cm}\times 
\left(
1+\varphi_{1}(\lambda_{i})-\lambda_{i} \hrho_{i}
\right)\left(
1+\varphi_{1}(\lambda_{j})-\lambda_{j}\hrho_{j}
\right)\ 
\nonumber
\\
&& \hspace{-3.5cm}\times \exp\left[-\lambda_{i} \hrho_{i}-\lambda_{j} \hrho_{j}+\varphi_{0}(\lambda_{i})+\varphi_{0}(\lambda_{j})\right]
.\end{eqnarray}
This leads to the definition of the first sample-bias function,
\begin{equation}
b_{\rm s1}(\hrho_{i})=\frac{1}{P(\hrho_{i})}
\int
\frac{\dd\lambda}{2\pi\ii}
\left(
1+\varphi_{1}(\lambda)-\lambda \hrho_{i}
\right)
\exp\left[-\lambda \hrho_{i}+\varphi_{0}(\lambda)\right]
,\end{equation}
which can be re-expressed in terms of the density-bias function defined in Eq. (\ref{biasdefinition}) and the derivative of $P(\hrho_{i})$ with respect to $\hrho_{i}$
\begin{equation}
b_{\rm s1}(\hrho_{i})=b(\hrho_{i})+1+\frac{\partial\log(P(\hrho_{i}))}{\partial\log \hrho_{i}}.
\end{equation}
In this case, the covariance matrix elements are then expected to be given by
\begin{equation}
\Cov(\hrho_{i},\hrho_{j})=b_{\rm s1}(\hrho_{i})P(\hrho_{i})\,\xis\,b_{\rm s1}(\hrho_{j})P(\hrho_{j}).
\label{NaiveCovs1}
\end{equation}
Remarkably, $b_{\rm s1}(\rho)$ can entirely be expressed in terms of $b(\rho)$.
For the sake of completeness, we also consider the case of $\rhob_{i}=\rho_{i}-(\rho_{s}-1)$. In this case, it is easy to show that
\begin{eqnarray}
P(\rhob_{i})&=& \int 
\frac{\dd\lambda_{i}}{2\pi\ii}
\exp\left[-\lambda_{i}\rhob_{i}+\varphi(-\lambda_{i},\lambda_{i})\right]\\
P(\rhob_{i},\rhob_{j})&=&\int 
\frac{\dd\lambda_{i}}{2\pi\ii}
\frac{\dd\lambda_{j}}{2\pi\ii}\ 
\nonumber
\\
&& \hspace{-1.0cm}\times \exp\left[-\lambda_{i}\rhob_{i}-\lambda_{j}\rhob_{j}+\varphi(-\lambda_{i}-\lambda_{j},\lambda_{i},\lambda_{j})\right]
\label{joinrhobPDF}
.\end{eqnarray}
Following the same approach as for the previous case, the leading-order expression in $\xis$ of the connected joint
PDF is
\begin{eqnarray}
P(\rhob_{i},\rhob_{j})-P(\rhob_{i})\,P(\rhob_{j})&=&
\xis\,\int
\frac{\dd\lambda_{i}}{2\pi\ii}\, \frac{\dd\lambda_{j}}{2\pi\ii}
\nonumber
\\
&& \hspace{-3.5cm}\times 
\left(
\varphi_{1}(\lambda_{i})-\lambda_{i}
\right)\left(
\varphi_{1}(\lambda_{j})-\lambda_{j}
\right)\ 
\nonumber
\\
&& \hspace{-3.5cm}\times 
\exp\left[-\lambda_{i} \rhob_{i}-\lambda_{j} \rhob_{j}+\varphi_{0}(\lambda_{i})+\varphi_{0}(\lambda_{j})\right].
\end{eqnarray}
It leads to the definition of the second sample-bias function,
\begin{equation}
b_{\rm s2}(\rhob_{i})=b(\rhob_{i})+\frac{\partial\log(P(\rhob_{i}))}{\partial\rhob_{i}},
\end{equation}
so that 
\begin{equation}
\Cov(\rhob_{i},\rhob_{j})=b_{\rm s2}(\rhob_{i})P(\rhob_{i})\,\xis\,b_{\rm s2}(\rhob_{j})P(\rhob_{j}).
\label{NaiveCovs2}
\end{equation}
The three bias functions are therefore closely related. Although the density-bias function $b(\rho)$ cannot be derived from the shape
of $P(\rho)$ alone, as mentioned before, the relations between $b(\rho)$ and either $b_{s1}(\rho)$ and $b_{s2}(\rho)$
depend on the PDF alone. Furthermore, the two relative density bias functions obey the following consistency relations:
\begin{eqnarray}
\int\dd \rho_{i}\,  b_{s\#}(\rho_{i})P(\rho_{i})&=&0\\
\int\dd \rho_{i}\, \rho_{i}\ b_{s\#}(\rho_{i})P(\rho_{i})&=&0.
\end{eqnarray}
The second relation is at variance with the corresponding relation (\ref{bconsistency2}) for the density-bias function. It indicates that for typical values of $\rho P(\rho)$, the sample bias functions, $b_{s\#}(\rho)$,  are likely to be smaller than the density-bias function $b(\rho)$.

\subsection{Structure of the covariance matrix}

The consequences of these formulae on the structure of the covariance matrix are illustrated below with the help of the Rayleigh-Levy flight model. Fig. \ref{DPDF_SetA_0p5} compares the results
from exact derivations of the covariance matrix with these prescriptions. The diagonal parts of the covariance matrices are well accounted for by these formulae. The root mean square of the measured local density PDF in particular exhibits the expected density dependence, at least for mild values of the density.

In all the formulae (\ref{NaiveCov}, \ref{NaiveCovs1}, \ref{NaiveCovs2}), the expression of the covariance exhibits a simple structure, as 
it is factorizable in the two densities. This implies,
for instance, that the reduced covariance matrix
\begin{equation}
\Cov_{\rm reduced}(\rho_{i},\rho_{j})=\frac{\Cov(\rho_{i},\rho_{j})}{\sqrt{\Cov(\rho_{i},\rho_{i}) \Cov(\rho_{j},\rho_{j})}}
\end{equation}
has an extremely simple structure: it is given by the sign of the product of the bias functions (i.e., $\sign[b(\rho_{i})b(\rho_{j})]$, $\sign[b_{s1}(\rho_{i})b_{s1}(\rho_{j})]$ , and $\sign[b_{s2}(\rho_{i})b_{s2}(\rho_{j})]$  for the three different measurement strategies).   This leads to the butterfly-like structure in the plotted matrices, as illustrated in Fig. \ref{NumRedCovariance}. 
This simple form betrays the fact that the density covariance is only poorly known. To be more specific, formulae (\ref{NaiveCov}, \ref{NaiveCovs1}, \ref{NaiveCovs2})  give only a single eigendirection of the covariance matrix (namely $b(\rho_{i})P(\rho_{i})$) and the amplitude of a single eigenvalue associated with it. The numerical calculations suggest that it is the leading one when $\xis$ does not vanish, as  illustrated on Fig. \ref{FirstEigenvect}.  These formulae do not offer any indication of the amplitude of the covariance in orthogonal directions, however. Taken at face value, they imply that the other  eigenvalues all vanish, preventing the covariance matrix from being invertible. These formulae therefore cannot be used alone to model the covariance for practical purposes, and complementary contributions have to be derived from other (and a priori subdominant) effects. 

\subsection{Beyond leading-order effects}

In the previous subsection, we identified the long-distance leading contributions. As mentioned before, this leads to only limited information of the covariance structure. This difficulty is even more acute for covariances evaluated in numerical experiments
consisting of a collection of independent samples, each of them with periodic boundary conditions (this does not have to be so, but it is often the case in practice). By construction, the mean correlation function within the sample then vanishes, $\xis\to 0,$ making the term we have computed identically zero. All these considerations indicate that further contributions need to be identified. The identification of the next-to-leading order effects in Eq. (\ref{jointpdflinear}) is difficult to do a priori, however:
\begin{itemize}
\item One natural next-to-leading contribution is obtained by taking into account second-order terms in $\xi(d)$ in Eq. (\ref{jointpdflinear}), that is, 
by considering doubles lines between cells in a diagrammatic representation. This would induce a term
of about $\xi(d)^{2}$ , whose average never vanishes\footnote{In the minimal tree model, it is possible to compute these terms in the so-called mean-field approximation (see appendix), but they do not lead to a positive definite covariance matrix and therefore cannot be the sole, or dominant, contribution to the covariances.}. As shown in the appendix, these contributions can be formally derived in the context of the hierarchical models. This leads to correction terms that can be organized in a sum of factorized terms. Therefore, although it can indeed provide corrective terms to the covariance matrix, only a limited number of eigendirections can be generated.
\item Other  contributions naturally come from proximity effects due to the fact that cells are finite, and could even overlap, which makes the expansion in $\xi(d)$ ineffective. In a diagrammatic point of view, they are due to the fact that many more diagrams contributed when cells are too close. This has dramatic effects for overlapping cells. For hierarchical models, an approximate form can be used to help model these effects, which we use below.
\item Finally, effects due to the fact that discrete tracers are used in count-in-cells statistics might also play a role at short distances. They are also tentatively modeled below.
\end{itemize}
In the following, we propose some modeling of these effects and explore how they depend on the properties of the survey.

\subsubsection{Joint PDF at short distances}

There are no general forms for the joint PDF at close distance. The hierarchical models suggest the following form
(derived from the saddle point approximation, which is valid for moderate values of $\xib$ and of the density contrast), however:
\begin{eqnarray}
P_{\shortd}(\rho_{i},\rho_{j})\dd\rho_{i}\dd\rho_{j}&=&P(\rho_{m}) 
\nonumber
\\
&& \hspace{-3.5cm}\times 
\exp\left[-\frac{\delta_{\rho}^{2}}{\rho_{m}^{\alpha}\Delta_{\xi}(d)}\right]\frac{\dd\rho_{m}\dd\delta_{\rho}}{\sqrt{\pi\rho_{m}^{\alpha}\Delta_{\xi}(d)}}
\label{shortdistjPDF}
,\end{eqnarray}
where $\rho_{m}=(\rho_{i}+\rho_{j})/2$ and $\delta_{\rho}=(\rho_{i}-\rho_{j})/2,$ and 
where $\alpha$ is model-dependent parameter. In other words, the PDF of the difference between $\rho_{i}$ and $\rho_{j}$ can 
be described by a simple Gaussian with a known width
driven by the expression of $\Delta_{\xi}(d)\equiv \xib-\xi_{12}(d),$ provided it is small compared to $\xib$. 
We note that $\Delta_{\xi}(d)$ obviously vanishes at $d=0$, it then leads to a Kronecker $\delta$ function at zero separation as expected,  and generically scales like $d^{2}$ at short distances\footnote{This limited form would induce a minimum contribution to $\Cov(\rho_{i},\rho_{i})$ given by $\Delta_{\rho_{i}}/\mP(\rho_{i}),$ where $\Delta_{\rho_{i}}$ is the bin size in density.}.
Interestingly, for the minimal tree model the form (\ref{shortdistjPDF}) is exact for $\alpha=1$ (see appendix). 
In general, this is also the expected form based on the saddle point approximation (valid when $\xib$ is small) for generic hierarchical 
tree models. The value of $\alpha$ can be identified from small-order cumulants, 
\begin{equation}
\alpha=\frac{2}{3}S_{3}
,\end{equation}
where $S_{3}$ is the reduced third-order cumulant,
\begin{equation}
S_{3}=\frac{\left<\delta^{3}\right>}{\left<\delta^{2}\right>^{2}}.
\end{equation}
This form is probably not very accurate in general. It can be used to model the impact of close distances to the covariance matrix, however,
as shown below.

\subsubsection{Poisson noise and minimum separation}

A further contribution to this joint PDF can come from discrete effects that arise because the density is evaluated from the counting of discrete tracers (as explored in \cite{1996ApJ...470..131S} or more recently in \cite{2021MNRAS.500.3631R}). In this subsection, we assume that the density corresponds to the density obtained after application of a top-hat filter and that tracers are Poisson realizations of continuous fields \citep[although it is possible to encounter sub- or over-Poissonnian noises;][]{2018PhRvD..98b3508F}. The use of other filters can be explored but would require specific developments that we do not pursue here. Within such hypotheses then, the joint distribution of counts-in-cells $N_{i}$  is given by the convolution of the joint density PDF, $\mP(\left\{\rho_{i}\right\})$, in the continuous limit convolved by Poisson counts-in-cells as 
\begin{equation}
\mP(\{N_{i}\})=\int\Pi_{i} \dd \rho_{i}\,P_{\Poiss}(N_{i};\Nb_{i}\rho_{i})\,\mP(\left\{\rho_{i}\right\})
,\end{equation}
where $P_{\Poiss}(N;\Nb)$ is more precisely the probability of having $N$ tracers in a cell where the mean density of tracers is $\Nb$.
In practice, $\Nb_{i}$ is given by $nV_{i}$ , where $n$ is the number density of tracers and $V_{i}$ is the volume of the cell $V_{i}$.

Discrete effects would then induce further scatter between the estimated values of $\rho_{i}$ and $\rho_{j}$. The latter are given by Poisson noise induced by the nonoverlapping parts of the cells, as shown in \cite{1996ApJ...470..131S}, further contributing to the scatter.
The scatter in the difference in the number of points is 
\begin{equation}
\sigma^{2}_{\Poiss}=\frac{2}{\Nb}\rho_{m}\,f_{e}(d)\label{PoissContr}
.\end{equation}
It can be incorporated as a contribution to the variance of the PDF of $\delta_{\rho}$ of the form
\begin{equation}
\sigma^{2}_{\Poiss}=\frac{1}{2 \Nb}\rho_{m}\,f_{e}(d)\label{PoissContr}
,\end{equation}
where $f_{e}(d)$ is the fraction of the volume of each cell that does not overlap with the other as a function of the cell distance.
For short distances (i.e., for about $d\lesssim R$), it is in the 2D case given by
\begin{equation}
f_{e}(d)=\frac{2d}{\pi\,R}.
\end{equation}
The expression (\ref{PoissContr}) is then a priori to be added to the variance term that appears in Eq. (\ref{shortdistjPDF})
so that the total variance for the density difference reads
\begin{equation}
\sigma_{\delta_{\rho}}^{2}(d)=\frac{1}{2}\rho^{\alpha}_{i}\Delta_{\xi}(d)+\frac{1}{2\Nb}\rho_{i}\,f_{e}(d)\label{FulldeltaVariance}
.\end{equation}

Equation (\ref{shortdistjPDF}) then fully encodes the fact that nearby cells are likely to have similar densities.  This encodes, for instance, 
that nearby cells are within the same haloes. This contribution is expected to enhance the covariance terms. It shows that the amount of information is limited at small scales: there is therefore a minimum separation between cells smaller than which no gain in precision is expected of PDF measurements. The minimum distance depends on the bin size: $d_{\min}$ is the distance such that the densities in two cells separated by less than $d_{min}$ are almost  certainly in the sale density bin. $d_{\min}$ therefore depends on the bin width $\Delta_{i}$.
From eq. (\ref{FulldeltaVariance}), it is possible to infer this value. We desire 
\begin{equation}
\sigma_{\delta_{\rho}}^{2}(d_{\min})\ll \Delta_{i}^{2}.
\end{equation}

This suggests that a minimum distance between cells can be adopted, given by
\begin{equation}
d_{\min \Poiss}=\frac{\pi\Delta_{i}^{2}}{{\Nb}}\,R.
\end{equation}
The other upper limit comes on $d$ from the expression of $\Delta_{\xi}$ as a function of $d$. The latter depends on both the shape of the power spectrum and on the filter that is used. In general (e.g., for a Gaussian filter), $\Delta_{\xi}(d)$ scales like $d^{2}/R^{2}$ , where $R$ is the filtering radius, with a coefficient $c_{n_{s}}$ that depends on the power spectrum index $n_{s}$ and is proportional to its amplitude. Top-hat filters have different analytical properties. We give here the formal expression of $\Delta_{\xi}(d)$ at 2D for a power-law spectrum of index $n_{s}$,
\begin{eqnarray}
\frac{\Delta_{\xi}(d)}{\xib}&\!\!\!\!\!=&\!\!\!\!\!-\frac{2^{n_s\!-\!1} \Gamma \left(1\!-\!\frac{n_s}{2}\right) \Gamma \left(2\!-\!\frac{n_s}{2}\right) \Gamma \left(\frac{1}{2}
   \left(n_s\!-\!1\right)\right) }{\sqrt{\pi } \Gamma \left(\frac{1}{2}\!-\!\frac{n_s}{2}\right) \Gamma
   \left(\frac{3}{2}\!-\!\frac{n_s}{2}\right) \Gamma \left(\frac{n_s}{2}+1\right)}
\left(\frac{d}{R}\right)^{1\!-\!n_s}\nonumber\\
&\!\!\!\!\!\approx&\!\!\!\!
0.72\,\frac{d^{3/2}}{R^{3/2}} \hbox{  for  }n_{s}=0.5.
\end{eqnarray}
This is the situation we encounter below in the numerical experiments we perform. This leads to the following
form:
\begin{equation}
d_{\min \halo}=\left(\frac{\Delta_{i}^{2}}{0.72\,\xib}\right)^{2/3}\,R\label{dminhalo}.
\end{equation}
It is to be noted that it can be in practice a rather short distance, shorter than the filtering scale $R$. For instance, for a bin width of $1/4$, a variance of about unity, $d_{\min \halo}$ is about $R/5$. 

Equation (\ref{shortdistjPDF}), together with the expressions of the bias functions described above, is the main results of this paper.
We illustrate below how they can be used to compute the covariance matrices.

\section{Hierarchical models}

In order to illustrate the previous findings, we make use of toy models for which explicit computations can be made.
\subsection{General formalism}
Hierarchical models are a class of non-Gaussian fields whose correlation 
functions follow the so-called hierarchical ansatz,
\begin{equation}
\xi_{p}(\vr_{1},\dots,\vr_{p})=\sum_{t\in\trees}Q_{p}(t)\,\prod_{\lines\in t}\xi(\vr_{i},\vr_{j})
,\end{equation}
where the sum is made over all possible trees that join the $p$ points (diagram without loops), and the tree value 
is obtained by the product of a fixed weight (that depends only on the tree topology) and the product of the two-point
correlation functions for all pairs that are connected together in the given tree. This construction ensures that the
average p-point function, $\overline{\xi_{p}}$, scales like the $\xib^{p-2}$ , where $\xib$ is the averaged two-point function.
More precisely, there are $S_{p}$ parameters such that
\begin{equation}
\overline{\xi_{p}}=S_{p}\,\xib^{p-2}.
\end{equation}
The precise value of the $S_{p}$ parameters depend on the $Q_{p}$ parameters and on the averages of the product of 
$\xi(r_{ij})$ functions. A very good approximation is to assume that the average of the products of this function is given 
by the product of these averages. Then the $S_{p}$ coefficients depend solely on $Q_{p}$,
\begin{equation}
S_{p}=\sum_{t}Q_{p}(t).
\end{equation}

\subsection{The (minimal) tree model}

The tree models are based on a further assumption on the  $Q_{p}$ parameters. It is basically assumed that 
tree expressions can be computed locally\footnote{Perturbation theory results do not exactly follow this construction as vertices are then dependent on the geometry of the incoming lines. However, in this case, $Q_{p}$ values are indeed obtained from a product of vertices.}
, that is,
\begin{equation}
Q(t)=\prod_{\vertices\in t} \nu_{p}
,\end{equation}
where $\nu_{p}$ is a weight attributed to all vertices with $p$ incoming lines ($\nu_{0}=\nu_{1}=1$ for completion).
In this formalism, the vertex generating function is generally introduced, 
\begin{equation}
\zeta(\tau)=\sum_{p}\frac{\nu_{p}}{p!}\tau^{p}
.\end{equation}
The minimal tree model is a model in which $\nu_{2}$ alone does not vanish. In the minimal model\footnote{it is minimal in the sense that it can be shown that $\nu_{2}$ cannot be smaller than $1/2$ in the 
strongly nonlinear regime (Peebles, 1980).} , its value is fixed and is
given by $\nu_{2}=1/2,$ so that
\begin{equation}
\zeta_{\RL}(\tau)=(1+\tau/2)^{2}.
\end{equation}
Together with the Gaussian case (which corresponds to $\zeta(\tau)=1+\tau$), this is the only case for which we are sure that it can be effectively built
(in the sense that other models may be unphysical).

In this model, it is possible to build the cumulant generating function of the local density. For the one-point case, assuming the mean-field approximation, the CGF is given by
\begin{equation}
\varphi(\lambda)=\lambda\left[\zeta(\tau)-\frac{1}{2}\tau\zeta'(\tau)\right]
\end{equation}
with
\begin{equation}
\tau=\lambda\,\xib\,\zeta'(\tau).\label{taueq}
\end{equation}
This is not the result of large deviation principle calculations, but of mere combinatorics, although it leads to the same formal 
transformation between the CGF and the vertex-generating function. In case of the minimal model, Eq. (\ref{taueq}) takes a simple form that can be easily solved.
We finally have
\begin{equation}
\varphi(\lambda)=\frac{\tau(\lambda)}{\xib},\ \ \ \tau(\lambda)=\frac{\lambda\xib}{1-\lambda\xib/2}.
\end{equation}
The one-point PDF of the density can then be computed explicitly (see appendix),
\begin{equation}
P(\rho)=\frac{4}{\xib^{2}}\exp\left[-\frac{2}{\xib}(1+\rho)\right] \ _{0}F_{1}\left(2,\frac{4\rho}{\xib}\right)
\label{LFpdf}
,\end{equation}
as can the density-bias function,
\begin{equation}
b(\rho)=\frac{\ _{0}F_{1}\left(1,\frac{4\rho}{\xib}\right)}{\ _{0}F_{1}\left(2,\frac{4\rho}{\xib}\right)}-\frac{2}{\xib},\end{equation}
where $\xib$ is the averaged two-point correlation function within the cell.

\subsection{Rayleigh-Levy flight model}

\begin{figure}
   \centering
 \includegraphics[width=7cm]{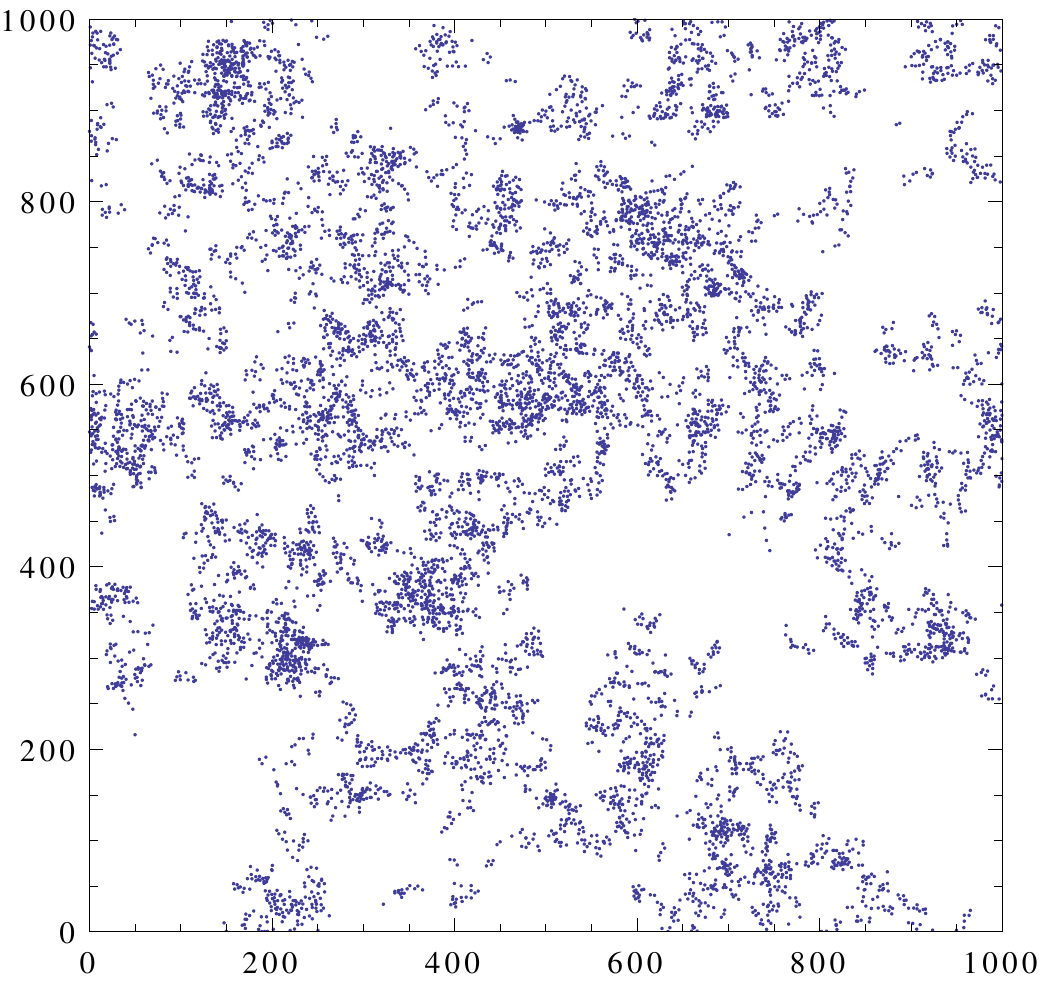}
   \caption{Example of a realization of a Rayleigh-Levy walk. Points mark the end point of each displacement. They are clearly correlated.}
   \label{samplepoints}
\end{figure}

The minimal tree model can be implemented with Rayleigh-Levy random walks (or rather Rayleigh-Levy flights, as described in Peebles 1980).  This is a Markov random walk 
where the PDF of the step length $\ell$ follows a power law,
\begin{equation}
P(\ell)\sim\frac{1}{\ell^{\alpha}}
,\end{equation}
with a regularizing cutoff at small separation, and where $\alpha$ satisfies
\begin{equation}
0 < \alpha < 2.
\end{equation}
The sample points are then all the step points reached by the walker.  

More precisely, the cumulative distribution function of step of length $\ell$ is 
\begin{eqnarray}
P(>\ell_{0})&=&1,\\
P(>\ell)&=&\left(\frac{\ell_{0}}{\ell}\right)^{\alpha}\ \ \hbox{for}\ \ \ell>\ell_{0}
,\end{eqnarray}
where $\ell_{0}$ is a small-scale parameter. The two- and higher-order correlation functions can then be explicitly computed.
Starting with a first point at position $\vr_{0}$, the density of the subsequent point (first descendant) at position $\vr$ 
is given by
\begin{eqnarray}
f_{1}(\vr)&=&\frac{\alpha}{2\pi}\frac{\ell_{0}^{\alpha}}{\vert\vr-\vr_{0}\vert^{2+\alpha}}\ \ \hbox{in 2D space};\\
f_{1}(\vr)&=&\frac{\alpha}{4\pi}\frac{\ell_{0}^{\alpha}}{\vert\vr-\vr_{0}\vert^{3+\alpha}}\ \ \hbox{in 3D space}.\end{eqnarray}
In the following, the dimension of space is denoted $D$.
The density of the descendants, assuming there are an infinity of them, of a point at position $\vr_{0}$ is then given by a series
of convolutions,
\begin{equation}
f(\vr_{0},\vr)=f_{1}(\vr)+\int\dd^{D}\vr_{1}\ f_{1}(\vr-\vr_{1})f_{1}(\vr_{1}-\vr_{0})+\dots
,\end{equation}
with subsequent convolutions and where the integral is done in the whole space. Defining the Fourier transform of $f_{1}(\vr)$ as $\psi(k)$,
\begin{equation}
\psi(k)=\int\dd^{D}\vr\ f_{1}(\vr)\,e^{-\ii\vk.\vr},
\end{equation}
which is then a function of $k$ only, it is easy to see that 
\begin{eqnarray}
f(\vr_{0},\vr)&=&\int\frac{\dd^{D}\vk}{(2\pi)^{D}}\ e^{\ii\vk.(\vr-\vr_{0})}\left[\psi(k)+\psi^{2}(k)+\dots\right]\nonumber\\
&=&\int\frac{\dd^{D}\vk}{(2\pi)^{D}}\ e^{\ii\vk.(\vr-\vr_{0})}\,\frac{1}{1-\psi(k)}
,\end{eqnarray}
where we take advantage of the expression of convolutions in Fourier space and their resummations. The two-point correlation function is then given by two possible configurations: a neighbor can either be an ascendant or a descendant, so that the two-point correlation functions between positions $\vr_{1}$ and $\vr_{2}$ are given by
\begin{eqnarray}
\xi_{2}(\vr_{1},\vr_{2})&=&\frac{1}{n}\left[f(\vr_{1},\vr_{2})+f(\vr_{2},\vr_{1})\right]\nonumber\\
&=&\frac{1}{n} \int\frac{\dd^{D}\vk}{(2\pi)^{D}}\ e^{\ii\vk.(\vr_{2}-\vr_{1})}\,\frac{2}{1-\psi(k)}\label{exactxiRL}
,\end{eqnarray}
where $n$ is the number density of points in the sample that can be associated with a typical length $\ell_{n,}$
\begin{equation}
n=\frac{1}{\ell_{n}^{D}}.
\end{equation}

At large scale, this expression causes the power spectra to be power laws. They scale like $k^{-\alpha}$ , and
the resulting two-point correlation function then takes the form in the large separation limit,
\begin{eqnarray}
\xi_{\alpha,\,2D}(r)&=&\frac{\alpha}{\pi }\ r^{\alpha -2} \ell_{0}^{-\alpha }\ell_{n}^{2}\label{axiRL2D}\\
\xi_{\alpha,\,3D}(r)&=&\frac{\left(1-\alpha ^2\right) \tan \left(\frac{\pi  \alpha }{2}\right)}{\pi ^2}\ r^{\alpha -3} \ell_{0}^{-\alpha}\ell_{n}^{3}
\label{axiRL3D}
.\end{eqnarray}

It is to be noted, however, that this expression does not take into account the boundary conditions, in particular if they are assumed to be periodic. This case is examined in some detail in the next paragraph. It is to be noted, however, that in this case, the function $\xi(r)$ has a more complex form. It is in particular no more isotropic.

Higher-order correlation functions can also be computed in this model: $n$ points are correlated when they are embedded in a chronological sequence that can be run in one direction or the other. 
Thus the three-point function is simply given by 
\begin{equation}
\xi_{\alpha}(\vr_{1},\vr_{2},\vr_{3})=\frac{1}{n^{2}}\left[f(\vr_{1},\vr_{2})f(\vr_{2},\vr_{3})+...\right]
\label{xi3expression1}
,\end{equation}
with five other terms obtained by all combinations of the indices. Expressing the result in terms of the two-point function, we have 
\begin{eqnarray}
\xi_{\alpha}(\vr_{1},\vr_{2},\vr_{3})&=&\frac{1}{2}\left[\xi_{\alpha}(\vr_{1},\vr_{2})\xi_{\alpha}(\vr_{2},\vr_{3})+\right.\nonumber\\
&&\hspace{-2cm}\left.\xi_{\alpha}(\vr_{2},\vr_{3})\xi_{\alpha}(\vr_{3},\vr_{1})+\xi_{\alpha}(\vr_{3},\vr_{1})\xi_{\alpha}(\vr_{1},\vr_{2})
\right]
\label{xi3expression2}
,\end{eqnarray}
corresponding to a tree structure with $\nu_{2}=1/2$. 

Higher-order correlation functions can be computed similarly. They follow a tree structure in the sense above, with $\nu_{2}=1/2$ and $\nu_{p}=0$ for $p\ge 3$.

\subsection{Periodic boundary conditions}

We briefly explore here the case of periodic boundary conditions. Then the multipoint density field $g^{PBC}({\vr_{i}})$ for periodic boundary conditions can be expressed in terms of the former density field $g({\vr_{i}})$ as sums of all copies of the sample, that is,
\begin{equation}
g^{\PBC}(\{\vr_{i}\})=\sum_{\vn_{i}}g(\{\vr_{i}+\vn_{i}L\})
,\end{equation}
where $\vn_{i}$ are vectors whose components are integers, $\vn_{i}=(n_{i}^{x},n_{i}^{y},n_{i}^{z})$ and the sums run over all integer values for all $i$; $L$ is the size of the sample (assumed to be the same in all directions).

When it is applied in this context, we can construct the $n-$point density function out of the density function $f$ computed previously. Thus the
two-point density function is given by
\begin{equation}
g^{\PBC}(\vr_{1},\vr_{2})=n^{\PBC}\sum_{\vn_{12}}f(\vr_{1},\vr_{2}-\vr_{1}+\vn_{12}L)
,\end{equation}
where $\vn_{12}=\vn_{2}-\vn_{1}$ and $n^{\PBC}$ is the resulting one-point (and therefore homogeneous) density in the sample. This expression is therefore written in terms of the function
\begin{equation}
f^{\PBC}(\vr_{0},\vr)= \sum_{\vn}f(\vr_{0},\vr-\vr_{0}+\vn L).
\end{equation}

We can now compute its expression in terms of the power spectra, or more specifically, the function $\psi(k)$ defined previously.
We have
\begin{equation}
f^{\PBC}(\vr_{0},\vr)=
\int\frac{\dd^{D}\vk}{(2\pi)^{D}}\ e^{\ii\vk.(\vr-\vr_{0})}\,\sum_{\vn}e^{\ii\vn.\vk\,L}\frac{1}{1-\psi(k)},
\end{equation}
and the latter sum ensures that the contributing wave modes $\vk$ are only those that are periodic in all three directions, that is,
those whose components are multiples of $2\pi/L$ so that
\begin{equation}
f^{\PBC}(\vr_{0},\vr)=\sum_{\vn^{*}}\frac{1}{L^{D}}\ e^{2\pi \ii\ \vn.(\vr-\vr_{0})/L}\,\frac{1}{1-\psi(k_{n})}
,\end{equation}
with
\begin{equation}
k_{n}=(\vn.\vn)^{1/2}\frac{2\pi}{L}
,\end{equation}
and where the sum is all over possible integer triplets but $\vn=(0,0,0)$. The two-point correlation function is now given by
\begin{eqnarray}
\xi_{\alpha}^{\PBC}(\vr_{1},\vr_{2})=\frac{1}{n^{\PBC}}\left[f^{\PBC}(\vr_{1},\vr_{2})+f(\vr_{2},\vr_{1})\right]
.\end{eqnarray}
A similar result can be obtained for the three-point correlation function with 
\begin{eqnarray}
\xi_{\alpha}^{\PBC}(\vr_{1},\vr_{2},\vr_{{3}})&=&\nonumber\\
&&\hspace{-2cm}\frac{1}{(n^{\PBC})^{2}}\left[f^{\PBC}(\vr_{1},\vr_{2})\ f^{\PBC}(\vr_{2},\vr_{3})+\dots\right].
\end{eqnarray}
As a consequence, the functional relation between the three-point correlation function and the two-point correlation function is left unchanged. This is also the case at all orders.

\subsection{Covariance matrix of the minimal tree model}

Remarkably, 
in case of the minimal tree model, the derivation of the CGF can also be made for multiple cells, and in particular, for two cells. Its expression is derived in the appendix. It takes the form 
\begin{equation}
\varphi(\lambda_{1},\lambda_{2})=
\frac{\lambda_{1}+\lambda_{2}+(\xi_{12}-\xib)\lambda_{1}\lambda_{2}}{1-(\lambda_{1}+\lambda_{2})\,\xib/2-\lambda_{1}\lambda_{2}\,(\xi_{12}^{2}-\xib^{2})/4}.\label{MeanFPhi21}
\end{equation}
In this case, it is then possible to expand its expression in powers of $\xi_{12}$ for distant cells or in powers 
of $(\xib-\xi_{12})$ for close cells, and in both cases, closed forms can be obtained to any order. It leads to the possibility of computing the joint PDF for any configuration (see the appendix for details) and finally to evaluate the covariance matrix directly. This is even possible for any of the thre sets of variables we consider, $\{\rho_{i}\}$, $\{\hrho_{i}\}$, or $\{\rhob_{i}\}$.

We performed these computations for the minimal tree model with a power-law behavior $\xi(r)\sim r^{-1.5}$ ($\alpha=0.5$), a 2D survey with a size of $200^{2}$ pixels, and a top-hat smoothing radius of $4.25$ pixels. The amplitude of the correlation function was fixed to give $\xib=1.09$ at the smoothing scale. It precisely corresponds to the setting of the numerical simulations of Rayleigh-Levy flights we also performed, as described in the next section. It allows us to compare the two approaches. These analytic results have two limitations: the results are based on the mean-field approximation for the computation of the two-variable GCF, and the covariance elements are computed ignoring the bin sizes (i.e., by evaluating the expression of the covariance for their central values). Although in most cases this should not be an issue, it still might have a non-negligible impact when the PDF varies rapidly with the density.

\section{Simplified models of the covariance matrix}

The purpose of this section is then to propose two levels of modeling of the covariance matrix based on the previous results and to compare these propositions with results of either the full analytic results presented before or with the results of numerical experiments based on Rayleigh-Levy flights.

\subsection{Modeling the covariance matrix}

More specifically, we considered two approximate forms for the full covariance. The first approximation is fully analytic. It makes use of the
large-scale contributions and those from the short distance expression (\ref{shortdistjPDF}). It reads as the sum of the two contributions
\begin{eqnarray}
\Cov(\rho_{i},\rho_{j})&=&b_{\#}(\rho_{i})P(\rho_{i})\xis b_{\#}(\rho_{j})P(\rho_{j})
\nonumber
\\
&& \hspace{-.5cm}+
\int_{0}^{r_{\max}} \dd r_{d}\,P_{s}(r_{d})\, P_{\shortd}(\rho_{i},\rho_{j},r_{d}).
\label{CovExpAp1}
\end{eqnarray}
In this expression, the only free parameter is $r_{\max}$. This is indeed a crucial parameter as it determines  to a large extent the amplitude of the short-distance effects. In the following, we take $r_{\max}=R$, that is, the filtering scale. It is found to give a good result
for the 2D case and for $n_{s}=-1/2,$ but this choice is likely to depend on the shape of the power spectrum. In general, this formula is intended to give a good account of the general properties of the covariance matrix, it cannot provide reliable quantitative results a priori.

The other form we propose is intended to be much more precise quantitatively. Is is given by the following expression:
\begin{equation}
\Cov(\rho_{i},\rho_{j})=b_{\#}(\rho_{i})P(\rho_{i})\xis b_{\#}(\rho_{j})P(\rho_{j})
+\Cov^{\PBC}(\rho_{i},\rho_{j})
\label{CovExpAp2}
,\end{equation}
where $\Cov^{\PBC}(\rho_{i},\rho_{j})$ is the expression of the covariant matrix for periodic boundary conditions. It is obtained here simply by replacing $P(\rho_{i},\rho_{j},\xib,\xi_{12}(r_d))$ by $P(\rho_{i},\rho_{j},\xib-\xis,\xi_{12}(r_d)-\xis)$ before integrating over $r_{d}$ so that the averaged joint correlations vanish identically. The rationale for this proposition is that $\Cov^{\PBC}(\rho_{i},\rho_{j})$ could be more easily estimated from specific numerical experiments. In both cases, the short-distance contributions are the same for the three types of observables $\rho_{i}$, $\hrho_{i}$ , and $\rhob_{i}$.
These forms are then compared to numerical results.

\subsection{Numerical experiments with the Rayleigh-Levy flight model}

\begin{figure}
   \centering
 \includegraphics[width=7cm]{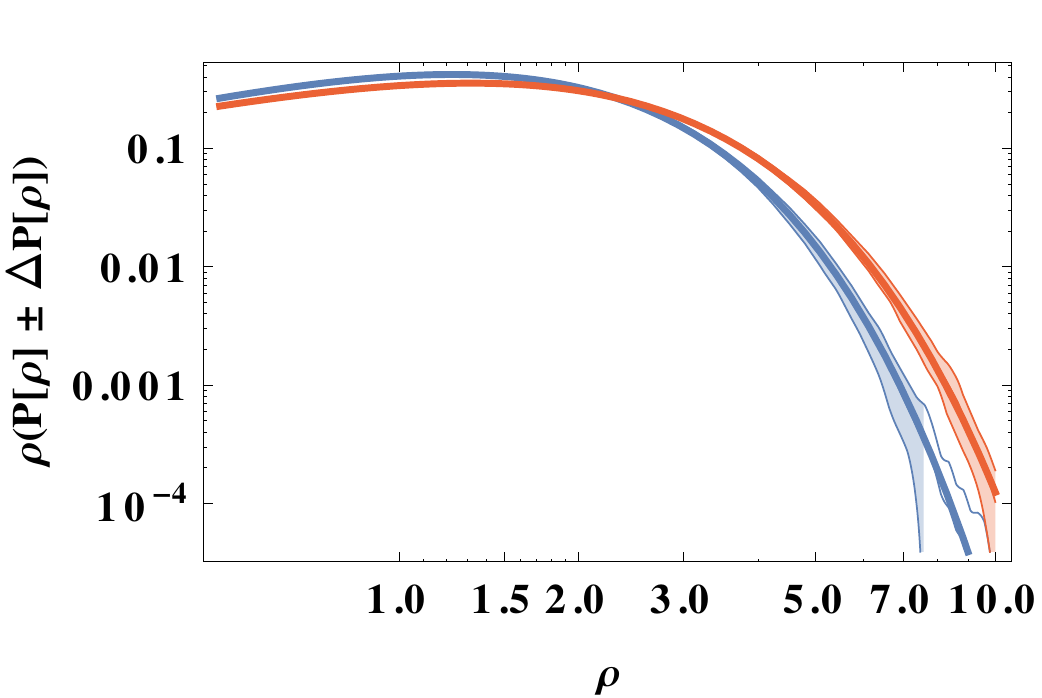}
 \hspace{.5cm}
  \includegraphics[width=7cm]{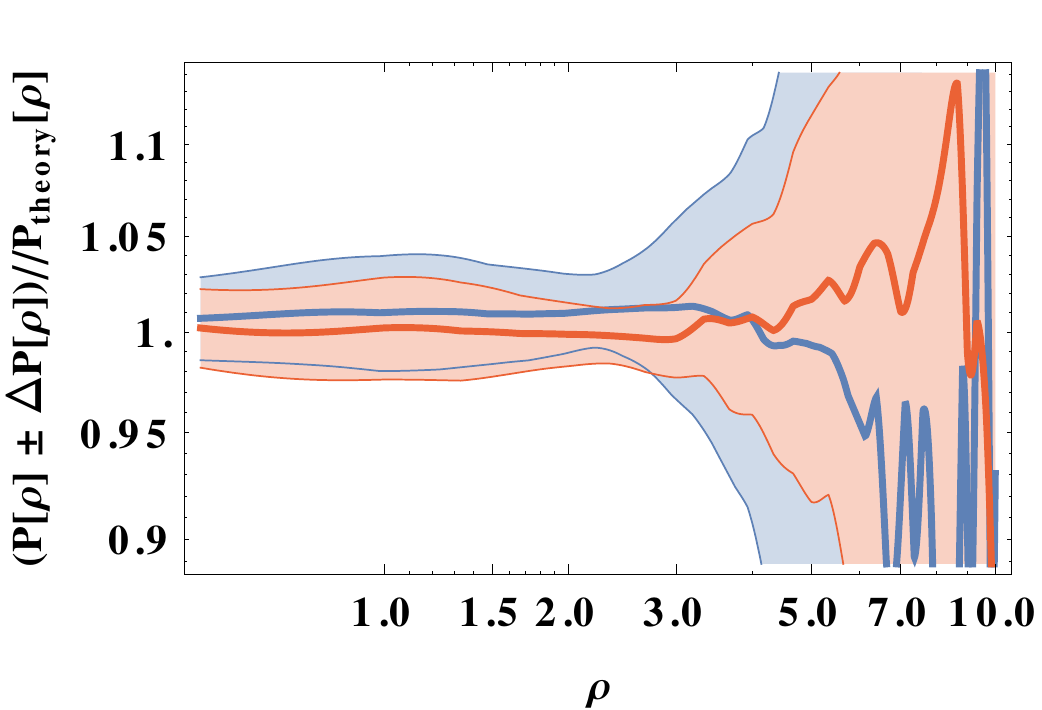}
   \caption{One-point density PDF obtained with top-hat filters compared with the theoretical predictions, Eq. (\ref{LFpdf}). The values of $\xib$ are 0.8 and 1.09 for the blue and red curves, respectively, corresponding to two different values of $l_{0}$. The bottom panel shows the residuals. Departure from theory might be due to binning and/or to the finite number of samples.}
   \label{densityPDF-TH}
\end{figure} 

\begin{figure*}[tbp]
   \centering
 \includegraphics[width=5.5cm]{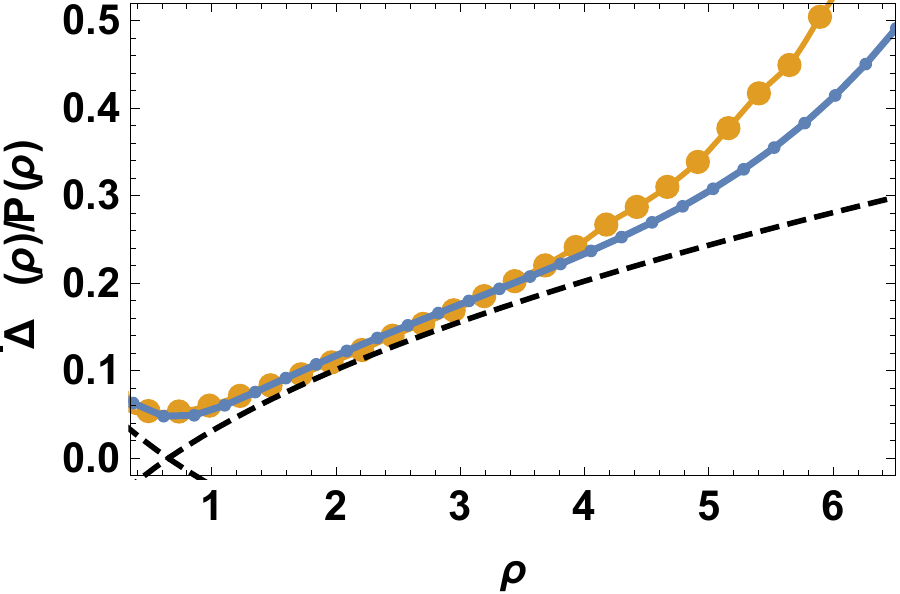}
 \includegraphics[width=5.5cm]{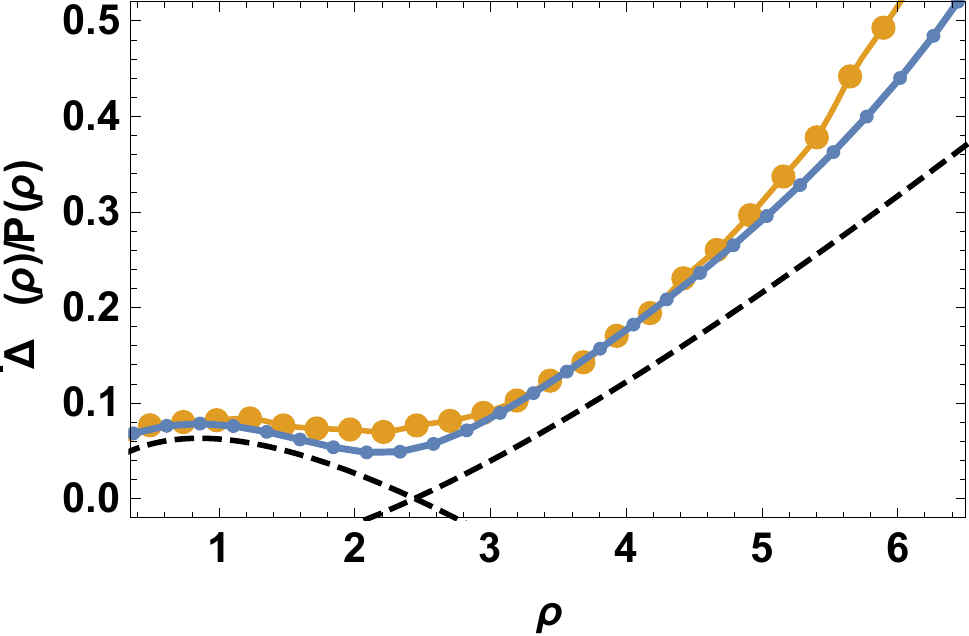}
 \includegraphics[width=5.5cm]{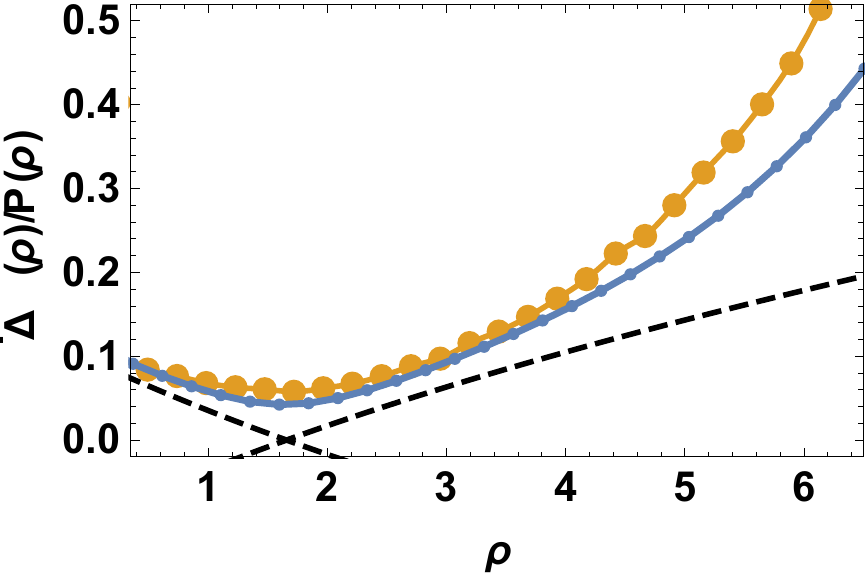}
   \caption{Measured variance of the density PDF, i.e., diagonal elements of the covariance matrix, in sets $\mA$  for $\alpha=0.5$ and different prescription of the measured density. From left to right, raw density $\rho_{i}$, scaled density $\hrho_{i}$ , and scaled density $\rhob_{i}$. The blue dots and solid lines are from the mean-field analytical expressions, and the large gold symbols are from the numerical simulations. The dashed black lines are what is expected from the large-scale leading contribution. The variance at cell scale is about $1.09,$ and the variance at sample scale, $\xib_{s}$, is about $0.09$.}
   \label{DPDF_SetA_0p5}
\end{figure*}

\begin{figure*}[tbp]
\centering
 \includegraphics[width=5.5cm]{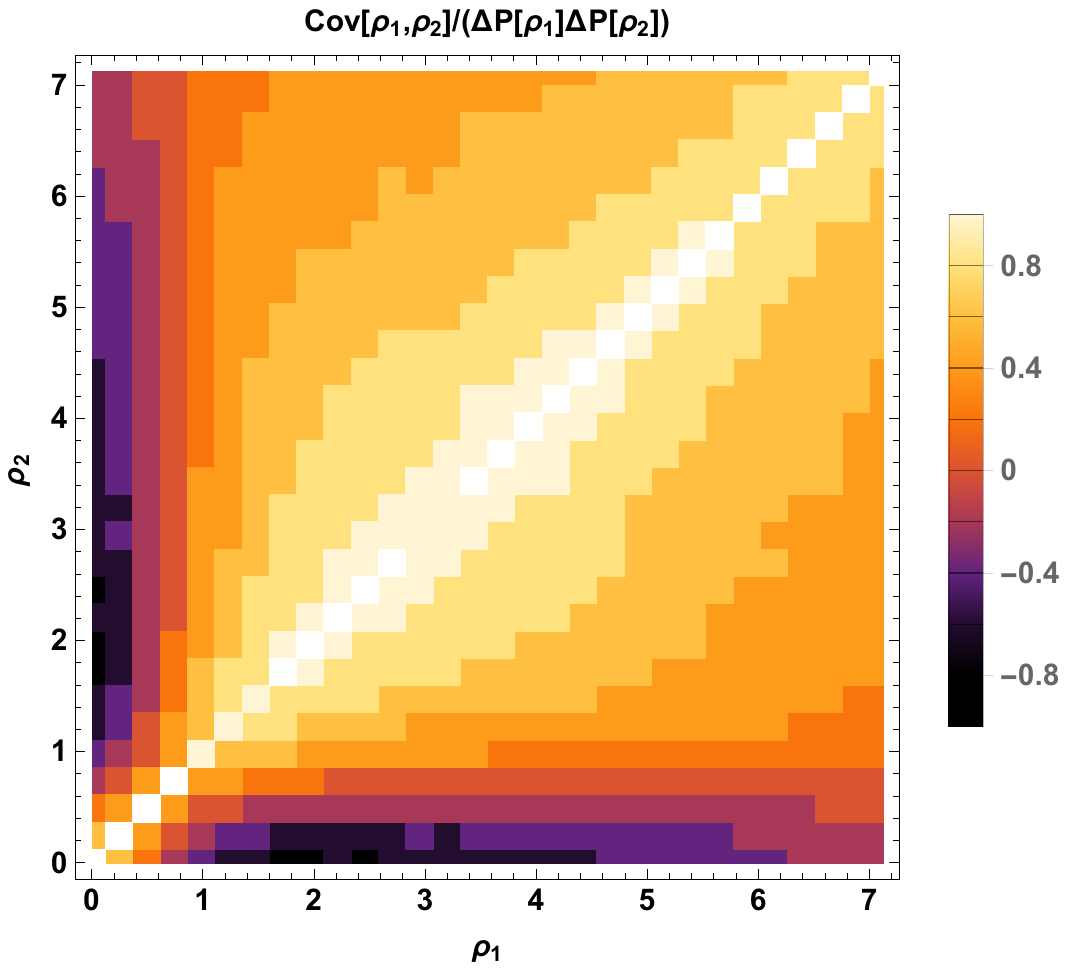}
 \includegraphics[width=5.5cm]{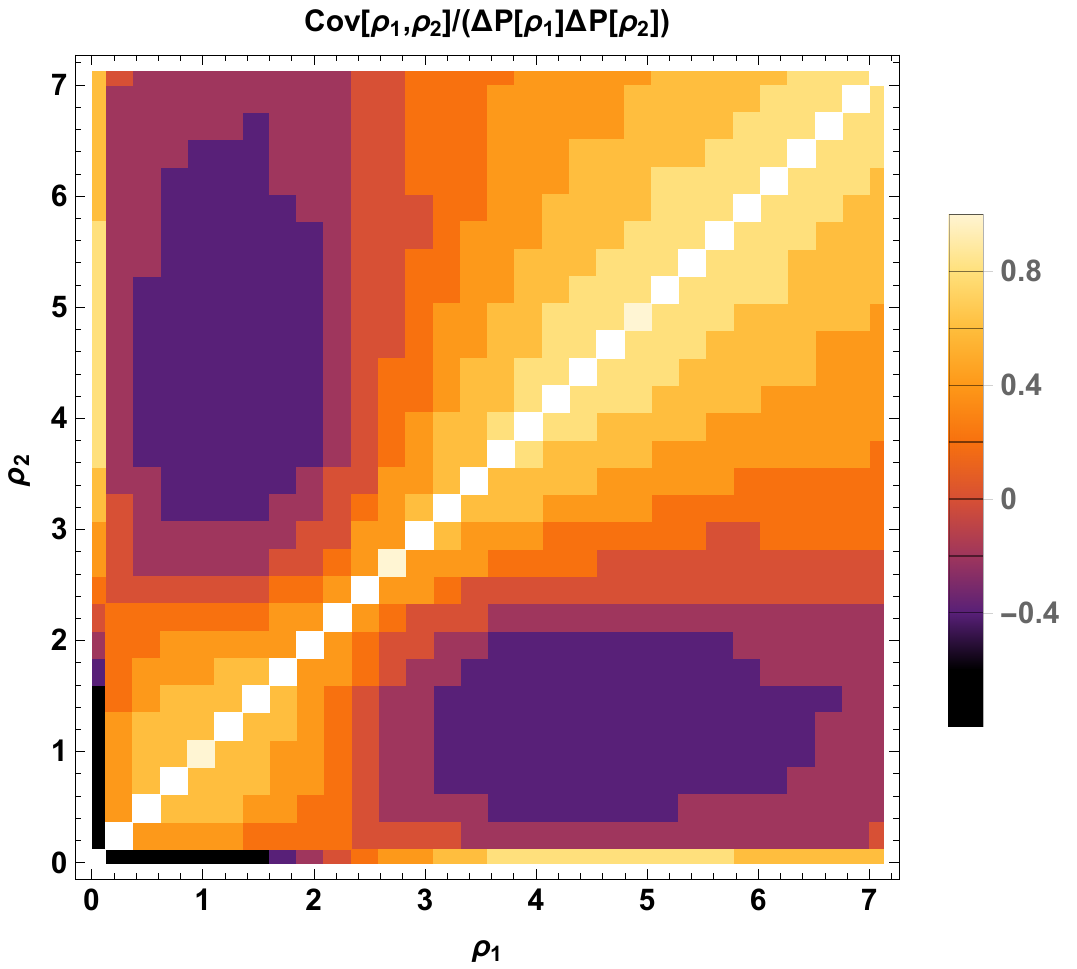}
 \includegraphics[width=5.5cm]{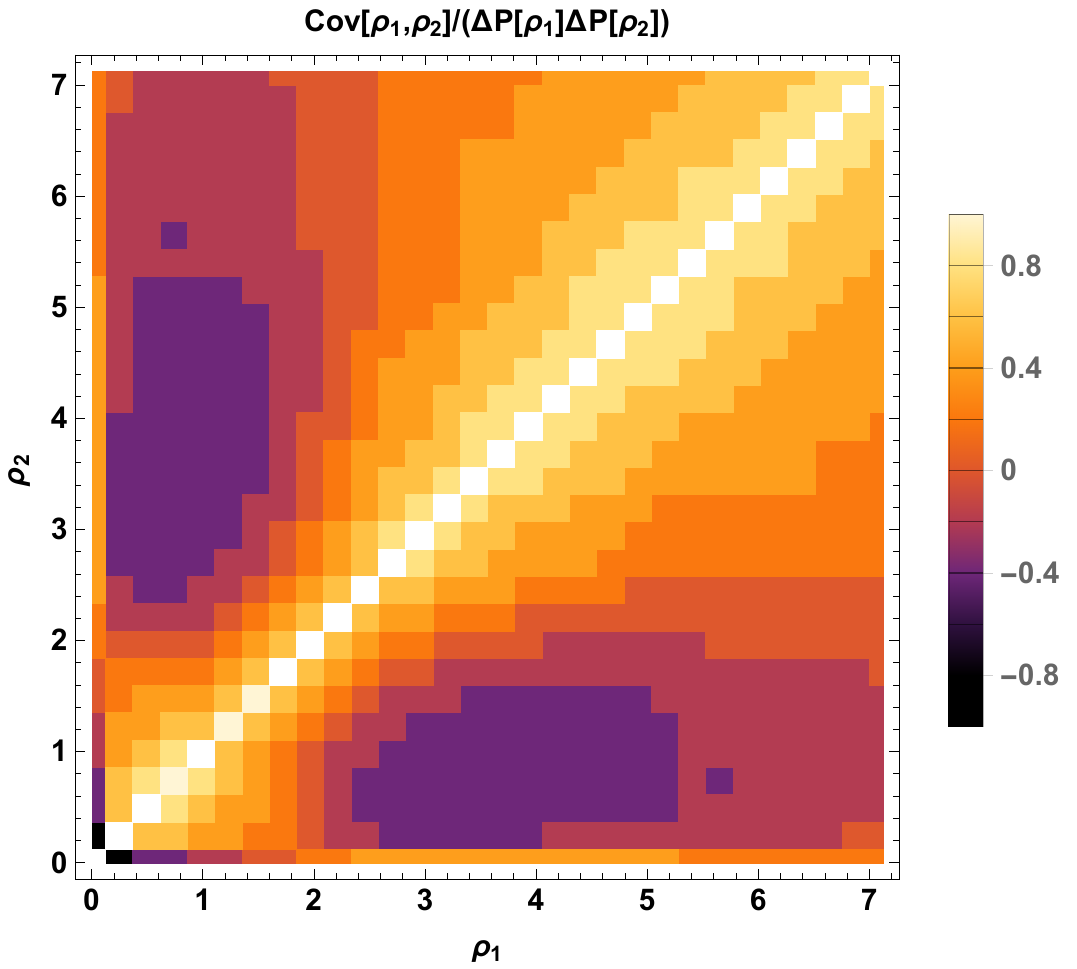}
   \caption{Resulting  reduced covariance matrix for the three types of observables for set $\mA$. The covariance matrix is dominated by its leading eigenvalue and direction, leading to this typical butterfly shape of the reduced covariance matrix.
   \label{NumRedCovariance}}
\end{figure*}

\begin{figure}
   \centering
 \includegraphics[width=7cm]{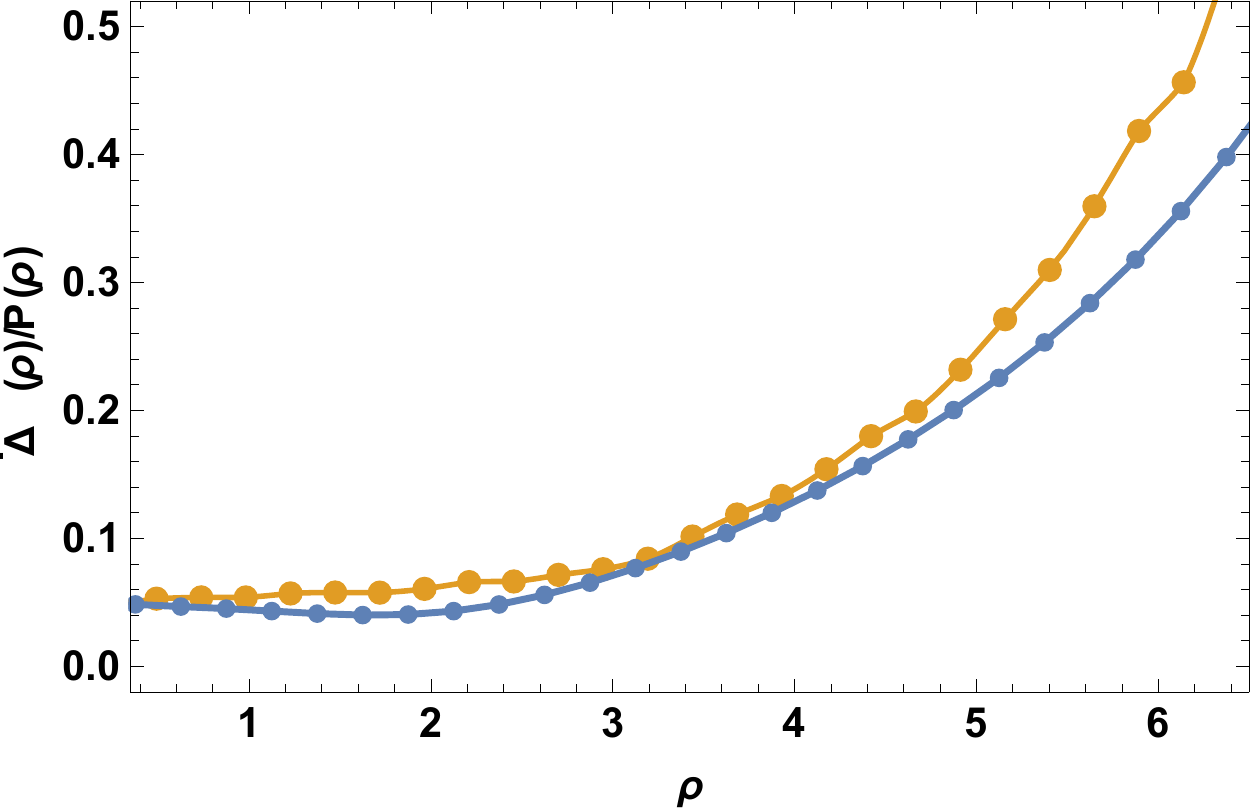}
   \caption{Measured variance of the density PDF obtained for set $\mB$). Symbols are the same as in Fig. \ref{DPDF_SetA_0p5}.}
   \label{DPDF_SetB_0p5}
\end{figure}

A series of experiments of 2D walks with a large number of samples were performed as described below. We restricted our analysis to  
$\alpha=0.5$ with $l_{0}=0.003$ pixel size (the dependence on $l_{0}$ was tested as illustrated on Fig. \ref{densityPDF-TH}, where $l_{0}=.006$ was also used, but the analyses were made for a fixed value of $l_{0}$). Fig. \ref{samplepoints} illustrates how points are distributed in these samples. The point distribution does not show the filamentary structure of realistic cosmological simulations. It exhibits the presence of concentrated halos surrounded by empty regions, however, which are reminiscent of the structure of the largest matter concentrations of the cosmic web.

Two different setting were employed to explore different aspects of the results that were found:
\begin{itemize}
\item Set $\mA$: 1600 samples extracted from a single numerical realization (with periodic boundary conditions) with a size of $8000 \times 8000$ pixels$^{2}$ containing $64 \times 10^{6}$ points. Each sample then has $200 \times 200$ pixels$^{2}$  containing an average of $200^{2}$ points each. For this set of samples, the average and covariance of the PDF were extracted following the three procedures mentioned before: either the density was taken with respect to the mean density of the realization, with respect to the density of each sample, or by subtracting the sample density. It therefore corresponds to an evaluation of the mean and covariance of the PDF of $\rho_{i}$, $\hrho_{i}$ , and $\rhob_{i}$ , respectively.
\item Set $\mB$: 1600 samples, each with periodic boundary conditions, with a size of $200 \times 200$ pixels$^{2}$ containing $200^{2}$ points each. By construction, the average two-point function in the sample, $\xib_{s}$ , vanishes in this case, and covariance is entirely due to proximity effects. 
\end{itemize}

In each case, the local density was obtained after a filtering procedure. The point positions were first pixelized, that is, each point was attributed to a pixel so that the mean number of points per pixel was one. The field was then filtered by a (quasi) circular top-hat functions. In practice, the number of pixels in the window function was 57. This makes the effective smoothing radius about 4.25 in pixel units. The resulting density was then measured at each pixel location. Their histograms were then computed after density binning. To avoid large undue discrete effects, the bin width was chosen to be a multiple of $1/57,$ and in order to ensure that the requirement (\ref{dminhalo}) was met at the pixel distance, we chose a bin width of about $1/4$, more precisely, of $14/57$.

Fig. \ref{densityPDF-TH} shows the resulting PDF as measured in the simulations and how it compares to the theoretical prediction, Eq. (\ref{LFpdf}), for two different choices of $l_{0}$. The expected scaling for $\xib$ is recovered. The measured PDF also follows theoretical predictions for a wide range of probabilities remarkably well. It gives us confidence in the whole procedure and in the approaches used to compute PDFs in this model. The detailed comparisons were made for $l_{0}=0.003,$ leading to $\xib=1.09,$ and a sample density variance in sets $\mA$ given by $\xib_{s}=0.09$. 

The measured variance of the density PDF is obtained from 1600 samples in each case. The resulting shapes are presented in Figs. \ref{DPDF_SetA_0p5} and \ref{DPDF_SetB_0p5} for the different cases, density in a supersample realization, and in samples with periodic boundary conditions. The results show the comparison between results obtained in the numerical simulations with yellow symbols, and results derived from the analytic prescriptions as blue dots, based on the mean-field approximation. The agreement between the two is very good. The overall shape of the variance and its dependence on the density is well reproduced. Discrepancies can be observed for densities above 4 or 5, however, where the theoretical predictions are seen to underestimate the results. The reasons for these discrepancies are not clear at this stage. A possible explanation might be the finite number of samples that is used to infer the variances\footnote{Although the number of samples is large, the number of haloes contained in each sample is finite leading to discretization errors in the estimate of the covariance. Estimate of the minimal number of realizations required to make such estimates is beyond the scope of this paper.}. The variance of the density PDF is also compared with the large-scale contributions (\ref{NaiveCov}),
(\ref{NaiveCovs1}), and (\ref{NaiveCovs2}) for set $\mA$ depending on the cases (at this order, the covariance vanishes for set $\mB$). It shows that this formula captures some features of the variance (especially at low and moderate densities), but does not account for all. This is also illustrated in Fig. \ref{NumRedCovariance}, which shows the reduced covariance. The fact that the covariance is determined to a large extent by its leading large-scale contribution leads to values of the reduced covariance close to 1 or -1, leading to these butterfly patterns. Proximity effects, not captured in these forms, also contribute to the covariances at a significant level, however. This is already apparent in Fig. \ref{DPDF_SetA_0p5}.

\begin{figure*}[tbp]
   \centering
 \includegraphics[width=5.5cm]{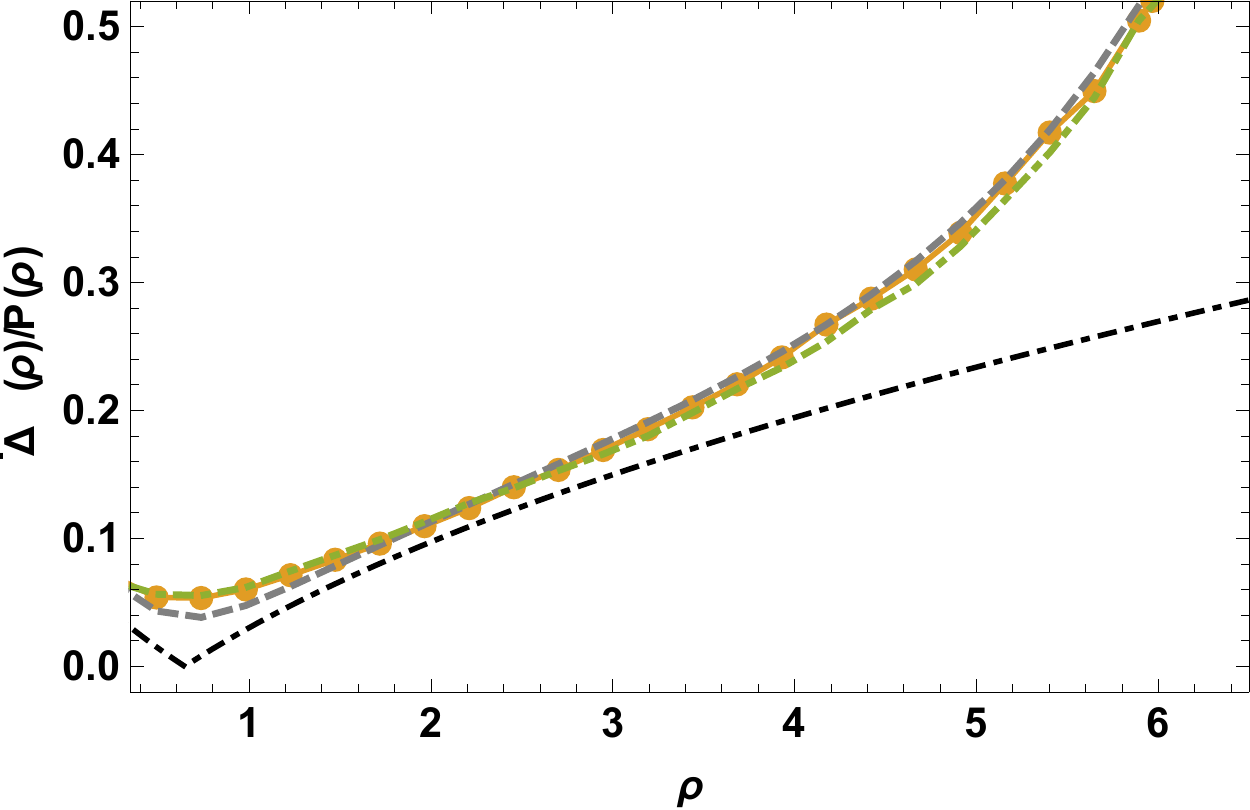}
 \includegraphics[width=5.5cm]{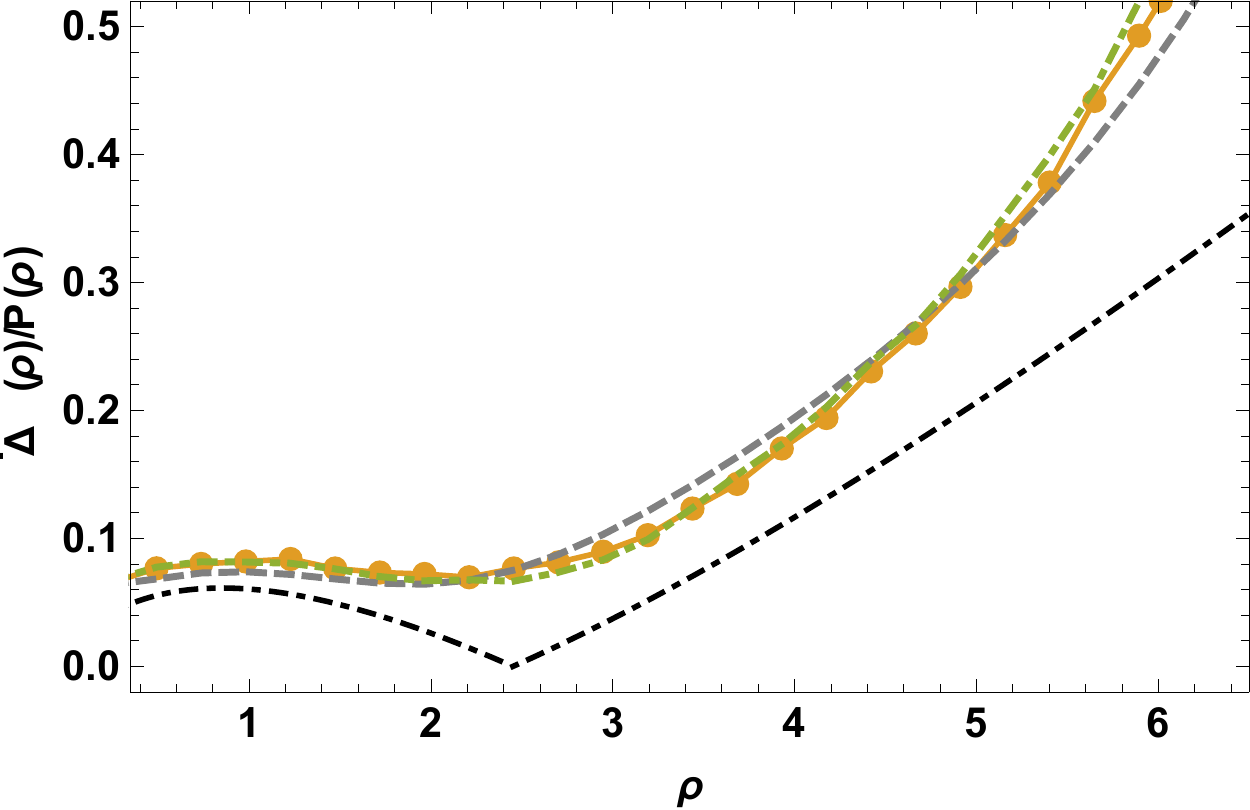}
 \includegraphics[width=5.5cm]{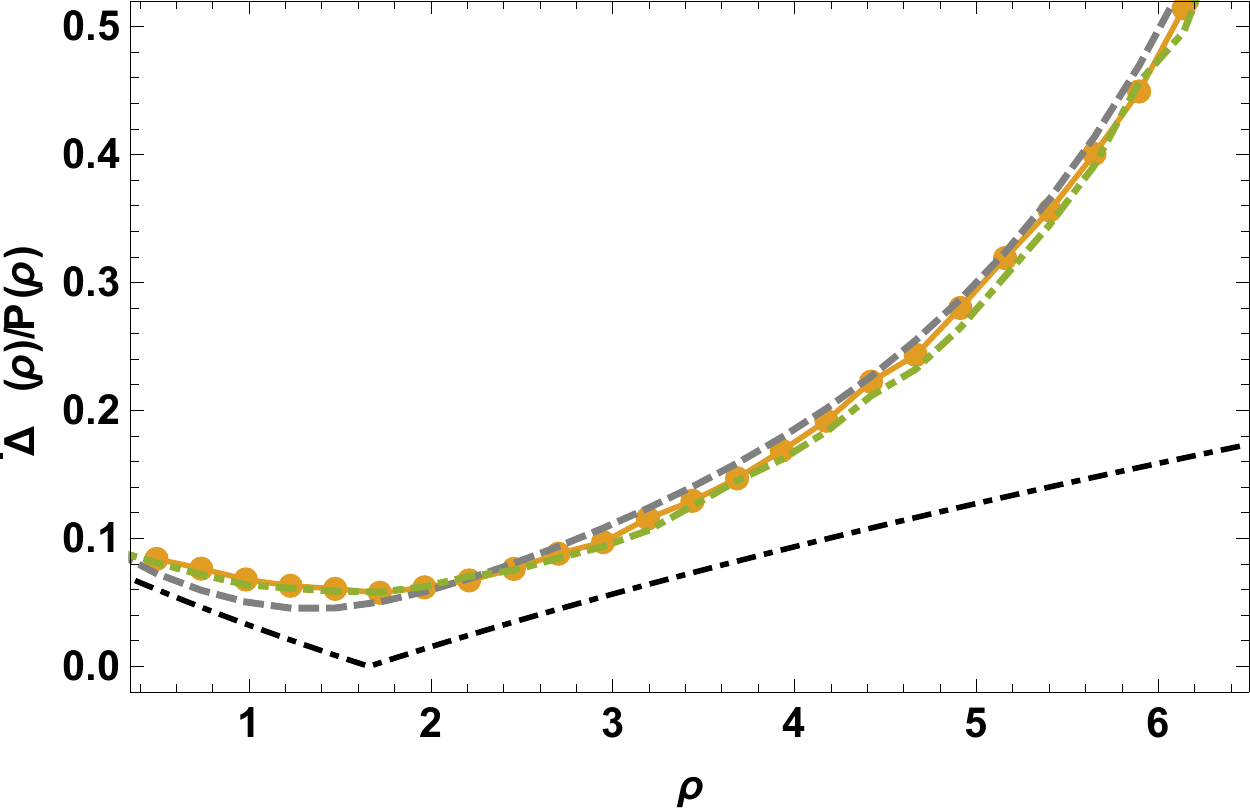}
   \caption{Measured variance of the density PDF, i.e., diagonal elements of the covariance matrix, in sets $\mA$ and comparisons with proposed approximate forms. The yellow line and symbols are the results obtained in the numerical experiments. The dot-dashed line is the prediction derived from relation (\ref{CovExpAp2}), and the dashed gray line shows the prediction from Eq. (\ref{CovExpAp1}). The dot-dashed black lines correspond to the large-scale contributions.}
   \label{DPDF_ApForm_0p5}
\end{figure*}

\begin{figure*}[tbp]
\centering
 \includegraphics[width=5.5cm]{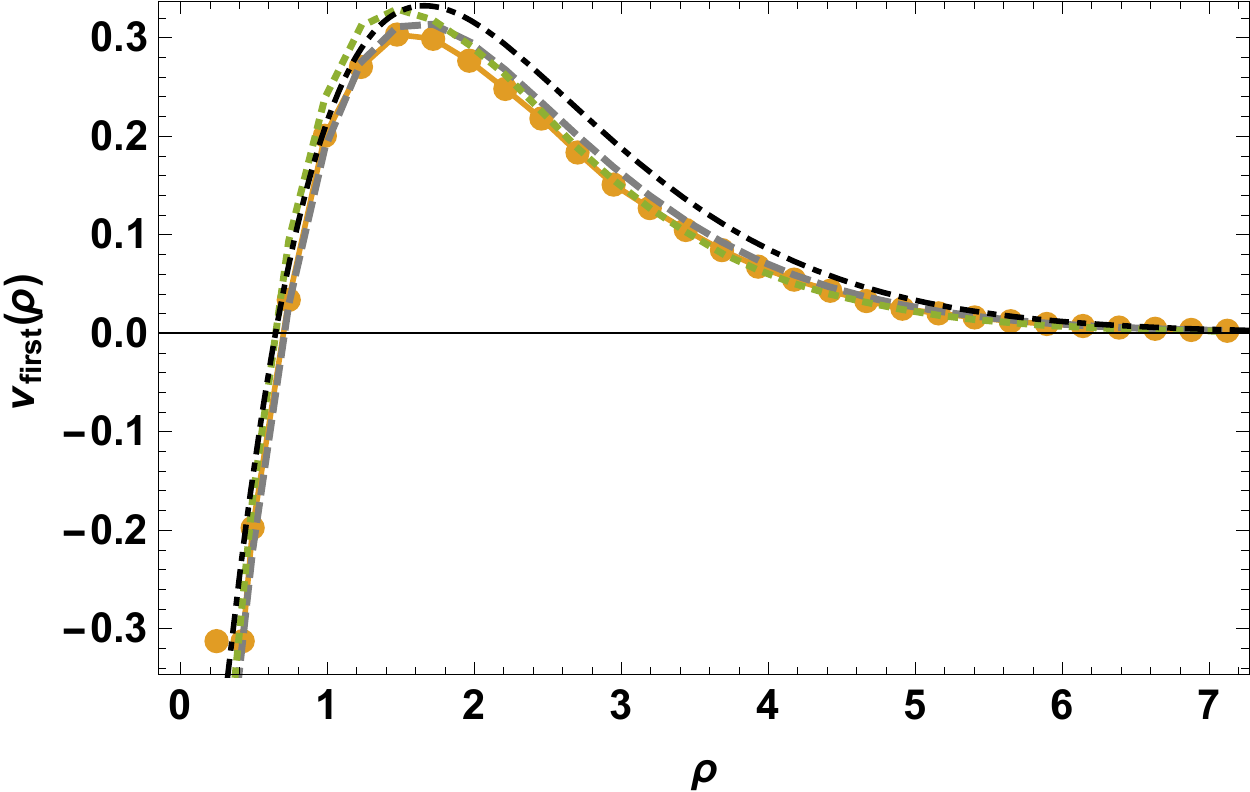}
 \includegraphics[width=5.5cm]{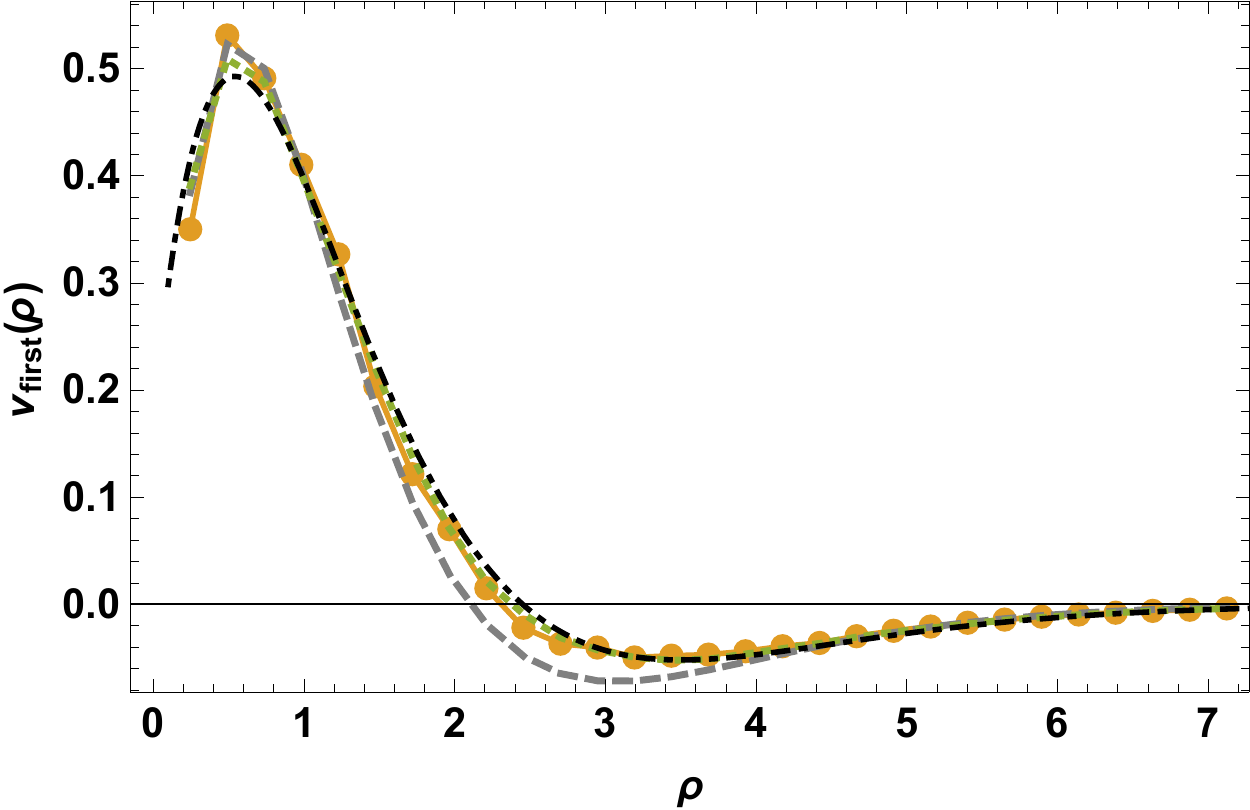}
 \includegraphics[width=5.5cm]{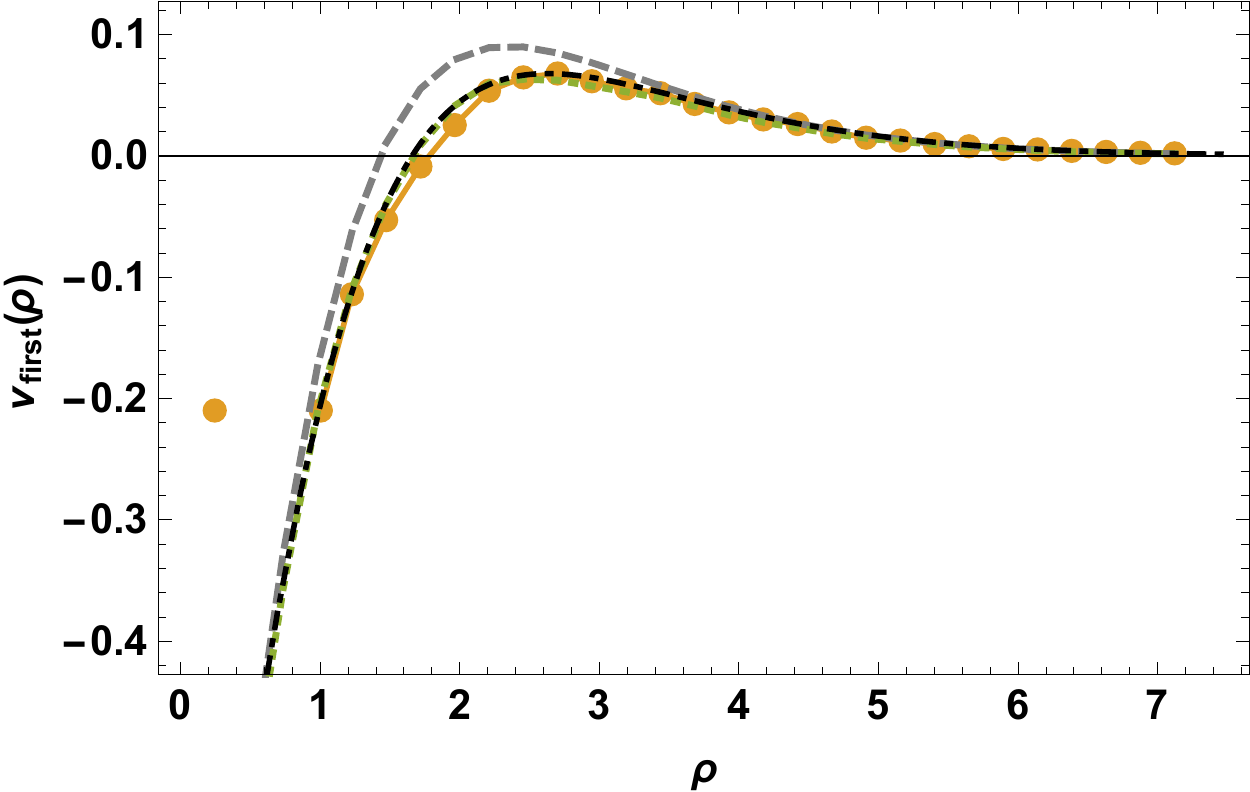}
  \caption{Behavior of the first eigenvector with the same color-coding as in Fig. 6. The dashed black lines are the large-scale prediction, $b_{\#}(\rho_{i}) P(\rho_{i})$ appropriately normalized. The size of the data vector is 30.} 
\label{FirstEigenvect}
\end{figure*} 

\subsection{Testing models of covariance matrices}

\begin{figure*}[tbp]
\centering
 {\includegraphics[width=5.5cm]{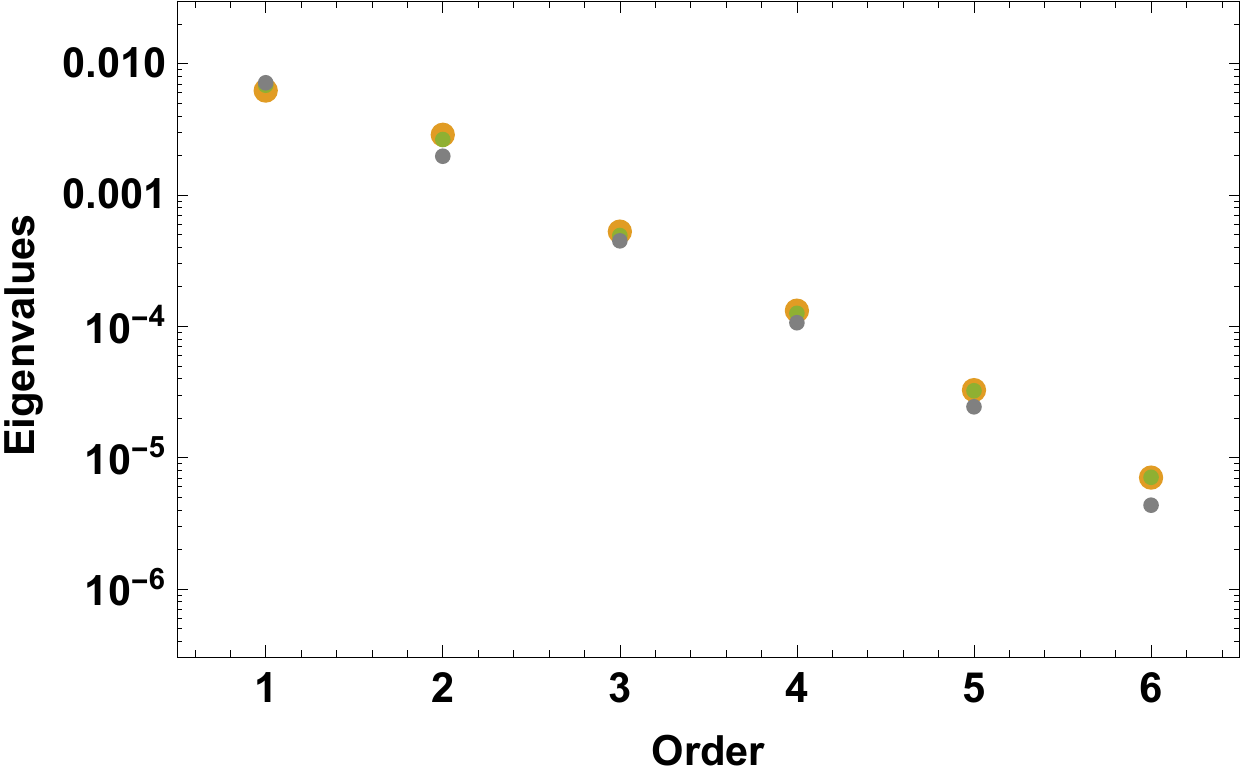}
 \includegraphics[width=5.5cm]{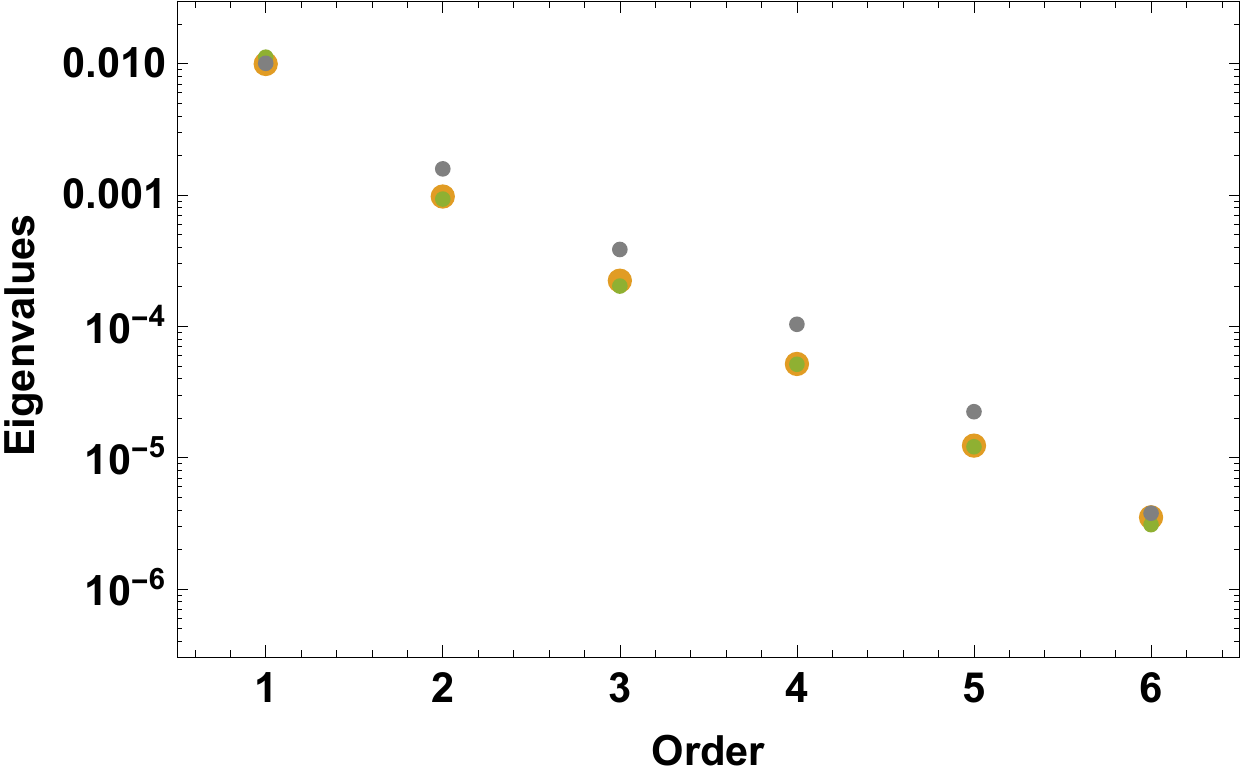}
 \includegraphics[width=5.5cm]{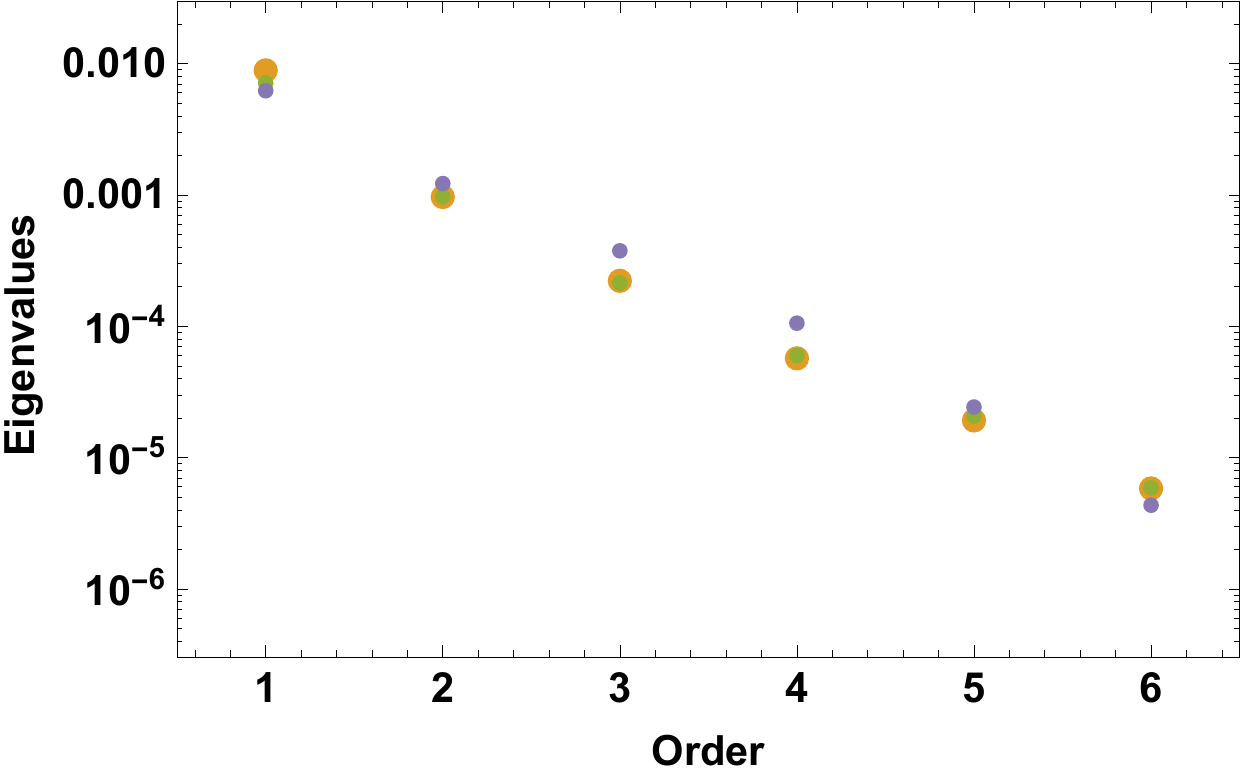}\vspace{.3cm}}
 \includegraphics[width=5.5cm]{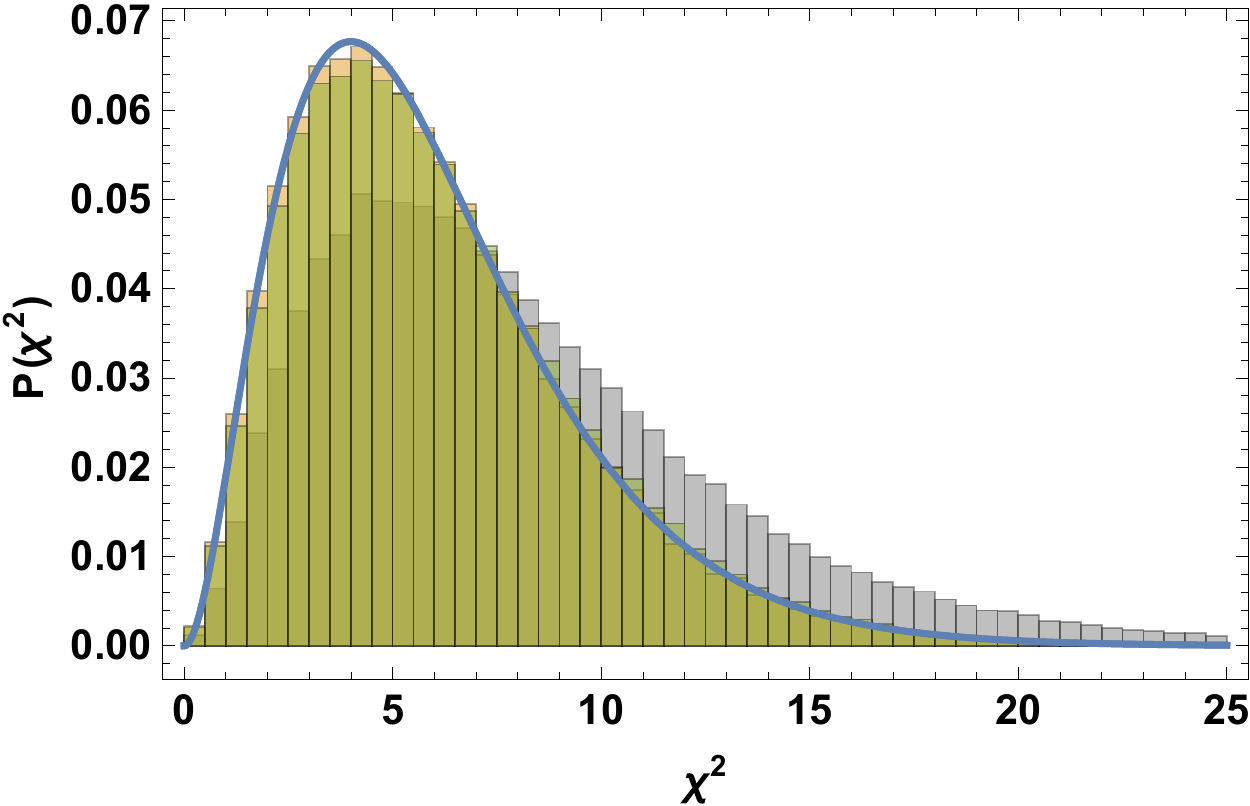}
  \includegraphics[width=5.5cm]{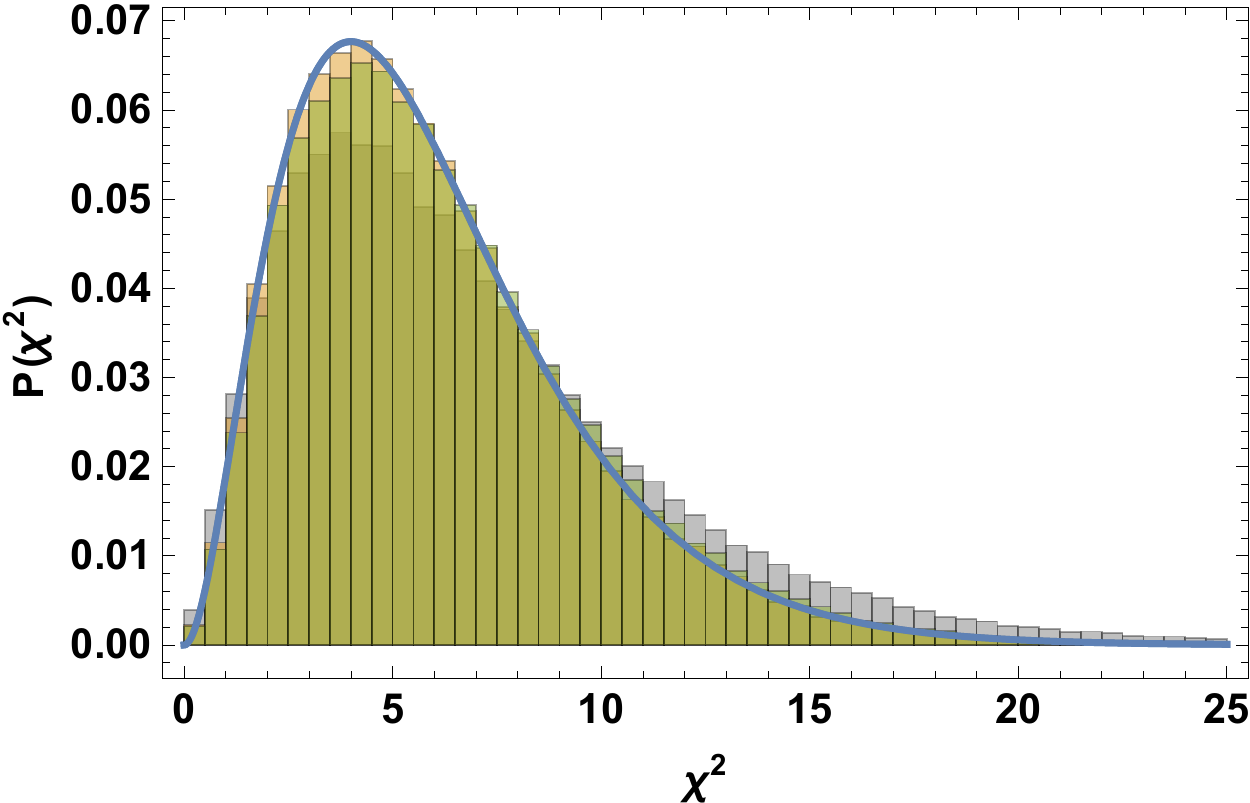}
 \includegraphics[width=5.5cm]{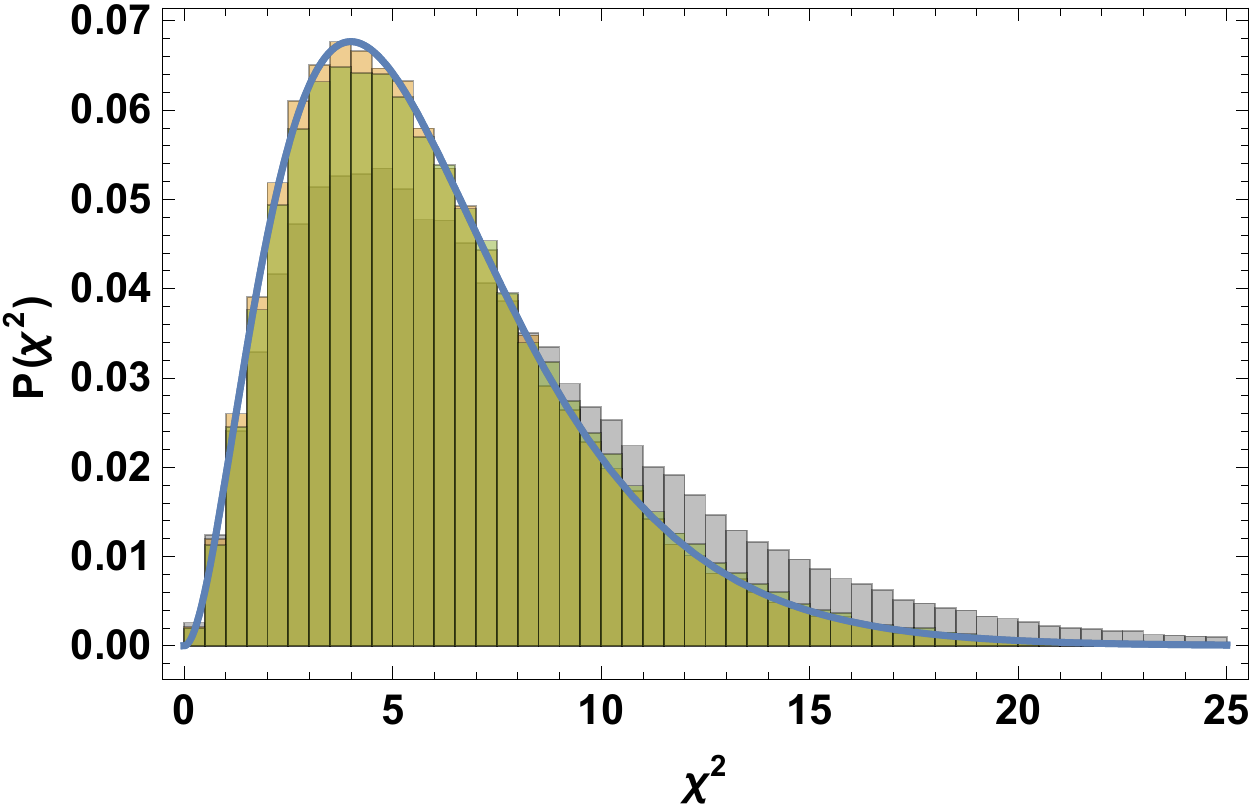}
   \caption{
Performances of the approximate forms of the covariance matrix in terms of rigenvalues and $\chi^{2}$-distributions.
Top 
panel: Eigenvalues of the covariance matrices (rebinned into six bins) compared to what can be obtained from the proposed approximate forms; same color-coding as for Fig. \ref{DPDF_ApForm_0p5}. 
The $\chi^{2}$ distributions are shown in the bottom panel. Model (\ref{CovExpAp2}) reproduces the very same $\chi^{2}$ distributions. Model (\ref{CovExpAp1}), in gray, is not as accurate and tends to slightly overestimate the $\chi^{2}$. 
This latter behavior is amplified when a larger number of bins is used.
   \label{ChiSquareTests}}
\end{figure*}

Expressions (\ref{CovExpAp1}) and (\ref{CovExpAp2}) are precise propositions to show how the large-scale contributions can be completed to account for the full form of the covariance. The comparisons between the predicted form and those obtained from the numerical experiments are explored in detail at different levels and using the following criteria:
\begin{itemize}
\item amplitude of the PDF variance,
\item density dependence of the first eigenvalue of the covariance matrix,
\item amplitude of the eigenvalues of the covariance matrix, and
\item resulting $\chi^{2}$ distribution of a set of data vectors drawn from the original covariance. 
\end{itemize}
These comparisons are shown in figures \ref{DPDF_ApForm_0p5} to \ref{ChiSquareTests}. For model (\ref{CovExpAp2}), the term 
$\Cov^{\PBC}(\rho_{i},\rho_{j})$ is taken from the measured covariance of set $\mB$. Figs. \ref{DPDF_ApForm_0p5}  and \ref{FirstEigenvect} show that these two prescriptions give a good account of the leading behavior of the covariance matrix. The conclusion is quite sensitive for the choice of $r_{\max}$ for prescription (\ref{CovExpAp1}). On the other hand, there is no free parameter that can be adjusted for model (\ref{CovExpAp2}). Interestingly, Fig. \ref{FirstEigenvect} shows that the PDF variance also departs significantly from the large-scale term. The first eigenvector reproduces the functional form of the large-scale density-bias functions very faithfully.

The last two criteria are designed to verify that the reconstructed covariances also capture the subleading behavior of the matrix and can eventually be safely inverted and used as a model of likelihood. To avoid numerical uncertainties and make the comparison tractable, we chose to reduce the binning to six bins (through a rebinning of the histograms and densities ranging from $0.5$ to $6.5$). The resulting eigenvalues are shown in the top panel of Fig. \ref{ChiSquareTests}. It shows that the eigenvalues decrease rapidly in amplitude, suggesting that the eigendirections are well sequenced and that the approximate form captures their values rather accurately. Form (\ref{CovExpAp2}) in particular reproduces all six eigenvalues almost exactly.

Finally, $\chi^{2}$ distributions were computed from a set of random values $P_{i} ^{\ex}$ drawn in each case from a Gaussian likelihood built from the measured covariance (with six bins). The values of $\chi^{2}(P_{i} ^{\ex})$  were then computed for each data vector, and their histogram was computed from each of the proposed models (including the original model for reference),
\begin{equation}
\chi_{\model}^{2}(P_{i} ^{\ex})=\frac{1}{2}\sum_{ij}\mN_{\model}^{ij}P_{i} ^{\ex}P_{j} ^{\ex}
,\end{equation}
where $\mN_{\model}^{ij}$ is the inverse of the covariance matrix, either computed from Eq. (\ref{CovExpAp1}) or from Eq. (\ref{CovExpAp2}). For the original model, the expected distribution of the $\chi^{2}$ values is then expected to be precisely that of a $\chi^{2}$ distribution with six degrees of freedom. This is indeed what is almost exactly obtained for model (\ref{CovExpAp2}). Results obtained from prescription (\ref{CovExpAp1}) are not quite as good. This is expected as the short-distance effects are estimated rather crudely in Eq. (85). The performance of this prescription deteriorates when the dimension of the data vector (i.e., the number of bins) increases.

\section{Conclusions and lessons}

\begin{figure}
\centering
\includegraphics[width=6.5cm]{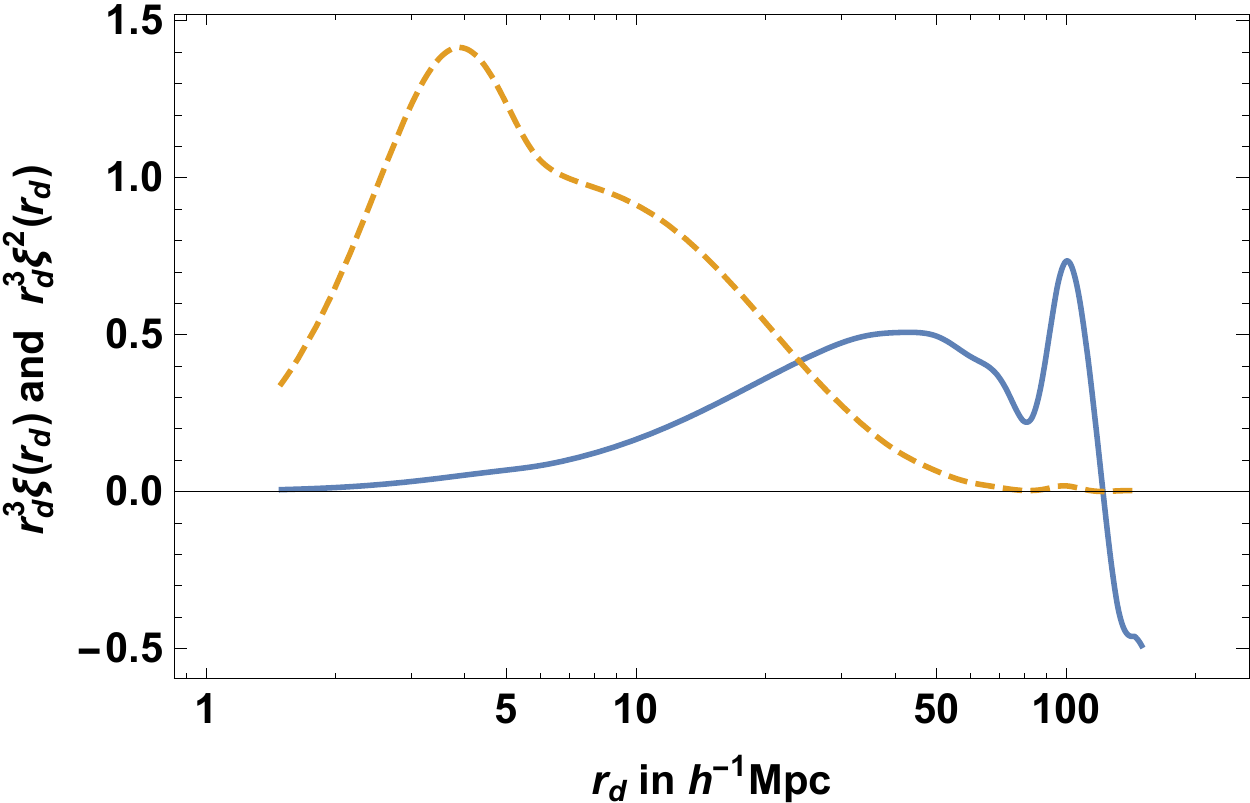}
\includegraphics[width=6.5cm]{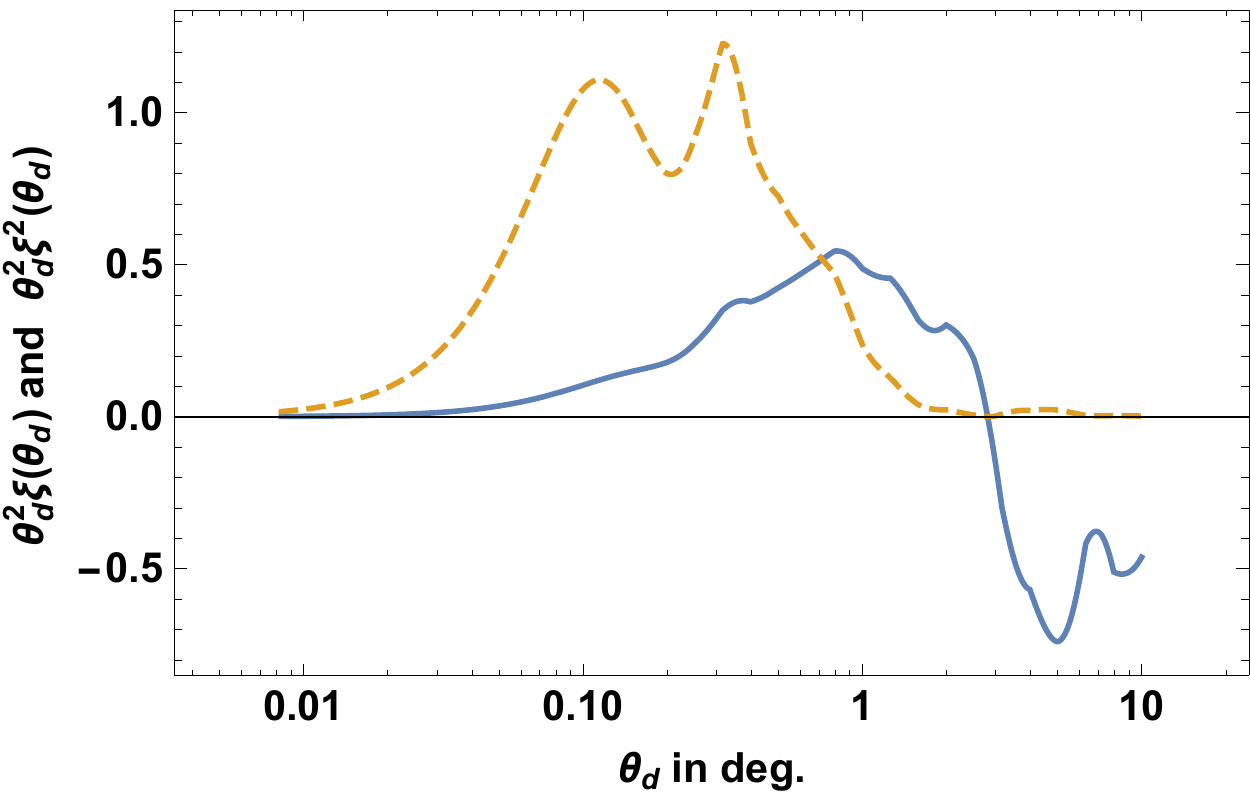}
   \caption{Scale dependence of the matter correlation functions for  a realistic cosmological model \citep[cosmological parameters derived from Plank,][]{2020A&A...641A...6P} for the 3D density and the projected density (for a uniformly sampled survey with a depth of about 800 $h^{-1}$Mpc between $z=0.75$ and $z=1.25$). The top panel shows $r_{d}^{3}\xi(r_{d})$ (solid blue line)  and $r_{d}^{3}\xi^{2}(r_{d})$ (dashed red line) for the 3D density field, and 
 the bottom panel shows $\theta^{2}_{d}\xi(\theta_{d})$ and $\theta^{2}_{d}\xi^{2}(\theta_{d})$ for the projected density. In both cases, the average value of the first moment of the two-point correlation function is dominated by large-distance contributions, whereas short-distance contributions dominate the second moment, assuming survey sizes of about 100 $h^{-1}$Mpc or above. 
   \label{xider_moments}}
\end{figure} 
 
We presented key relations that give the large-scale behavior of the joint PDF, and hence the leading behavior of the covariance matrix of the density PDF. 
These contributing terms do not give the expression of a covariance matrix that can be used to build a likelihood function, 
however, as it is not invertible. Further significant contributions are found to be due to small separation effects, and an approximate form
is proposed in eq. (\ref{shortdistjPDF}). The latter is found to encapsulate most of the proximity effects, that is, it informs 
on the fact that nearby regions are likely to be correlated. They also give an indication on the minimal grid size that can be used 
the maximum bin size that can be used without information loss for a given bin width. 

We then used a toy model for which numerical experiments can easily be performed and for which the exact PDF and large-scale covariance  can be derived. It allows us to evaluate the efficiency of approximate schemes precisely. The conclusions of these comparisons are listed below.\begin{itemize}
\item The theoretical forms Eqs. (\ref{NaiveCov}, \ref{NaiveCovs1}, and \ref{NaiveCovs2}) give the leading-order expression of the covariance elements when supersample effects are taken into account. It gives an accurate prediction of the leading eigenvalue and eigendirection of the covariance matrix.
\item Whether subdominant effects can be accounted for by subsequent terms depends on the behavior of the two-point function: if the r.m.s. of the two-point function is dominated by large separations, then next-to-leading-order effect need to be taken into account; otherwise, short-distance effects will be the dominant contributor. 
\item In case short-distance effects dominate, the covariance matrix can be accessed from small simulations provided the relevant dominant large-scale contributions are added. 
\item This suggests that in realistic situations, the supersample effects, that is, the effects due to modes whose wavelength is larger than the size of the survey, have limited impact on the structure of the covariance matrix and that they can be  captured by the only leading large-scale contribution. This is supported by a further analysis of the behavior of $\xi(r_{d})$ in realistic cosmological settings. For the standard model of cosmology \citep[as derived from cosmic microwave background observations,][]{2020A&A...641A...6P}, the behavior of the matter correlation function can be derived. This is illustrated in Fig. \ref{xider_moments}, which illustrates the scales that are the main contributors to the first two moments of the two-point correlation function. Whether in 2D or in 3D, the first moment is dominated by large-scale contributions, whereas the second moment is dominated by small-scale contributions.
\item In the context of this study, we  assumed that the measured $P_{i}$  were Gaussian distributed. Although it is difficult to assess the accuracy of this hypothesis, the structure uncovered in section \ref{sec:PDFcovariances} can be used to make such an attempt. In tree models, higher-order expressions of the joint density PDFs are expected to preserve the tree structure; see \citet{1999A&A...349..697B}. The connected part of the three-point joint density PDF is then expected to take the form
\begin{eqnarray}
&&\Cov(\rho_{i},\rho_{j},\rho_{k})=\nonumber\\
&&b_{2}(\rho_{i})P(\rho_{i})\,\xib_{s}\,b(\rho_{j})P(\rho_{j})\,\xib_{s}\,b(\rho_{k})P(\rho_{k})+\sym,\end{eqnarray}
where $b_{2}(\rho)$ is the two-line bias function of amplitude similar to $b^{2}(\rho)$. This implies in particular that the third-order cumulant is about $b(\rho)^{4}P(\rho)^{3}\xis^{2}$, much smaller than $\left[b(\rho)^{2}P(\rho)^{2}\xis\right]^{3/2}$, making the distribution of the measured values of $P(\rho)$ (quasi-) Gaussian distributed. There might be some combination of $\rho_{i}$ and values of $\xis$ , however, for which a higher-order term could play a role in the expression of the likelihood function. 
\end{itemize}

For the application of these formulae in practical cases, some limitations have to be noted. We list them below.
\begin{itemize} 
\item In the proposed form, the fact that in practice, PDFs are generally measured  on a grid, that is, on a finite set of locations, is not taken into account. For instance, the exclusion of nonoverlapping cells is not considered. this is expected to introduce additional noise in the PDF estimates. The covariance matrix for these constructions cannot then be derived from general formulae (\ref{keyrelationCov}) even when the integral in $r_{d}$ is restricted above a given threshold.
\item Relation (\ref{shortdistjPDF})  has been derived in a specific regime  (using saddle point approximations) for tree hierarchical models.  They are expected to capture the phenomenon at play for ``typical'' values of the densities, but they may not perform so well in the rare event tails (the exception being the minimum model, for which it is exact). Further checks of the validity of (\ref{shortdistjPDF}) should therefore certainly be done.
\item The general formulae (\ref{NaiveCov}, \ref{NaiveCovs1}, and \ref{NaiveCovs2}) are valid for any type of filtering schemes, even for a compensated filter. This is not the case for relation (\ref{shortdistjPDF}). The proximity effects for compensated filters ought to be considered specifically.
\item Prescription (\ref{CovExpAp2}) is found to give a very precise account of the properties of the covariance matrix. It is based on the proposition that large-scale (supersample) effects can be added separately from the proximity effects and that the latter can be evaluated with small-scale mocks in which supersample effects are absent (with periodic boundary conditions). This is not an exact result, however,. It relies in particular on the fact that the r.m.s. of the $\xi_{s}$ is dominated by scales much smaller than the sample size.
\item  Prescription (\ref{CovExpAp1}) is less solid. It can be used for a quick assessment of the different contributing terms, or to build fully invertible covariance matrices, but it is unlikely to give reliable predictions at the $\chi^{2}$ level.
\end{itemize}
In all cases, prescriptions (\ref{CovExpAp1}) and (\ref{CovExpAp2}) can be the starting point of a more precise evaluation of the covariance from specific numerical experiments that can complement its evaluation following the approach presented in  \cite{2018MNRAS.473.4150F}, for instance. The authors also showed that some strategies could be adopted to limit the number of realizations required to reach a specific precision. This point is not discussed here. 

\begin{acknowledgements} 
The author of this article is indebted to Cora Uhlemann, Alex Gough, Oliver Friedrich, Sandrine Codis, Aoife Boyle and Alexandre Barthelemy for many comments and careful examination of the preparatory notes of this manuscript. 
 \end{acknowledgements}

\bibliographystyle{aa}
\bibliography{PDFCVariance}
\appendix
\section{Hierarchical tree models}

In hierarchical tree models, the $p$-point matter correlation functions are assumed to follow tree structures in the sense
described in the main text. They are thus entirely defined by the two-point functions $\xi(r)$ and the vertex-generating function
$\zeta(\tau)$. 
The exact generating function of multiple cell correlation functions can be built through simple transforms. We therefore consider a set of $n$ 
cells $V_{i}$. These cells can overlap.  The joint cumulants we consider are those of the average densities in cells $V_{i}$ that can be expressed in terms of spatial averages\footnote{The formulae are written here for a top-hat profile, but can be extended to arbitrary profiles.} of correlation functions,
\begin{eqnarray}
\langle\rho_{1}^{p_{1}}\dots\rho_{n}^{p_{n}}\rangle_{c}&=&
\int_{V_{1}}\frac{\dd\vx_{1,1}}{V_{1}}\dots
\int_{V_{1}}\frac{\dd\vx_{1,p_{1}}}{V_{1}}
\dots
\nonumber\\
&&\hspace{-2cm}
...
\int_{V_{n}}\frac{\dd\vx_{n,1}}{V_{n}}\dots
\int_{V_{n}}\frac{\dd\vx_{n,p_{n}}}{V_{n}}
\nonumber\\
&&\hspace{-2cm}\times\xi_{p}(\vx_{1,1},\dots,\vx_{1,p_{1}},\dots\vx_{n,1},\dots,\vx_{n,p_{n}})
.\end{eqnarray}
We then wish to build the cumulant-generating function,
\begin{equation}
\varphi(\lambda_{1},\dots,\lambda_{n})=
\sum_{p_{1},\dots,p_{n}}\langle\rho_{1}^{p_{1}}\dots\rho_{n}^{p_{n}}\rangle_{c}\,\frac{\lambda_{1}^{p_{1}}}{p_{1}!}\dots\frac{\lambda_{n}}{p_{n}!}
.\end{equation}
This function represents the generating function of (averaged) tree diagrams where $\lambda_{i}$ counts the number of points 
in each cells. 
As shown in \cite{1999A&A...349..697B}, this is obtained with the help of the intermediate function $\tau(\vx)$ solution of the consistency equation\footnote{It takes the very same form as the stationary equation in the context of the large deviation principle, although the results here do not correspond to this regime.},
\begin{equation}
\tau(\vx)=\sum_{j}\lambda_{j}\int_{V_{j}}\frac{\dd\vx}{V_{j}}\,\xi(\vx,\vx')\,\zeta'(\tau(\vx'))\label{treesumtau}
,\end{equation}
and then
\begin{eqnarray}
\varphi(\lambda_{1},\dots,\lambda_{n})&=&\sum_{j}\lambda_{j}
\int_{V_{j}}\frac{\dd\vx}{V_{j}}\zeta(\tau(\vx))
\nonumber\\
&&\hspace{-0cm}
-\frac{1}{2}\sum_{j}\lambda_{j}\int_{V_{j}}\frac{\dd\vx}{V_{j}}\tau(\vx)\zeta'(\tau(\vx))
\label{treesumphi}
.\end{eqnarray}
This is an exact result based on pure combinatorics.

For cases of interest, it is possible to do a mean-field approximation that consists of assuming that $\tau(\vx)$ is constant within each cell. 
We then have the system of equations for $\tau_{i}$,
\begin{equation}
\tau_{i}=\sum_{j}\lambda_{j}\xi_{ij}\zeta'(\tau_{j})
,\end{equation}
where
\begin{equation}
\xi_{ij}=
\int_{V_{i}}\frac{\dd\vx_{i}}{V_{i}}
\int_{V_{i}}\frac{\dd\vx_{i}}{V_{i}}
\xi(\vx_{i},\vx_{j})
\end{equation}
and 
\begin{equation}
\varphi(\lambda_{1},\dots,\lambda_{n})=
\sum_{j}\lambda_{j}\left[\zeta(\tau_{j})-\frac{1}{2}\tau_{j}\zeta'(\tau_{j})\right].
\end{equation}
\cite{1999A&A...349..697B} found this to be very accurate, and we extensively use this approximation in the following, in particular for the minimal tree model.

\section{Joint PDF, density-bias function in PT, and hierarchical tree models}

Here we consider the joint distribution of densities in two cells whose centers are at distance $d$. The calculation is based on the 
inverse Laplace transform of the joint cumulant-generating function $\varphi(\lambda_{1},\lambda_{2}),$
\begin{equation}
\varphi(\lambda_{1},\lambda_{2})=\sum_{p,q}\langle\rho_{1}^{p}\rho_{2}^{q}\rangle_{c}\,\frac{\lambda_{1}^{p}}{p!}\,\frac{\lambda_{2}^{q}}{q!}
,\end{equation}
where $\langle\rho_{1}^{p}\rho_{2}^{q}\rangle_{c}$ are the cumulants of the local density fields. They depend on the size and distance $d$
between the cells. We assume in particular that the density correlation function between cells $\xi(d)$ is small compared to unity and can serve as a small parameter. 

\subsection{Leading order in the mean-field approximation}
\label{biasmeanfield}
Expanding with respect to $\xi(d)$ then leads to the following form:
\begin{equation}
\varphi(\lambda_{1},\lambda_{2})=\varphi_{0}(\lambda_{1})+\varphi_{0}(\lambda_{2})+\xi(d)\varphi_{1}(\lambda_{1})\varphi_{1}(\lambda_{2})\label{uptophi1}
,\end{equation}
that is, a factorization of the linear term in $\xi(d)$. This can explicitly be shown in case of tree models (as described in the main text). This is also the case in perturbation theory. 

In case of the tree models, we have
\begin{equation}
\varphi_{1}(\lambda)=\frac{\tau(\lambda)}{\xib}
,\end{equation}
where $\xib$ is the averaged correlation function within the cells. This is obtained assuming a mean-field approximation. We expect subleading corrections when $d$ becomes comparable to the size of the cells.

\subsection{Extending the previous case to the three variable case}

In addition to the two variables $\rho_{1}$ and $\rho_{2}$ , we introduce here the variable $\rho_{s}$ , which is the density within the sample.
We assume that the correlation functions are built with the same model. Here the small parameter is the correlation between two cells $V_{1}$ and $V_{2}$  (at positions $\vx_{1}$ and $\vx_{2}$) and the correlation function at sample size. It is natural in the context we are interested in to assume that these two quantities are on the same order.

We further assume we can use the mean-field approximation for the two cells $V_{1}$ and $V_{2}$. This is not a priori the case for the density in the whole sample, however. We therefore derive the results without this approximation.
The general expression is then
\begin{eqnarray}
\varphi(\lambda_{s},\lambda_{1},\lambda_{2})&=&
\lambda_{s}\int\dd \vx_{s}\left(\zeta(\tau(\vx_{s}))-\frac{1}{2}\tau(\vx_{s})\zeta'(\tau(\vx_{s}))\right)
\nonumber\\
&&\hspace{-2.5cm}
+\lambda_{1}\left[\zeta(\tau_{1})-\frac{1}{2}\tau_{1}\zeta'(\tau_{1})\right]
+\lambda_{2}\left[\zeta(\tau_{2})-\frac{1}{2}\tau_{2}\zeta'(\tau_{2})\right]
,\end{eqnarray}
with the consistency relations
\begin{eqnarray}
\tau(\vx_{s})&=&\lambda_{s}\int\dd \vx'_{s}\,\xi(\vx_{s},\vx'_{s})\,\zeta'(\tau(\vx'_{s}))
\nonumber\\
&&\hspace{-.0cm}
+\lambda_{1}\xi(\vx_{s},\vx_{1})\zeta'(\tau_{1})
+\lambda_{2}\xi(\vx_{s},\vx_{2})\zeta'(\tau_{2})\\
\tau_{1}&=&\lambda_{s}\int\dd\vx_{s}\,\xi(\vx_{1},\vx_{s})\,\zeta'(\tau(\vx_{s}))
\nonumber\\
&&\hspace{-.0cm}
+\lambda_{1}\,\xib\,\zeta'(\tau_{1})
+\lambda_{2}\,\xi_{12}\,\zeta'(\tau_{2})\\
\tau_{2}&=&\lambda_{s}\int\dd\vx_{s}\,\xi(\vx_{2},\vx_{s})\,\zeta'(\tau(\vx_{s}))
\nonumber\\
&&\hspace{-.0cm}
+\lambda_{1}\,\xi_{12}\,\zeta'(\tau_{1})
+\lambda_{2}\,\xib\,\zeta'(\tau_{2})
.\end{eqnarray}
We therefore derive the expression of $\varphi(\lambda_{s},\lambda_{1},\lambda_{2})$ up to linear order jointly in 
$\xi(\vx_{s},\vx'_{s})$, $\xi(\vx_{s},\vx_{1})$,  $\xi(\vx_{s},\vx_{1})$ and  $\xi_{12}\equiv\xi(\vx_{1},\vx_{2})$. At zeroth order, $\tau(\vx_{s})$ vanishes and
\begin{equation}
\tau_{1}^{(0)}=\lambda_{1}\,\xib\zeta'(\tau_{1})
,\end{equation}
with a similar relation for $\tau_{2}^{(0)}$. At linear order, we have
\begin{eqnarray}
\tau^{(1)}(\vx_{s})&\!\!=\!\!&\lambda_{s}\int\dd \vx'_{s}\,\xi(\vx_{s},\vx'_{s})
\nonumber\\
&&\hspace{-1.cm}
+\lambda_{1}\xi(\vx_{s},\vx_{1})\zeta'(\tau_{1}^{(0)})
+\lambda_{2}\xi(\vx_{s},\vx_{2})\zeta'(\tau_{2}^{(0)})\\
\mF(\tau_{1}^{(0)})\tau_{1}^{(1)}&\!\!=\!\!&\lambda_{s}\int\dd\vx_{s}\,\xi(\vx_{1},\vx_{s})+\lambda_{2}\,\xi_{12}\,\zeta'(\tau_{2}^{(0)})\\
\mF(\tau_{2}^{(0)})\tau_{2}^{(1)}&\!\!=\!\!&\lambda_{s}\int\dd\vx_{s}\,\xi(\vx_{2},\vx_{s})+\lambda_{1}\,\xi_{12}\,\zeta'(\tau_{1}^{(0)})
,\end{eqnarray}
where
\begin{equation}
\mF(\tau)\equiv\left(1-\frac{\tau\zeta''(\tau)}{\zeta'(\tau)}\right).
\end{equation}
The resulting cumulant-generating function reads
\begin{eqnarray}
\varphi(\lambda_{s},\lambda_{1},\lambda_{2})&=&
\lambda_{s}
+\lambda_{1}\left[\zeta(\tau_{1}^{(0)})-\frac{1}{2}\tau_{1}^{(0)}\zeta'(\tau_{1}^{(0)})\right]
\nonumber\\
&&\hspace{-2.5cm}
+\lambda_{2}\left[\zeta(\tau_{2}^{(0)})-\frac{1}{2}\tau_{2}^{(0)}\zeta'(\tau_{2}^{(0)})\right]+\frac{\lambda_{s}}{2}\int\dd \vx_{s}\,\tau^{(1)}(\vx_{s})
\nonumber\\
&&\hspace{-2.5cm}
+\frac{\lambda_{1}}{2}\zeta'(\tau_{1}^{(0)})\mF(\tau_{1}^{(0)})\tau_{1}^{(1)}
+\frac{\lambda_{2}}{2}\zeta'(\tau_{2}^{(0)})\mF(\tau_{2}^{(0)})\tau_{2}^{(1)},
\end{eqnarray}
where the first three terms     are at zeroth order and the last three are at linear order. Using the previous expression, we obtain
\begin{eqnarray}
\varphi(\lambda_{s},\lambda_{1},\lambda_{2})&=&
\lambda_{s}+\varphi_{0}(\lambda_{1})+\varphi_{0}(\lambda_{2})
\nonumber\\
&&\hspace{-2cm}
+\frac{\lambda_{s}^{2}}{2}\int\dd\vx_{s}\dd\vx'_{s}\,\xi(\vx_{s},\vx'_{s})
+\lambda_{s}\int\dd\vx_{s}\,\xi(\vx_{s},\vx_{1})\,\varphi_{1}(\lambda_{1})
\nonumber\\
&&\hspace{-2cm}+\lambda_{s}\int\dd\vx_{s}\,\xi(\vx_{s},\vx_{2})\,\varphi_{1}(\lambda_{2})
+\varphi_{1}(\lambda_{1})\,\xi_{12}\,\varphi_{1}(\lambda_{2}).\label{3ptbias}
\end{eqnarray}
This relation is used to derive the expression of the sample bias functions in the next section.

\subsection{Second order in the mean-field approximation}

\begin{figure*}
   \centering
 \includegraphics[width=15cm]{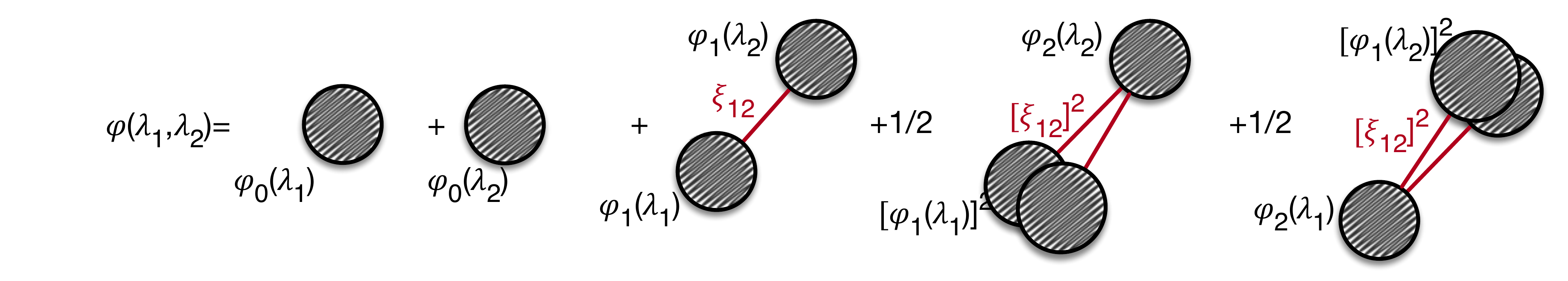}
   \caption{Diagrammatic visualization of the bias functions. The function $\varphi_{0}(\lambda)$ is the generating function of all trees within one cell, $\varphi_{1}(\lambda)$ of all trees within one cell with one external lines, and $\varphi_{2}(\lambda)$ with two external lines. The resulting connected diagrams up to second order in $\xi_{12}$ are thus those presented here. Two $\varphi_{2}$ generating functions cannot be conntected as that would induce a loop contribution. This reflects the underlying tree structure.}
   \label{biasfunctions1}
\end{figure*} 

\begin{figure}
   \centering
 \includegraphics[width=8.5cm]{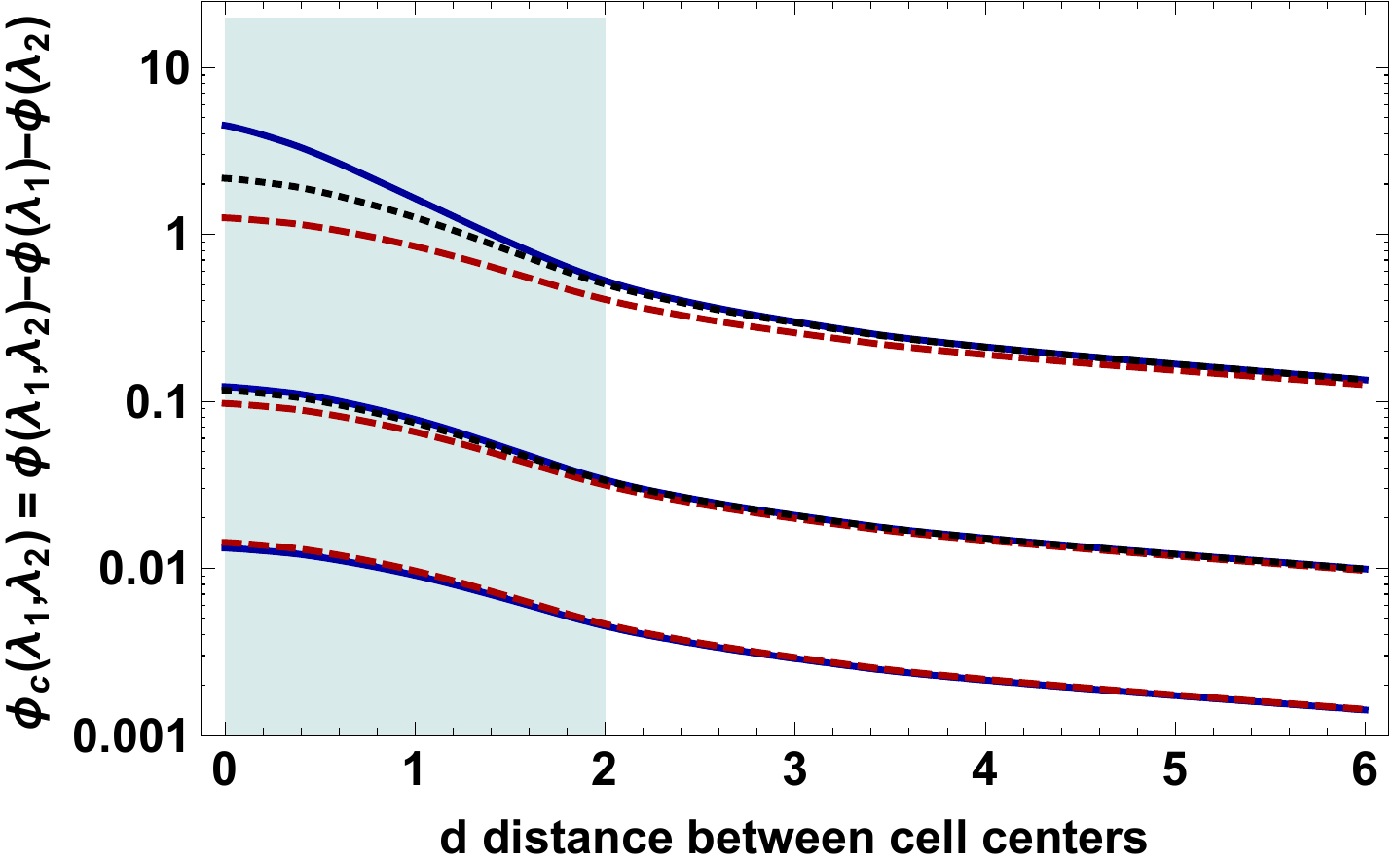}
   \caption{Joint CGF as a function of $d$ and for different values of $\lambda_{1},\lambda_{2}$: from bottom to top,
    $\lambda_{1}=\lambda_{2}=-0.1$,     $\lambda_{1}=\lambda_{2}=0.2$, $\lambda_{1}=\lambda_{2}=0.5$. The solid blue lines correspond to the two-cell mean-field expression, (\ref{MeanFPhi21}). The predictions given in Eqs. (\ref{uptophi1}) and (\ref{uptophi2}) are shown as dashed red lines and  dotted black lines. 
The shaded area is the region of overlapping cells.} 
   \label{phicdistance}
\end{figure} 

Results of Sect. \ref{biasmeanfield} can naturally be extended to any order in the cross-cell correlation function
in the context of the tree-hierarchical models (as illustrated on Fig \ref{biasfunctions1}). Up to second order, it takes the form
\begin{eqnarray}
\varphi(\lambda_{1},\lambda_{2})&=&
\varphi_{0}(\lambda_{1})+\varphi_{0}(\lambda_{2})+
\varphi_{1}(\lambda_{1})\,\xi_{12}\,\varphi_{1}(\lambda_{2})
\nonumber\\
&&\hspace{-1cm}
+\frac{1}{2}\varphi_{1}^{2}(\lambda_{1})\,\xi_{12}^{2}\,\varphi_{2}(\lambda_{2})
+\frac{1}{2}\varphi_{2}(\lambda_{1})\,\xi_{12}^{2}\,\varphi_{1}^{2}(\lambda_{2})\label{uptophi2}
,\end{eqnarray}
where the function $\varphi_{2}(\lambda)$ takes the form
\begin{equation}
\varphi_{2}(\lambda)=\frac{\lambda\zeta''(\tau)}{1-\tau\zeta''(\tau)/\zeta'(\tau)}.
\end{equation}
This last expression can be directly obtained through a perturbative expansion as presented in the previous subsection\footnote{A quicker approach is to view $\varphi_{0}(\lambda)$ as a function of the ``leaf weight'' , that is, the value of $\nu_{1}$; $\varphi_{1}(\lambda)$, $\varphi_{2}(\lambda)$ are then obtained by successive derivatives of $\varphi_{0}(\lambda)$ with respect to $\nu_{1}$.}.

In case of the minimal tree model, this perturbative expansion can be directly compared with exact mean-field results\footnote{it is possible to fully solve the consistency relations in case of two cells; the conclusions remain unchanged.}.
This is shown in Fig. \ref{phicdistance}. It shows that for a large regime in $\lambda,$ the relation (\ref{uptophi2})
provides a very accurate description of the joint cumulant-generating function down to a distance corresponding to overlapping cells. 
For overlapping cells, relation (\ref{uptophi2}) continues to be accurate except for high values of $\lambda$.
In general,
\begin{equation}
\varphi(\lambda_{1},\lambda_{2})\to\varphi_{0}(\lambda_{1}+\lambda_{2})
\end{equation}
when $d\to 0$ (more accurate results in case of the minimal tree model are given in Appendix D.).

Using (\ref{uptophi2}), we then derive corrective perturbative terms to the joint density PDF. More specifically, we have
\begin{eqnarray}
P(\rho_{i},\rho_{j})&=&P(\rho_{i})P(\rho_{j})\left[
1+b_{1}(\rho_{i})\,\xi_{12}\,b_{1}(\rho_{j})
\right.\nonumber\\
&&\hspace{-.5cm}\left.
+\frac{1}{2}b_{2}(\rho_{i})\,\xi_{12}^{2}\,b_{2}(\rho_{j})+
\frac{1}{2}b_{2}(\rho_{i})\,\xi_{12}^{2}\,q_{1}(\rho_{j})
\right.\nonumber\\
&&\hspace{-.5cm}\left.
+\frac{1}{2}q_{1}(\rho_{i})\,\xi_{12}^{2}\,b_{2}(\rho_{j})
\right]\label{jPDF2ndorder}
,\end{eqnarray}
where 
\begin{eqnarray}
b_{2}(\rho_{i})P(\rho_{i})&=&\int\frac{\dd\lambda}{2\pi\ii}\left[\varphi_{1}(\lambda)\right]^{2}\exp\left[-\lambda\rho_{i}+\varphi(\lambda)\right]\\
q_{1}(\rho_{i})P(\rho_{i})&=&\int\frac{\dd\lambda}{2\pi\ii}\,\varphi_{2}(\lambda)\exp\left[-\lambda\rho_{i}+\varphi(\lambda)\right]
.\end{eqnarray}
For a sample with periodic boundary conditions, the average of $\xi_{12}$ vanishes, which a priori makes the other terms the leading contributors to the covariance elements. Equation  (\ref{jPDF2ndorder}) can be written as a sum of symmetric factorized terms,
\begin{eqnarray}
P(\rho_{i},\rho_{j})&=&P(\rho_{i})P(\rho_{j})\left[
1+b_{1}(\rho_{i})\,\xi_{12}\,b_{1}(\rho_{j})
\right.\nonumber\\
&&\hspace{-.5cm}\left.
+\frac{1}{2}\left(b_{2}(\rho_{i})+q_{1}(\rho_{i})\right)\,\xi_{12}^{2}\,\left(b_{2}(\rho_{j})
+q_{1}(\rho_{j})\right)
\right.\nonumber\\
&&\hspace{-.5cm}\left.
-\frac{1}{2}q_{1}(\rho_{i})\,\xi_{12}^{2}\,q_{1}(\rho_{j})
\right]\label{jPDF2ndorderp}
,\end{eqnarray}
showing the eigenstructure of the resulting matrix and showing that it defines three different eigendirections at most.

\subsection{Relative density joint PDFs and bias functions}

We wish to compute the joint PDF of the density when expressed in terms of the survey average density $\rho_{s}$.
In order to do so, we consider the joint density $P(\rho_{s},\rho_{i},\rho_{j}),$ where $\rho_{s}$ is the density in the sample and $\rho_{i}$ and
$\rho_{j}$ are the densities in two cells at distance $d$.

We wish to compute the joint probability distribution function of $\hrho_{i}$ and $\hrho_{j}$ , defined as 
\begin{equation}
\hrho_{i}=\frac{\rho_{i}}{\rho_{s}},
\end{equation}
and the joint distribution functions of $\rhob_{i}$ and $\rhob_{j}$ , defined as 
\begin{equation}
\rhob_{i}=\rho_{i}-\rho_{s}+1.
\end{equation}
From these changes of variables, we have
\begin{equation}
P(\hrho_{i})=\int \dd\rho_{s}\,\rho_{s}\,P(\rho_{s},\hrho_{i}\rho_{s})
\end{equation}
and
\begin{equation}
P(\hrho_{i},\hrho_{j})=\int \dd\rho_{s}\,\rho_{s}^{2}\,P(\rho_{s},\hrho_{i}\rho_{s},\hrho_{j}\rho_{s}).
\end{equation}
Similarly, we also have
\begin{equation}
P(\rhob_{i})=\int \dd\rho_{s}\,P(\rho_{s},\rhob_{i}+\rho_{s}-1)
\end{equation}
and
\begin{equation}
P(\rhob_{i},\rhob_{j})=\int \dd\rho_{s}\,P(\rho_{s},\rhob_{i}+\rho_{s}-1,\rhob_{j}+\rho_{s}-1).
\end{equation}
We continue the calculations by expressing the joint PDF with the help of inverse Laplace transforms,
\begin{eqnarray}
P(\rho_{s},\rho_{i})&=&\int \frac{\dd\lambda_{s}}{2\pi\ii}\, \frac{\dd\lambda}{2\pi\ii}\,
\nonumber\\
&&\hspace{-.5cm}\times
\exp\left[
-\lambda_{s}\rho_{s}-\lambda\rho_{i}+\varphi(\lambda_{s},\lambda)
\right]\\
P(\rho_{s},\rho_{i},\rho_{j})&=&\int \frac{\dd\lambda_{s}}{2\pi\ii}\, \frac{\dd\lambda_{1}}{2\pi\ii}
\, \frac{\dd\lambda_{2}}{2\pi\ii}\,
\nonumber\\
&&\hspace{-1.5cm}\times
\exp\left[
-\lambda_{s}\rho_{s}-\lambda_{1}\rho_{i}-\lambda_{2}\rho_{j}+\varphi(\lambda_{s},\lambda_{1},\lambda_{2})
\right].
\end{eqnarray}
As a result,
\begin{eqnarray}
P(\hrho_{i})&=&
\int 
\frac{\dd\lambda_{s}}{2\pi\ii}\,  \frac{\dd\lambda}{2\pi\ii}
\frac{1}{(\lambda_{s}+\lambda\hrho_{i})^{2}}
\exp\left[\varphi(\lambda_{s},\lambda)
\right]\\
P(\hrho_{i},\hrho_{j})&=&
\int 
\frac{\dd\lambda_{s}}{2\pi\ii}\,  \frac{\dd\lambda_{1}}{2\pi\ii}
\, \frac{\dd\lambda_{2}}{2\pi\ii}
\frac{1}{(\lambda_{s}+\lambda_{1}\hrho_{i}+\lambda_{2}\hrho_{j})^{3}}\,
\nonumber\\
&&\hspace{-.5cm}\times
\exp\left[\varphi(\lambda_{s},\lambda_{1},\lambda_{2})
\right]
\end{eqnarray}
after integration over $\rho_{s}$. The latter expressions can be expressed as 
\begin{eqnarray}
P(\hrho_{i})&=&\int 
\frac{\dd\lambda}{2\pi\ii}
\left[
\frac{\partial\varphi}{\partial\lambda_{s}}
\right]_{\big\vert_{\lambda_{s}=-\lambda\hrho_{i}}}
\hspace{-.5cm}
\exp\left[\varphi(-\lambda\hrho_{i},\lambda_{i})\right]\\
P(\hrho_{i},\hrho_{j})&=&\int 
\frac{\dd\lambda_{1}}{2\pi\ii}\, \frac{\dd\lambda_{2}}{2\pi\ii}
\left[
\left(
\frac{\partial\varphi}{\partial\lambda_{s}}
\right)^{2}+\frac{\partial^{2}\varphi}{\partial\lambda_{s}^{2}}
\right]_{\big\vert_{\lambda_{s}=-\lambda_{1}\hrho_{i}-\lambda_{2}\hrho_{j}}}\hspace{-1cm}
\nonumber\\
&&\hspace{-.5cm}\times
\exp\left[\varphi(-\lambda_{1}\hrho_{i}-\lambda_{2}\hrho_{j},\lambda_{1},\lambda_{2})\right]
.\end{eqnarray}
In a similar manner, we can obtain the form of the joint PDF for $\{\rhob_{i}\}$,
\begin{eqnarray}
P(\rhob_{i})&=&\int 
\frac{\dd\lambda}{2\pi\ii}
\exp\left[-\lambda\rhob_{i}+\varphi(-\lambda,\lambda)\right]\\
P(\rhob_{i},\rhob_{j})&=&\int 
\frac{\dd\lambda_{1}}{2\pi\ii}\, \frac{\dd\lambda_{2}}{2\pi\ii}\,
\nonumber\\
&&\hspace{-1.5cm}\times
\exp\left[-\lambda_{1}\rhob_{i}-\lambda_{2}\rhob_{j}+\varphi(-\lambda_{1}-\lambda_{2},\lambda_{1},\lambda_{2})\right]
.\end{eqnarray}
We then use the relation (\ref{3ptbias}) to compute the form of these functions.

Noting that the expressions
$\int\dd\vx_{0}\dd\vx'_{0}\,\xi(\vx_{0},\vx'_{0})$, $\int\dd\vx_{0}\,\xi(\vx_{0},\vx_{1})$ take all the same averaged value when integrated over the sample, which we note $\xis$, then at linear order in $\xis$,
\begin{eqnarray}
\frac{\partial\varphi}{\partial\lambda_{s}}&=&1+\xis\left[\lambda_{s}+\varphi_{1}(\lambda_{1})+\varphi_{1}(\lambda_{2})\right].\\
\frac{\partial^{2}\varphi}{\partial\lambda_{s}^{2}}&=&\xis.
\end{eqnarray}
At the same order, we then have
\begin{equation}
\left(
\frac{\partial\varphi}{\partial\lambda_{s}}
\right)^{2}+\frac{\partial^{2}\varphi}{\partial\lambda_{s}^{2}}
=
1+\xis\left[
1+2\left(
\lambda_{s}+\varphi_{1}(\lambda_{1})+\varphi_{1}(\lambda_{2})
\right)
\right].
\end{equation}

Combining both the expressions of $P(\hrho_{i})$ and $P(\hrho_{i},\hrho_{j})$ and expanding all terms at linear order in $\xis$
, we obtain
\begin{eqnarray}
P(\hrho_{i},\hrho_{j})-P(\hrho_{i})P(\hrho_{j})&=&
\int
\frac{\dd\lambda_{1}}{2\pi\ii}\, \frac{\dd\lambda_{2}}{2\pi\ii}
\nonumber\\
&&\hspace{-3.5cm}\times
\left[
1+\xis\left(
1+\varphi_{1}(\lambda_{1})-\lambda_{1}\hrho_{i}
\right)\left(
1+\varphi_{1}(\lambda_{2})-\lambda_{2}\hrho_{j}
\right)
\right]\ 
\nonumber
\\
&& \hspace{-3.5cm}\times
\exp\left[-\lambda_{1}\hrho_{i}-\lambda_{2}\hrho_{j}+\varphi_{0}(\lambda_{1})+\varphi_{0}(\lambda_{2})\right].
\end{eqnarray}
This leads to the definition of the first sample bias function,
\begin{equation}
b_{\rm s1}(\hrho_{i})=\frac{1}{P(\hrho_{i})}
\int
\frac{\dd\lambda}{2\pi\ii}
\left(
1+\varphi_{1}(\lambda)-\lambda \hrho_{i}
\right)
\exp\left[-\lambda \hrho_{i}+\varphi_{0}(\lambda)\right]
,\end{equation}
which can be re-expressed in terms of the density-bias function defined in Eq. (\ref{biasdefinition}) and the derivative of $P(\hrho_{i})$ with respect to $\hrho_{i} $
\begin{equation}
b_{\rm s1}(\hrho_{i})=b(\hrho_{i})+1+\frac{\dd\log(P(\hrho_{i}))}{\dd\log \hrho_{i}}.
\end{equation}

The second sample-bias function can be obtained in a similar manner. We indeed have
\begin{eqnarray}
P(\rhob_{i},\rhob_{j})-P(\rhob_{i})P(\rhob_{j})&=&
\int
\frac{\dd\lambda_{1}}{2\pi\ii}\, \frac{\dd\lambda_{2}}{2\pi\ii}
\nonumber
\\
&& \hspace{-3.5cm} \times
\left[
1+\xis\left(
\varphi_{1}(\lambda_{1})-\lambda_{1}
\right)\left(
\varphi_{1}(\lambda_{2})-\lambda_{2}
\right)
\right]\ 
\nonumber
\\
&& \hspace{-3.5cm} \times
\exp\left[-\lambda_{1}\rhob_{i}-\lambda_{2}\rhob_{j}+\varphi_{0}(\lambda_{1})+\varphi_{0}(\lambda_{2})\right]
,\end{eqnarray}
which eventually leads to\begin{equation}
b_{\rm s2}(\rhob_{i})=b(\rhob_{i})+\frac{\dd\log(P(\rhob_{i}))}{\dd\rhob_{i}}.
\end{equation}

\subsection{Response to a change in amplitude in $\xi$}

A close notion related to the density-bias function is how the PDF is changed when the parameters of the simulations are changed. In particular for
tree models, the statistical properties are entirely determined by the amplitude of the two-point function, for instance, at cell size.
This dependence can be made explicit by writing Eq. (\ref{Prho}) as
\begin{equation}
P(\rho_{i},\xib)=\frac{1}{\xib}\int\frac{\dd \hlambda}{2 \pi \ii}\,\exp\left(-\hlambda\frac{\rho_{i}}{\xib}+\frac{1}{\xib}\psi(\hlambda)\right)
\end{equation}
after the change of variable and function,
\begin{equation}
\hlambda=\xib\lambda,\ \ \psi(\hlambda)=\xib\varphi(\lambda),
\end{equation}
where then the expression $\psi(\hlambda)$ does not depend on $\xib$ (only on the functional form of $\zeta$).
It follows that
\begin{eqnarray}
-\xib\frac{\partial P(\rho_{i},\xib)}{\partial\xib}&=&P(\rho_{i},\xib)+\rho\frac{\partial P(\rho_{i},\xib)}{\partial\rho_{i}}
\nonumber
\\
&& \hspace{-.5cm}
+\int\frac{\dd \lambda}{2 \pi \ii}\,\varphi(\lambda)\exp\left(-\hlambda\frac{\rho_{i}}{\xib}+\varphi(\lambda)\right).
\end{eqnarray}

This expression can be used to defined the function $b_{\xi}(\rho_{i})$ as 
\begin{equation}
b_{\xi}(\rho_{i})=-\frac{\partial \log P(\rho_{i},\xib)}{\partial\log\xib}.
\end{equation}
It appears that $b_{\xi}(\rho_{i})$ is very similar to $b_{\rm s1}(\rho_{i}),$ but the two are not equal in general.

\subsection{Close cell results}
\subsubsection{CGF for 2 close cells}

Saddle point approximation and close cell results.
In case of two cells, the general system in the mean-field approximation leads to
\begin{eqnarray}
\tau_{1}&=&\lambda_{1}\xib\zeta'(\tau_{1})+\lambda_{2}\xi_{12}\zeta'(\tau_{2})\\
\tau_{2}&=&\lambda_{2}\xib_{12}\zeta'(\tau_{1})+\lambda_{2}\xib\zeta'(\tau_{2})
\end{eqnarray}
and
\begin{eqnarray}
\varphi(\lambda_{1},\lambda_{2})&=&\lambda_{1}\left(\zeta(\tau_{1})-\frac{1}{2}\tau_{1}\zeta'(\tau_{1})\right)
\nonumber
\\
&& \hspace{-.5cm}
+\lambda_{2}\left(\zeta(\tau_{2})-\frac{1}{2}\tau_{2}\zeta'(\tau_{2})\right).
\end{eqnarray}
We are interested here in the behavior of $\varphi(\lambda_{1},\lambda_{2})$ when the two cells are close, that is, 
when $\xi_{12}\to\xib$. When $\xi_{12}=\xib$, $\tau_{1}$ and $\tau_{2}$ are also equal, making 
$\varphi(\lambda_{1},\lambda_{2})$ a sole function of $\lambda_{1}+\lambda_{2}$ and therefore forcing the joint PDF to be proportional to $\Dirac(\rho_{i}-\rho_{j})$. To be more precise, in this regime, $\xi_{12}\to\xib$,  
$\drho\equiv(\rho_{i}-\rho_{j})/2$ is expected to be distributed with a width of about $\Delta^{1/2}_{\xi}=(\xib-\xi_{12})^{1/2}$. This suggests that in this limit, the difference $\lambda_{1}-\lambda_{2}$ should be treated as a large quantity of about $(\xib-\xi_{12})^{-1/2}$. The limit behavior of the joint CGF can then be explicitly computed in terms of 
\begin{eqnarray}
\lambda&=&\lambda_{1}+\lambda_{2},\\
\mu&=&\lambda_{1}-\lambda_{2},\\
\Delta_{\xi}&=&\xib-\xi_{12}.
\end{eqnarray}
In this limit, we obtain
\begin{eqnarray}
\tau&=&\lambda\,\xib\,\zeta'(\tau)+\mu\,\xib\,\zeta''(\tau)\,\delta\tau,\\
\delta\tau&=&\frac{\mu}{2}\,\Delta_{\xi}\,\zeta'(\tau),
\end{eqnarray}
leading to
\begin{equation}
\tau=\lambda\,\xib\,\zeta'(\tau)+\frac{\mu^{2}\xib}{4}\,\Delta_{\xi}\,\left(\zeta'(\tau)^{2}\right)'
\end{equation}
and
\begin{eqnarray}
\varphi(\lambda_{1},\lambda_{2})&=&\lambda\left(\zeta(\tau)-\frac{1}{2}\tau\zeta'(\tau)\right)
\nonumber
\\
&& \hspace{-.5cm}
+\frac{\mu^{2}}{4}\,\Delta_{\xi}\,
\left(\zeta'(\tau)^{2}-\frac{1}{2}\tau
\left(\zeta'(\tau)^{2}\right)'
\right).
\end{eqnarray}
The joint PDF of $\rhom=(\rho_{i}+\rho_{j})/2$ and $\drho$ then reads
\begin{eqnarray}
P(\rhom,\drho)=
\int
\frac{\dd\lambda}{2\pi\ii}\, \frac{\dd\mu}{2\pi\ii}
\exp\left[-\lambda\rhom-\mu\drho+\varphi(\lambda,\mu)\right]
\label{Prhomdrho}
,\end{eqnarray}
for which there is in general no  closed form. We then need to rely on approximation schemes to complete
the calculations.

\subsubsection{Saddle point approximation}

One of the approximations that can be used to evaluate Eq. (\ref{Prhomdrho}) is to use the saddle point approximation.
It has been used in the literature to compute the PDF \citep[see][]{1989A&A...220....1B,1992ApJ...392....1B,2002A&A...382..412V,2016MNRAS.460.1598C}. It is a priori valid when $\xib$ is small (and not for too 
high values of the density).
In this approach, the expression under the exponential is approximated by a quadratic form at its minimum. In practice, 
the latter is obtained from the resolution of the system
\begin{eqnarray}
\frac{\partial \varphi(\lambda,\mu)}{\partial\lambda}&=&\rhom\\
\frac{\partial \varphi(\lambda,\mu)}{\partial\mu}&=&\drho
,\end{eqnarray}
which leads to the implicit or explicit values of $\lambda$, $\mu$, and $\tau$ at the saddle point position (we hereafter 
denote this with the subscript $s$),
\begin{equation}
\rhom=\zeta(\tau_{s}),\ \ \drho=\mu_{s}\,\frac{\Delta_{\xi}}{2}\left(\zeta'(\tau_{s})\right)^{2}.
\end{equation}
It is to be noted here that the value of $\tau_{s}$ is independent of $\drho$. At the saddle point position, we then have
\begin{equation}
-\lambda_{s}\rhom_{s}-\mu_{s}\drho+\varphi(\lambda_{s},\mu_{s})=
-\frac{\drho^{2}}{\Delta_{\xi}\left[\zeta'(\tau_{s})\right]^{2}}-\frac{\tau_{s}^{2}}{2}.
\end{equation}
This then suggests that the joint PDF is given by the product of the one-point PDF of $\rhom$ and a Gaussian distribution
of $\drho$ of width $\Delta^{1/2}_{\xi}\zeta'(\tau_{s})/\sqrt{2}$. For usual models, as described above, $\zeta'(\tau_{s})^{2}$ scales
like a power of $\rhom$ so that one suggested form for the joint PDF is the following:
\begin{equation}
P(\rhom,\drho)=P(\rhom)\exp\left(-\frac{\drho^{2}}{\Delta_{\xi}\rho_{m}^{\alpha}}\right)\frac{1}{\sqrt{\pi\Delta_{\xi}\rhom^{\alpha}}}.
\end{equation}
Interestingly, the value of $\alpha$ can be related to the reduced skewness of the density field from the computation of
$\left<\drho^{2}\rho_{m}\right>_{c}$ , and in the context of tree hierarchical models, it leads to
\begin{equation}
\alpha=\frac{2}{3}S_{3}.
\end{equation}
The validity of this form clearly ought to be checked. Its simplicity nonetheless offers a good grasp of the contribution 
of close cells to the covariance matrix.

\section{Minimal tree model}

In the previous section, general formulae were given. 
The aim of this section is to account for more precise results obtained in the case of a specific hierarchical model, namely the minimal tree model, as described below. It can then serve as a toy model for the construction of the approximate form for the covariance matrix.
We first recall that this model describes the Rayleigh Levy flights model.

\subsection{One-point results in the mean-field approximation}

The Rayleigh-Levy flight model makes it possible to build synthetic samples whose statistical properties follow the minimal model, that is, a hierarchical model with the following vertex-generating function:
\begin{equation}
\zRL(\tau)=1+\tau+\frac{1}{4}\tau^{2}.
\end{equation}

In the one-cell mean-field approximation, we have the equation
\begin{equation}
\tau=\lambda\xib\zRL'(\tau)
,\end{equation}
which can be solved in
\begin{equation}
\tau=\frac{\lambda\xi}{1-\lambda\,\xi/2}
,\end{equation}
which leads to the following expression for the CGF:\begin{equation}
\varphi(\lambda)=\frac{\lambda}{1-\lambda\,\xib/2}.
\end{equation}

The one-point PDF of the density can then be computed explicitly. It takes the form in the continuous limit of
\begin{eqnarray}
P(\rho)&\!\!=&\!\!\int\frac{\dd\lambda}{2\pi\ii}\,e^{-\lambda\rho+\varphi(\lambda)}
\nonumber\\
&\!\!=&\!\!e^{-\frac{2}{\xib}}\,\Dirac(\rho)+\frac{4}{\xib^{2}}e^{-\frac{2}{\xib}(1+\rho)} \ _{0}F_{1}\left(2,\frac{4\rho}{\xib}\right)
.\end{eqnarray}
For this particular model, the void probability distribution (VPF) is nonzero even in the 
continuous limit. We recall here that the general expression of the VPF is given by $\exp(\varphi(-\Nb)),$ which for the minimal model 
leads to
\begin{equation}
P_{0}=\exp\left(-2/\xib\right)
\end{equation}
when $\Nb\to\infty$.

The density-bias function can also be computed explicitly. For the minimal model, we have $\varphi_{1}(\lambda)=\varphi(\lambda)$
so that
\begin{eqnarray}
b_{1}(\rho)&=&\frac{1}{P(\rho)}\int\frac{\dd\lambda}{2\pi\ii}\varphi_{1}(\lambda)\,
e^{-\lambda\rho+\varphi(\lambda)}\nonumber\\
&=&\frac{\ _{0}F_{1}\left(1,\frac{4\rho}{\xib}\right)}{\ _{0}F_{1}\left(2,\frac{4\rho}{\xib}\right)}-\frac{2}{\xib}
\end{eqnarray}
for $\rho>0$.
For this model, the fact that $\varphi_{1}(\lambda)=\varphi(\lambda)$ implies that
\begin{equation}
b_{\rm s1}(\rho)=b_{\xi}(\rho).\label{MinimalIdentitybxi}
\end{equation}
This means that in the case of the minimal model, the density-bias function can be extracted from the functional form
of the one-point PDF as
\begin{equation}
b(\rho)P(\rho)=\left[-1-\frac{\dd}{\dd\log\rho}-\frac{\dd}{\dd\log\xib}\right]P(\rho,\xib)\label{MTbiasIndentity}
.\end{equation}
This is a somewhat remarkable identity (which can be extended to higher orders, as shown below.)

In this case, the second-order expansion leads to the form $\varphi_{2}(\lambda)$ given by
\begin{equation}
\varphi_{2}(\lambda)=\frac{1}{2}\varphi_{1}(\lambda)
,\end{equation}
and we note that $\varphi_{p}(\lambda)$ all vanish for $p\ge 3$.

\subsection{Two-cell results in the mean-field approximation}

These mean-field calculations can be extended to the two-cell case. In this case, we have the system
\begin{eqnarray}
\tau_{1}&=&\lambda_{1}\,\xib\,\zRL'(\tau_{1})+\lambda_{2}\,\xi_{12}\,\zRL'(\tau_{2})\\
\tau_{2}&=&\lambda_{1}\,\xi_{12}\,\zRL'(\tau_{1})+\lambda_{2}\,\xib\,\zRL'(\tau_{2})
\end{eqnarray}
when the two cells are of the same size.
This leads to the following expression for the joint CGF:
\begin{equation}
\varphi(\lambda_{1},\lambda_{2})=
\frac{\lambda_{1}+\lambda_{2}+(\xi_{12}-\xib)\lambda_{1}\lambda_{2}}{1-(\lambda_{1}+\lambda_{2})\,\xib/2-\lambda_{1}\lambda_{2}\,(\xi_{12}^{2}-\xib^{2})/4}.\label{MeanFPhi21}
\end{equation}
Remarkably, this expression can be written in terms of the one-cell CGF as
\begin{equation}
\varphi(\lambda_{1},\lambda_{2})=\frac{\varphi(\lambda_{1})+\varphi(\lambda_{2})+\xi_{12}\,\varphi(\lambda_{1})\varphi(\lambda_{2})}{1-\xi_{12}^{2}\,\varphi(\lambda_{1})\varphi(\lambda_{2})/4}.\label{MeanFPhi22}
\end{equation}
This opens the possibility of computing the joint PDF to any order of $\xi_{12}$. The calculation of this expansion is made simple
by the following observations:
The corrective terms will make intervene functions of the forms\begin{equation}
b_{n}(\rho)\,P(\rho)\equiv \int\frac{\dd \lambda}{2 \pi \ii}\left[\varphi(\lambda)\right]^{n}\ \exp\left(-\lambda\rho+\varphi(\lambda)\right).
\end{equation}
We further note that
\begin{equation}
\frac{\dd}{\dd \lambda}\varphi(\lambda)=\left(\frac{\xib}{2}\right)^{2}\left[\psi(\lambda)\right]^{2}
\end{equation}
with
\begin{equation}
\psi(\lambda)=\varphi(\lambda)+\frac{2}{\xib}.
\end{equation}
We the define $c_{n}(\rho)$ as
\begin{equation}
c_{n}(\rho)\,P(\rho)\equiv \int\frac{\dd \lambda}{2 \pi \ii}\left[\psi(\lambda)\right]^{n}\ \exp\left(-\lambda\rho+\varphi(\lambda)\right).
\end{equation}
We then have on one side
\begin{equation}
b_{n}(\rho)=\sum_{q=0}^{n}C_{n}^{q}\left(-\frac{2}{\xib}\right)^{q}c_{n-q}(\rho)
,\end{equation}
and on the other side
\begin{equation}
\rho\,c_{n}(\rho)=\left(\frac{\xib}{2}\right)^{2}\left(c_{n+2}(\rho)+nc_{n+1}(\rho)\right)
,\end{equation}
which derives from the fact that
\begin{equation}
\rho\, c_{n}(\rho)=\int\frac{\dd \lambda}{2 \pi \ii}\exp\left(-\lambda\rho\right)\frac{\dd}{\dd \lambda}
\left\{\left[\psi(\lambda)\right]^{n}\exp\left(\varphi(\lambda)\right)\right\}
\end{equation}
after integration by parts. As a result, the expression of the join PDF to any order can be written as polynomials making intervene 
$P(\rho_{1})$, $b(\rho_{1})$ $P(\rho_{2})$ and $b(\rho_{1})$ only.

\subsection{Perturbative expansion for close cells}

Another interesting result is when the cell centers are close (so that cells overlap), as described above. In this case,
the limit behavior of the joint CGF is given by
\begin{equation}
\varphi(\lambda_{1},\lambda_{2})=\frac{\lambda+\mu^{2}\,\Delta_{\xi}/4}{1-\lambda\,\xib/2-\mu^{2}\,\Delta_{\xi}\,\xib/8}
,\end{equation}
with
\begin{eqnarray}
\lambda&=&\lambda_{1}+\lambda_{2},\\
\mu&=&\lambda_{1}-\lambda_{2},\\
\Delta_{\xi}&=&\xib-\xi_{12}.
\end{eqnarray}
It is then remarkable to see the result can be expressed with the sole one-cell CGF,
\begin{equation}
\varphi(\lambda_{1},\lambda_{2})=\varphi(\lambda+\mu^{2}\,\Delta_{\xi}/4).
\label{CGFlm}
\end{equation}
In other words, the GFC of the variables $\rhom=(\rho_{1}+\rho_{2})/2$ and $\drho=(\rho_{1}-\rho_{2})/2$ is given by Eq. (\ref{CGFlm}).
It is possible to compute the joint PDF,
\begin{eqnarray}
P(\rhom,\drho)&=&\int\frac{\dd\lambda}{2\pi\ii}\int\frac{\dd\mu}{2\pi\ii}
\nonumber
\\&& \hspace{-.5cm}
\times \exp\left(
-\lambda\,\rhom-\mu\,\drho +\varphi(\lambda+\mu^{2}\,\Delta_{\xi}/4)\right)
,\end{eqnarray}
with the change of variable
\begin{equation}
\tilde\lambda=\lambda+\mu^{2}\,\frac{\Delta_{\xi}}{4}.
\end{equation}
The integral in $\tilde\lambda$ leads to the one-cell PDF of the density $\rhom$ , whereas the integral in
$\mu$ can be done explicitly as it is a quadratic form in $\mu$ , leading to a Gaussian distribution in $\drho$. The final
PDF is given by
\begin{equation}
P(\rhom,\delta\rho)=P(\rhom)\,
\frac{1}{\left[\pi\Delta_{\xi}\rhom\right]^{1/2}}\exp\left(-\frac{\drho^{2}}{\Delta_{\xi}\,\rhom}\right).\label{JPDFrdr}
\end{equation}
This shows that the joint PDF peaks for $\rho_{1}\sim\rho_{2}$ with a width that depends on the distance between the cells
through the difference $\xib-\xi_{12}$. Moreover, this form has no overlapping regime with the previous expansions of 
the joint PDF. It captures different aspects of the covariance calculations as listed below.
\begin{itemize}
\item The previous expression says that close cells contribute more specifically to the covariance when $\rho_{1}$ and $\rho_{2}$ are close. This suggests that Eq. (\ref{JPDFrdr}) contributes mostly to the near diagonal terms, whereas off diagonal terms could still be well described by perturbative expansions, as described before. 
\item As noted before, perturbative expansions are closely related to supersample effects. They encode the way in which the local densities are jointly correlated with long-wavelength modes. This is not the case in Eq. (\ref{JPDFrdr}). It rather captures how a rare event, such as a peak, can contribute to the covariance elements: if there is a peak somewhere, nearby cells are likely to have a similar density up to distances for which  $\xib-\xi_{12}$ remains small enough.
\end{itemize}

The above development can be pursued to any order in $\Delta^{1/2}_{\xi}$ provided the following recipe is applied:
\begin{equation}
\delta_{\rho}\sim\Delta^{1/2}_{\xi}\ \ \ \hbox{and}\ \ \ \mu\sim\frac{1}{\Delta^{1/2}_{\xi}}.
\end{equation}
Then the joint density can be computed to any order in $\Delta^{1/2}_{\xi}$, making use of
the very same expressions $b_{n}(\rho)$. 

\begin{figure}
\centering
 \includegraphics[width=7cm]{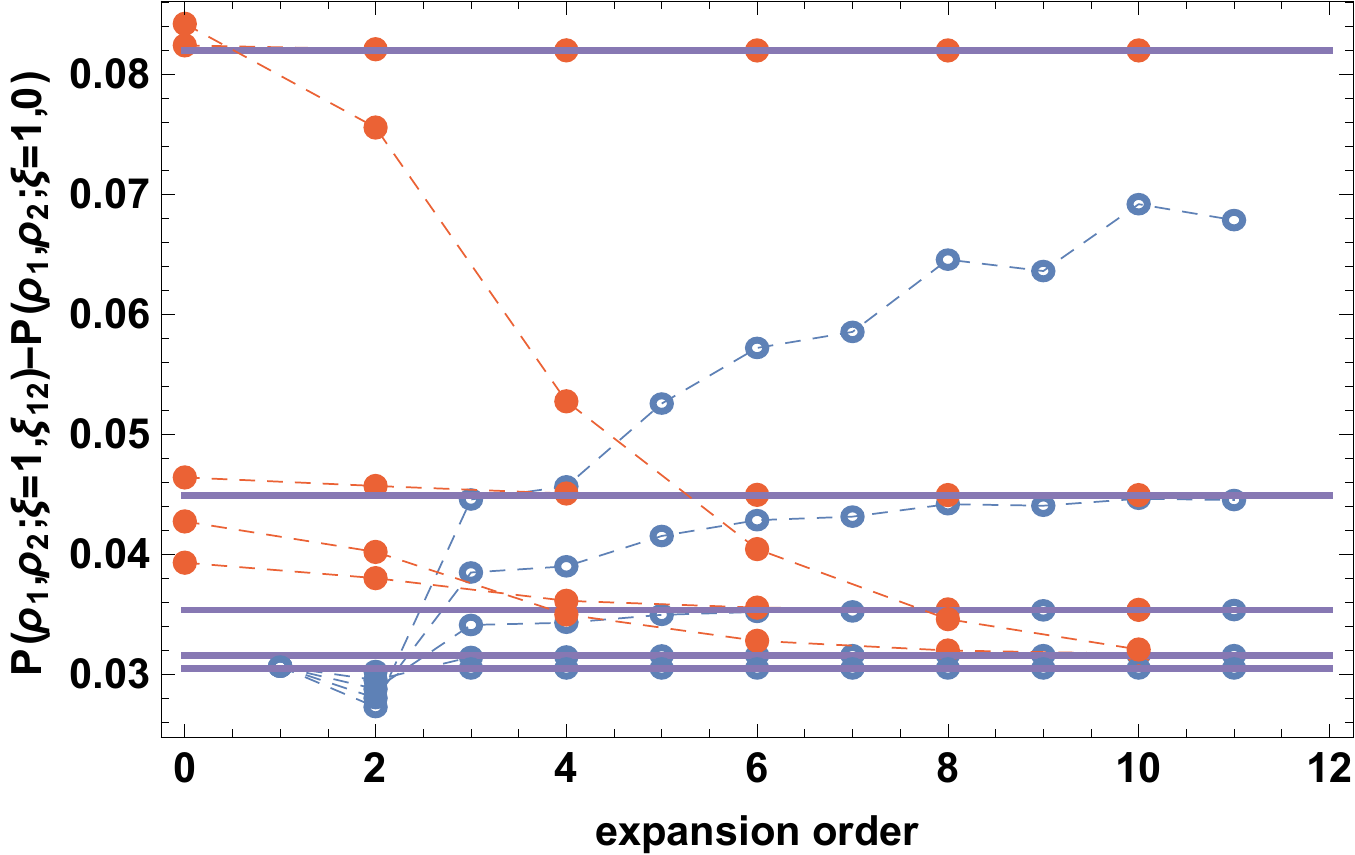}
 \hspace{1cm}
 \includegraphics[width=7cm]{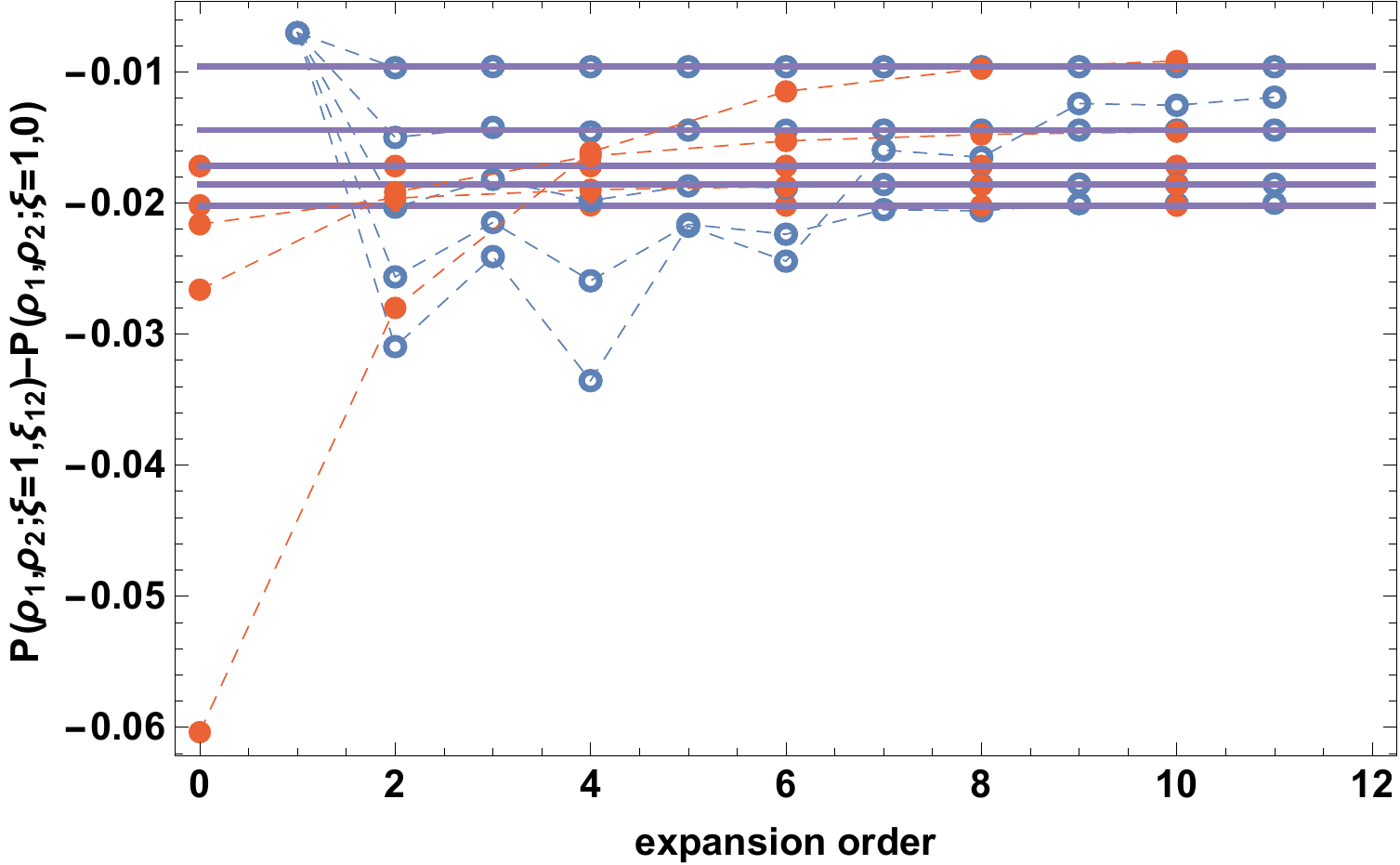}
   \caption{Performances of the perturbative expansions of the joint PDF $P_{\rm MF}(\rho_{1},\rho_{2};\xib,\xi_{12})$ in the mean-field 
   approximation either for the $\xi_{12}$ expansion (open blue dots) or the $\Delta^{1/2}_{\xi}$ expansion (red dots) up to 11th and 10th order, respectively. The comparisons are made for $\rho_{1}=\rho_{2}=2$ and $\xib=1$ (top panel) and for  $\rho_{1}=0.5,\ \rho_{2}=3.5$ and $\xib=1$ (bottom panel) and for $\xi_{12}$ equalling $0.1$, $0.3$, $0.5$, $0.7,$ and $0.9$.}
   \label{Convergence_jPDF}
\end{figure}

The next-to-leading order in  $\Delta^{1/2}_{\xi}$ is thus given by
\begin{eqnarray}
P^{(2)}(\rhom,\delta\rho)&=&P^{(0)}(\rhom,\delta\rho)
\nonumber\\
&&\hspace{-3cm}
\times\left\{
\frac{\delta _{\rho }^2}{2 \bar{\xi } \rho _m}-\frac{\rho _m \Delta _{\xi }}{\bar{\xi }^2}-\frac{\Delta _{\xi }}{4
   \bar{\xi }}
 -\frac{\Delta _{\xi }}{\bar{\xi }^2}-\frac{\delta _{\rho }^4}{4 \rho _m^3 \Delta _{\xi }}+\frac{3 \delta
   _{\rho }^2}{4 \rho _m^2}-\frac{3 \Delta _{\xi }}{16 \rho _m}
\right.
\nonumber\\
&&\hspace{-2.5cm}   
\left.   +\frac{\left(\bar{\xi } \rho _m \Delta _{\xi }-2 \bar{\xi } \delta _{\rho }^2+8 \rho _m^2 \Delta _{\xi }\right)}{8 \bar{\xi } \rho _m^2}
   \frac{_0{F}_1\left(1,\frac{4 \rho_m}{\bar{\xi }^2}\right) }{_0{F}_1\left(2,\frac{4 \rho_m}{\bar{\xi }^2}\right)}
\right\}
,\end{eqnarray}
and the expansion can be extended in any (even) order in $\Delta^{1/2}_{\xi}$. Fig. \ref{Convergence_jPDF}
illustrates the convergence properties of these expansions. Depending on $\xi_{12}/\xib$, either the expansion in 
$\xi_{12}$ or that in $\Delta^{1/2}_{\xi}$ gives a very accurate estimate of the joint PDF. It opens the way to computing the covariance matrix starting in the two-cell mean-field approximation (\ref{MeanFPhi21}).

\subsection{Construction of the theoretical covariance matrix for the minimal tree model}

The previous form can be used to compute the covariance matrix for the minimal tree model in simple implementations. It relies
on analytic forms for both the two-point 
cell correlation functions, which can formally be written as 
\begin{equation}
\xi_{12}(r)=\int\dd^{2}\vk \,W_{2D}^{2}(k\,R)\,J_{0}(k\,r)\,P(k)
\end{equation}
for a given power spectrum. We also make use of the form $P_{s}(r_{d})$ given in footnote 2 
to derive the PDF of cell distances. We then have all the required ingredients to compute  the elements of the covariance matrix in the mean-field approximation,
\begin{equation}
\Cov_{\rm MF}(\rho_{1},\rho_{2})=\int\dd r\,P_{d}(r)\,P_{\rm MF}^{(8)}(\rho_{1},\rho_{2};\xib,\xi_{12}(r L_{\rm sample}))
.\end{equation}
In practice, $P_{\rm MF}^{(8)}(\rho_{1},\rho_{2};\xib,\xi_{12})$ is computed from the 
eighth-order expansion either in $\xi_{12}$ when $\xi_{12}/\xib<0.4$  or in $\Delta_{\xi}$ when $\xi_{12}/\xib > 0.4$.
This is used to explore the detailed properties of the covariance matrix and the validity of approximate schemes.

\subsection{Joined PDF for relative densities}

The minimal model allows us also to pursue the computation of the joint PDF for the variables $\{\hrho_{i}\}$ 
or $\{\rhob_{i}\}$ in all regimes. The first step is to extend Eq. (\ref{3ptbias}) to a regime in which $\xi_{12}$ is not
assumed to be small. We find that
\begin{eqnarray}
\varphi(\lambda_{s},\lambda_{1},\lambda_{2})&=&
\lambda_{s}+\varphi(\lambda_{1},\lambda_{2})
+\frac{\lambda_{s}^{2}}{2}\int\dd\vx_{s}\dd\vx'_{s}\,\xi(\vx_{s},\vx'_{s})
\nonumber\\
&&
\hspace{-2cm}
+\lambda_{s}\int\dd\vx_{s}\,\xi(\vx_{s},\vx_{1})\,\varphi_{c}(\lambda_{1},\lambda_{2})
 \nonumber\\
&&
\hspace{-2cm}
+\lambda_{s}\int\dd\vx_{s}\,\xi(\vx_{s},\vx_{2})\,\varphi_{c}(\lambda_{2},\lambda_{1})
,\end{eqnarray}
where $\varphi_{c}(\lambda_{1},\lambda_{2})$ is given by
\begin{equation}
\varphi_{c}(\lambda_{1},\lambda_{2})=
\frac{\lambda_{1}+(\xi_{12}-\xib)\lambda_{1}\lambda_{2}/2}{1-(\lambda_{1}+\lambda_{2})\,\xib/2-\lambda_{1}\lambda_{2}\,(\xi_{12}^{2}-\xib^{2})/4}
\label{MeanFCPhi21}
,\end{equation}
and we can note that
\begin{equation}
\varphi_{c}(\lambda_{1},\lambda_{2})+\varphi_{c}(\lambda_{2},\lambda_{1})=\varphi(\lambda_{1},\lambda_{2}).
\end{equation}
At leading order in $\xi_{s}$, that is, when we assume that the density fluctuations at sample size are much smaller than at smoothing scale, this expression then reduces to 
\begin{equation}
\varphi(\lambda_{s},\lambda_{1},\lambda_{2})=
\lambda_{s}+\frac{\lambda_{s}^{2}}{2}\xi_{s}+\varphi(\lambda_{1},\lambda_{2})(1+\lambda_{s}\xi_{s}).\label{c2ptbias}
\end{equation}

We can then exploit this relation to compute the $P(\hrho_{i},\hrho_{j})$ and $P(\rhob_{i},\rhob_{j})$ from Eqs. (\ref{joinhrhoPDF})  and (\ref{joinrhobPDF}), respectively.
We then have at leading order in $\xi_{s}$
\begin{eqnarray}
P_{s1}(\hrho_{i},\hrho_{j})&=&P(\hrho_{i},\hrho_{j})+\xi_{s}\int
\frac{\dd\lambda_{1}}{2\pi\ii}\, \frac{\dd\lambda_{2}}{2\pi\ii}
\nonumber\\
&& \hspace{-1.5cm}\times\
\left(
1+2\lambda_{s}+\frac{1}{2}\lambda_{s}^{2}+(2+\lambda_{s})\varphi(\lambda_{1},\lambda_{2})
\right)\ 
\nonumber\\
&& \hspace{-1.5cm}\times
\exp\left[\lambda_{s}+\varphi(\lambda_{1},\lambda_{2})\right]_{\big\vert_{\lambda_{s}=-\hrho_{i}\lambda_{1}-\hrho_{j}\lambda_{2}}}
\end{eqnarray}
and
\begin{eqnarray}
P_{s2}(\rhob_{i},\rhob_{j})&=&P(\rhob_{i},\rhob_{j})\nonumber\\
&& \hspace{-1.5cm}+\xi_{s}\int
\frac{\dd\lambda_{1}}{2\pi\ii}\, \frac{\dd\lambda_{2}}{2\pi\ii}
\left(
\frac{1}{2}\lambda_{s}^{2}+\lambda_{s}\varphi(\lambda_{1},\lambda_{2})
\right)_{\big\vert_{\lambda_{s}=-\lambda_{1}-\lambda_{2}}}
\ 
\nonumber\\
&& \hspace{-1.5cm}\times
\exp\left[-\lambda_{1}\rhob_{i}-\lambda_{2}\rhob_{j}+\varphi(\lambda_{1},\lambda_{2})\right].
\end{eqnarray}

To complete the formal calculation of these expressions, we introduce the function 
\begin{eqnarray}
P_{b}(\rho_{i},\rho_{j},\xib,\xi_{12})&=&\int
\frac{\dd\lambda_{1}}{2\pi\ii}\, \frac{\dd\lambda_{2}}{2\pi\ii}
\varphi(\lambda_{1},\lambda_{2})\ 
\nonumber\\
&& \hspace{-1.5cm}\times
\exp\left[-\lambda_{1}\rho_{i}-\lambda_{2}\rho_{j}+\varphi(\lambda_{1},\lambda_{2})\right]
.\end{eqnarray}
We can first note that Eq. (\ref{MTbiasIndentity}) can be extended to 
\begin{eqnarray}
P_{b}(\rho_{i},\rho_{j},\xib,\xi_{12})&=&
\left(-2-\frac{\dd}{\dd\log\rho_{i}}\right.\nonumber\\
&& \hspace{-3.cm}
\left.-\frac{\dd}{\dd\log\rho_{j}}-\frac{\dd}{\dd\log\xib}-\frac{\dd}{\dd\log\xi_{12}}\right)
P(\rho_{i},\rho_{j},\xib,\xi_{12})
.\end{eqnarray}
This comes from the observation that
\begin{eqnarray}
P(\rho_{i},\rho_{j},\xib,\xi_{12},\eta)&\equiv&
\int
\frac{\dd\lambda_{1}}{2\pi\ii}\, \frac{\dd\lambda_{2}}{2\pi\ii}\ 
\nonumber\\
&& \hspace{-1.5cm}\times\exp\left[-\lambda_{1}\rho_{i}-\lambda_{2}\rho_{j}+\eta\varphi(\lambda_{1},\lambda_{2},\xib,\xi_{12})\right]
\nonumber
\end{eqnarray}
can also be written 
\begin{eqnarray}
P(\rho_{i},\rho_{j},\xib,\xi_{12},\eta)&=&\frac{1}{\eta^{2}}
\int
\frac{\dd\hlambda_{1}}{2\pi\ii}\, \frac{\dd\hlambda_{2}}{2\pi\ii}
\nonumber\\
&& \hspace{-1.5cm}\times
\exp\left[-\hlambda_{1}\frac{\rho_{i}}{\eta}-\hlambda_{2}\frac{\rho_{j}}{\eta}+\varphi\left(\hlambda_{1},\hlambda_{2},\frac{\xib}{\eta},\frac{\xi_{12}}{\eta}\right)\right]\nonumber\\
&=&\frac{1}{\eta^{2}}P\left(\frac{\rho_{i}}{\eta},\frac{\rho_{j}}{\eta},\frac{\xib}{\eta},\frac{\xi_{12}}{\eta},1\right)
\end{eqnarray}
and that
\begin{equation}
P_{b}(\rho_{i},\rho_{j},\xib,\xi_{12})=\frac{\partial}{\partial\eta}_{\big\vert_{\eta=1}}P(\rho_{i},\rho_{j},\xib,\xi_{12},\eta).
\end{equation}
The final expression of the PDF of the relative densities can then be obtained by noting that applying a multiplicative factor $\lambda_{i}$ to the moment-generating function is equivalent to the application of the operator $\partial/\partial\rho_{i}$ to the final expression.
this finally leads to the following forms:
\begin{eqnarray}
P_{s1}(\hrho_{i})&=&P(\hrho_{i})+\xi_{s}\,\left[\left(\hrho_{i}\frac{\partial}{\partial\hrho_{i}}+\frac{\hrho_{i}^{2}}{2}\frac{\partial^{2}}{\partial\hrho_{i}^{2}}\right)P(\hrho_{i})\right.\nonumber\\
&& \hspace{-1.cm}
\left.+\left(1+\hrho_{i}\frac{\partial}{\partial\hrho_{i}}\right)(b(\hrho_{i})P(\hrho_{i}))\right]\\
P_{s1}(\hrho_{i},\hrho_{j})&=&P(\hrho_{i},\hrho_{j})
\nonumber\\
&& \hspace{-2.cm}
+\xi_{s}\,\left[
\left(1+2\sum_{i}\hrho_{i}\frac{\partial}{\partial\hrho_{i}}+\frac{1}{2}\sum_{i}\hrho_{i}^{2}
\frac{\partial^{2}}{\partial\hrho_{i}^{2}}\right)P(\hrho_{i},\hrho_{j})
\right.\nonumber\\
&& \hspace{-2.cm}
\left.
+\left(2+\sum_{i}\hrho_{i}\frac{\partial}{\partial\hrho_{i}}\right)P_{b}(\hrho_{i},\hrho_{j})\right]
\end{eqnarray}
for the $\hrho_{i}=\rho_{i}/\rho_{s}$ and
\begin{eqnarray}
P_{s2}(\rhob_{i})&=&P(\rhob_{i})+\xi_{s}\,\left[\frac{1}{2}\frac{\partial^{2}}{\partial\rhob_{i}^{2}}P(\rhob_{i})
\right.\nonumber\\
&& \hspace{-.5cm}
\left.
+\frac{\partial}{\partial\rhob_{i}}(b(\rhob_{i})P(\rhob_{i}))\right]\\
P_{s2}(\rhob_{i},\rhob_{j})&=&
P(\rhob_{i},\rhob_{j})+\xi_{s}\,\left[\frac{1}{2}\sum_{i}\frac{\partial^{2}}{\partial\rhob_{i}^{2}}
P(\rhob_{i},\rhob_{j})
\right.\nonumber\\
&& \hspace{-.5cm}
\left.
+\sum_{i}\frac{\partial}{\partial\rhob_{i}}P_{b}(\hrho_{i},\hrho_{j})\right]
\end{eqnarray}
for $\rhob_{i}=\rho_{i}-\rho_{s}+1$. These relations can then be applied to the expressions of the joint density such as $P_{\rm MF}^{(8)}(\rho_{i},\rho_{j};\xib,\xi_{12})$  found in the previous subsection.

\end{document}